\documentclass[5p,lefttitle]{elsarticle}

\usepackage[english]{babel}
\usepackage{amssymb,amsthm,amsmath, amsbsy}
\usepackage{bm}
\usepackage{textcomp}
\usepackage[utf8]{inputenc}
\usepackage[T1]{fontenc}
\usepackage{hyperref}
\hypersetup{colorlinks = true, pdfauthor=author}
\usepackage{xcolor}
\definecolor{linkblue}{RGB}{33,150,209}
\usepackage[super]{nth}
\begin{document}

\hypersetup{linkcolor  = linkblue,citecolor  = linkblue,urlcolor   = linkblue}
\title{Concepts and use cases for picosecond ultrasonics with x-rays}

\author[1]{Maximilian Mattern}
\author[1]{Alexander von Reppert\corref{cor1}}
\ead{reppert@uni-potsdam.de}
\author[1,2]{Steffen Peer Zeuschner}
\author[1]{Marc Herzog}
\author[1,2,3]{Jan-Etienne Pudell}
\author[1,2]{Matias Bargheer}

\cortext[cor1]{correspondence to: Institut f\"ur Physik \& Astronomie,  Universit\"at Potsdam,  14476 Potsdam,  Germany}
\address[1]{Institut f\"ur Physik \& Astronomie,  Universit\"at Potsdam,  14476 Potsdam, Germany}
\address[2]{Helmholtz Zentrum Berlin, 12489 Berlin, Germany}
\address[3]{European XFEL,  22869 Schenefeld,  Germany}

\date{\today}

\newcommand{\etal}{et al.\ }
\newcommand{\ie}{i.e.,\ }
\newcommand{\eg}{e.g.\ }

\nonumnote{
\begin{tabular}{ll}
\multicolumn{2}{l}{\textbf{Abbreviations:}}\\
2TM  & Two-temperature model \\
AFM  & Antiferromagnetic \\
FEL  & Free-electron laser \\
FM   & Ferromagnetic \\
LCM  & Linaer chain model \\
NTE  & Negative thermal expansion \\
PM   & Paramagnetic \\
PSD  & Position-sensitive detector \\
PUX  & Picosecond ultrasonics with x-rays \\
RSM  & Reciprocal space map \\
RSS  & Reciprocal space slicing \\
TDBS & Time-domain Brillouin scattering  \\
TEM  & Transmission electron microscopy \\
UXRD & Ultrafast x-ray diffraction \\
 &  \\
\end{tabular}
}

\begin{abstract}
This review discusses picosecond ultrasonics experiments using ultrashort hard x-ray probe pulses to extract the transient strain response of laser-excited nanoscopic structures from Bragg-peak shifts. This method provides direct, layer-specific, and quantitative information on the picosecond strain response for structures down to few-nm thickness. 
We model the transient strain using the elastic wave equation and express the driving stress using Grüneisen parameters stating that the laser-induced stress is proportional to energy density changes in the microscopic subsystems of the solid, \ie electrons, phonons and spins. The laser-driven strain response can thus serve as an ultrafast proxy for local energy-density and temperature changes, but we emphasize the importance of the nanoscale morphology for an accurate interpretation due to the Poisson effect. The presented experimental use cases encompass ultrathin and opaque metal-heterostructures, continuous and granular nanolayers as well as negative thermal expansion materials, that each pose a challenge to established all-optical techniques.
\end{abstract}


\maketitle

\section{\label{sec:I_intro}Introduction} 
Picosecond ultrasonics subsumes the generation and detection of laser-induced nanoscopic strain pulses with GHz or even THz frequency components \cite{grah1989, mari1998}. Most studies of picosecond ultrasound have been and will be conducted with all-optical setups, which allow tackling both fundamental and applied research questions in light-matter interaction, condensed matter and material science \cite{tas1994, mari1998, tas1998, anto2006, mats2015b}. Picosecond ultrasonics experiments exploit that the energy deposition of an optical femtosecond pump-pulse creates a stress profile within an absorbing transducer layer. The subpicosecond rise of the stress gradient introduces a source term to the elastic wave equation, which is used to rationalize the occurrence of highly localized strain pulses \cite{thom1986, mats2015b, ruel2015}.

Many aspects of the strain propagation can be measured by reflection, transmission or scattering of optical probe pulses at interfaces and in the bulk. In the most straight-forward experiments, strain induced changes of the reflectivity 
witness the arrival of a strain pulse in the layer and the timing can be used to infer layer thicknesses or elastic properties \cite{grah1989, wrig1991b, anto2006,lejm2014}. In transparent materials, the detection process has been interpreted as time-domain Brillouin scattering (TDBS), which is observed as oscillations of the optical reflectivity at a frequency that is proportional to the sound velocity \cite{guse2014,boja2013,guse2018}. Although the development of all-optical probing schemes with ever-growing sensitivity has enabled studies of a multitude of picosecond ultrasonics effects such as strain pulses in heterostructures \cite{wrig1991b,wrig1995b}, shear waves \cite{peze2016}, acoustic solitons \cite{van2010} or even imaging of elastic properties \cite{daly2004, pere2020}, the reported strain signatures mostly remain on a qualitative level.

In ultrafast x-ray diffraction (UXRD) experiments the optical probe pulses are replaced by ultrashort hard x-ray pulses that are diffracted from the crystal structure in motion \cite{barg2006,lind2017, trig2021}. Similar to optical reflectivity, x-ray diffraction can detect propagating strain pulses in two ways: Strain in a material is heralded by Bragg peak shifts which encode the amplitude, sign and shape of the elastic wave \cite{lind2000,lars2002,haya2006}. Secondly, TDBS of x-rays \cite{boja2013} from the sum of a reciprocal lattice vector and a phonon wave vector can probe the phonon population and the elastic properties even for wave vectors close to the Brillouin zone boundary \cite{heni2016,dorn2019,shay2022}. For such studies the high degree of collimation of hard x-rays from synchrotron and free-electron laser (FEL) sources is highly beneficial. This is also true for the time-resolved detection of x-rays scattered from incoherent phonons by thermal diffuse scattering, which yields a time- and wavevector-resolved picture of the phonon population \cite{trig2010, trig2013, zhu2015, wall2018}. 

X-rays from laser-based plasma sources, in contrast, are sufficient to support the conceptually simple picosecond ultrasonics experiments that investigate Bragg peak shifts \cite{rose1999, barg2004, nico2011, quir2012, zeus2019}. These setups are particularly useful for broad diffraction peaks associated with thin films of imperfect crystallinity due to the relatively large divergence and bandwidth of the provided x-rays. For ultrathin layers, crystals consisting of light atoms, nanoparticles or advanced concepts using resonant diffraction, however, FELs are ideal as they pair high photon flux and sub-picosecond time resolution \cite{bost2016, abel2017, scho2019}. Synchrotron radiation sources typically lack temporal resolution, which can be improved via fs-pulse slicing schemes \cite{scho2000, beau2007}, x-ray streak camera setups \cite{lind2000, lars2002, enqu2010}, special short-pulse filling patterns \cite{jank2013,ross2021} or picosecond Bragg switches \cite{gaal2014, sand2019}, that all come at the cost of a reduction of the photon flux by multiple orders of magnitude.

In this article, we highlight the conceptual advantages of picosecond ultrasonics with x-rays (PUX) experiments that observe shifts of Bragg peaks of layered heterostructures. Each crystalline material scatters at a characteristic Bragg diffraction angle, and hence, the signal is often a layer-specific measure of the lattice distortions caused by the strain pulse propagation and heat flow. In other words, for each pump-probe delay, x-ray probe pulses can project the Bragg peak of each layer onto the detector with a peak shift that is proportional to the instantaneous strain averaged over one layer. 
The four central advantages of x-ray probing in picosecond ultrasonics are: (i) Diffraction yields quantitative strain values (ii) with layer specific information for (iii) strain pulses and quasi static strain. (iv) The x-ray probe penetrates metals and insulators irrespective of their optical properties. This permits a very flexible exploitation of dedicated strain-sensing layers that can be placed within the heterostructure and often naturally exist as buffer, contact or electrode layers. We elaborate and illustrate concepts for the analysis of PUX experiments that have emerged to utilize the quantitative access to the strain response for tracking the energy flow in laser-excited heterostructures. 

We structured the manuscript as follows: In Section~\ref{sec:II_PtCuNi_example} we introduce the measurement principle and observables of PUX experiments based on a representative example. We show that the layer for optical excitation can be separated from a dedicated probe layer, which is optimized for x-ray scattering. This facilitates the separation of the strain pulse traveling at the speed of sound and quasi-static strain that propagates via heat diffusion.

In Section~\ref{sec:III_model_theory} we reconsider aspects of the elastic wave equation that are important for a quantitative modeling of the picosecond strain response. It becomes evident that an in-plane motion introduces a transverse elastic stress that affects the out-of-plane strain response. Accordingly, constrained in-plane motion distinguishes the laser-induced thermal expansion response of a homogeneously excited continuous film from its unconstrained near equilibrium thermal expansion. We elaborate the direct relation of laser-induced energy and stress via a Gr\"uneisen parameter and emphasize that the quasi-static expansion is linearly proportional to the energy density that generates the stress. The concept of subsystem-specific Gr\"uneisen parameters is introduced to capture the stress contributions of different quasi-particle excitations to the strain response. 

In Section~\ref{sec:IV_examples} we present scenarios where ultrashort hard x-ray probe pulses excel at probing the strain response in laser-excited nanolayered heterostructures. At first,  we illustrate in \ref{sec:IV_a_tbfe}  how PUX tracks bipolar and unipolar picosecond strain pulses launched by an opaque transducer with transparent capping layers of various thicknesses. The second use case is the quantitative determination of the picosecond strain response in granular and continuous FePt thin films, that exhibits a strong dependence on the sample morphology as shown in \ref{sec:IV_b_fept}. The third use case demonstrates that the PUX experiments can distinguish the strain response of nanoscopic heterostructures that are thinner than the optical penetration depth as demonstrated on Au-Ni bilayer structures in \ref{sec:IV_c_auni}. This thin film scenario complements the introductory example where we access the electronic energy transfer between layers of an opaque Pt-Cu-Ni heterostructure discussed in Section~\ref{sec:II_PtCuNi_example}. 
The last use case in \ref{sec:IV_a_dy} is  the detection of ultrafast negative thermal expansion (NTE) exhibited by a rare-earth transducer, which launches unconventional picosecond strain pulses towards a buried detection layer. This example furthermore demonstrates the extraction of subsystem-specific Gr\"uneisen parameters from equilibrium thermal expansion data.

Section~\ref{sec:V_conclusion} discusses the advantages of large scale sources and provides an overview on related experimental schemes that utilize x-rays for picosecond ultrasonics.  

The appendix in Section~\ref{sec:VI_xray_diffraction} contains a brief discussion of the diffraction geometry and the relation between the diffraction angles and the reciprocal space coordinates. We revisit the concept of reciprocal space slicing (RSS) as a rapid data acquisition approach that is often sufficient and less time-consuming compared to the acquisition of reciprocal space maps (RSMs), which requires scanning the incidence angle of the x-rays on the sample. We explain the scaling factor that relates the Bragg peak shift in reciprocal space to the shift measured on an area or line detector in the RSS scheme and discuss scenarios when time-resolved RSMs are required for a proper strain assessment.

\section{\label{sec:II_PtCuNi_example}PUX experiments in a nutshell}
\begin{figure}[tbh!]
\centering
\includegraphics[width = 0.95\columnwidth]{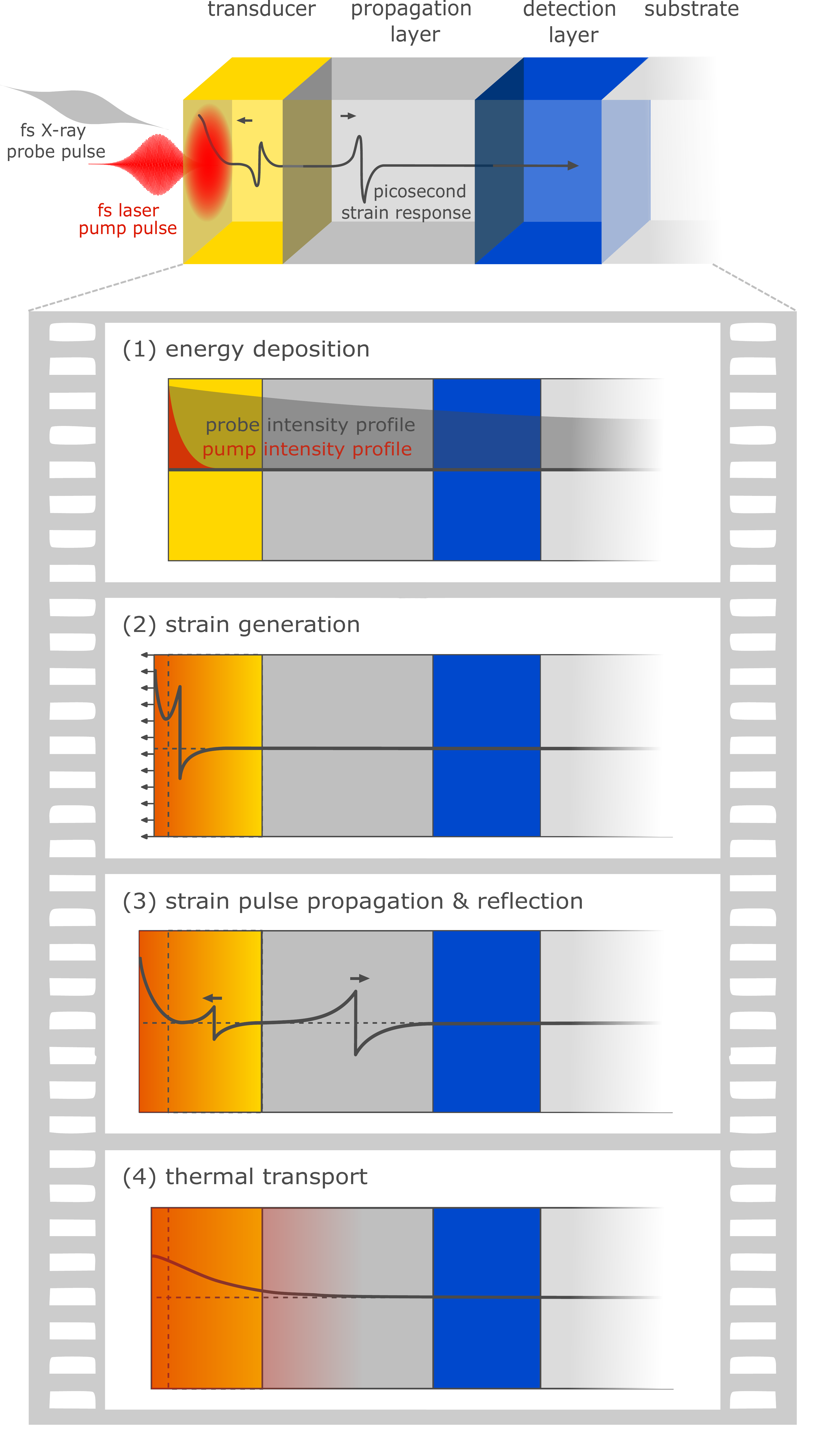}
\caption{\textbf{Generic series of events for picosecond ultrasonics:} A femtosecond laser pulse excites a metallic heterostructure that consists of multiple, often nanoscopically thin layers. The deposited energy creates a stress that drives a picosecond strain response consisting of strain pulses propagating at the speed of sound and a quasi-static thermal expansion which evolves via heat diffusion. The small skin depth of optical probes limits the probed volume to the near-surface region whereas hard x-rays often penetrate multiple microns or more into the bulk.}
\label{fig:II_concept_sketch}
\end{figure}

A typical picosecond ultrasonics experiment is schematically depicted in Fig.~\ref{fig:II_concept_sketch}, where we illustrate the generic series of events common to laser-excited heterostructures \cite{mats2015b}. The laser-induced strain response in this type of experiment contains information on all four conceptual steps that occur in response to the light-matter interaction between the femtosecond laser pulse and the absorbing transducer layer \ie the energy deposition profile (1), the strain generation from a laser-induced stress (2),  strain pulse propagation and reflection (3) and quasi-static strain concomitant with thermal transport (4). 

In applications using optical probe pulses, the main results of such experiments are often the echo time for thickness determination \cite{anto2006,lejm2014}, the acoustic transmission amplitude through an interface for measuring impedance mismatches \cite{tas1998, gros2017} or the oscillations in TDBS to determine the sound velocity \cite{thom1986b,lin1991,guse2018}. Here, we demonstrate that PUX provides information on film thicknesses, strain-pulse reflections and microscopic energy transfer processes within the heterostructure, that set the space and time-dependent stress driving the strain response. As indicated in Fig.~\ref{fig:II_concept_sketch} PUX experiments benefit from the large extinction length of hard x-rays (\eg $8\,\text{keV}$), which is typically on the order of few micrometers irrespective of the optical properties, \eg of metals, semiconductors and insulators. Therefore, the hard x-ray probe pulse can report on all layers of thicker heterostructures. More importantly, when each layer exhibits a different lattice constant, \ie has a characteristic Bragg diffraction angle, the probe is even layer-specific. 

In the remainder of this section, we discuss a representative PUX experiment on metallic heterostructures composed of Pt, Cu and Ni to exemplify the measurement principle and data evaluation. This example demonstrates the advantages of PUX experiments on a sample structure that is frequently used to study the effect of hot-electron pulses that are launched in a Pt layer and propagate through an opaque Cu stack towards a buried functional detection layer \cite{berg2016, fert2017, xu2017, berg2020}.

\begin{figure*}[tbh!]
\centering
\includegraphics[width = 1.00\textwidth]{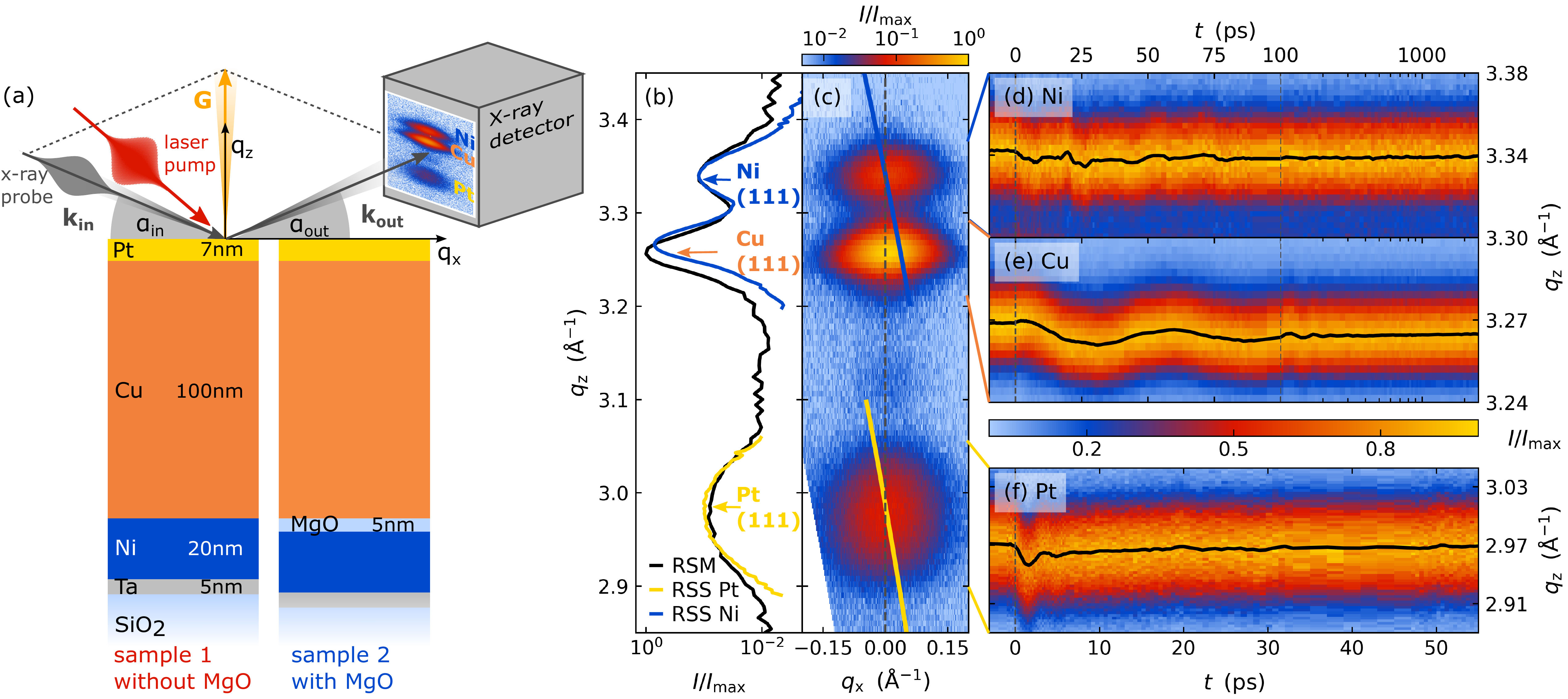}
\caption{\textbf{Extraction of layer-specific strain via x-ray diffraction exemplified for a Pt-Cu-Ni heterostructure:} (a) Sketch of the samples and diffraction geometry using a PSD. Scanning the angles $\alpha_\text{in}$ and $\alpha_\text{out}$ yields the RSM (c) around the Bragg peaks of Pt, Cu and Ni that are separated along the out-of-plane reciprocal coordinate $q_\text{z}$ (b). The yellow and blue lines denote the diffraction intensity distributions for the respective subset of the reciprocal space that is probed by the area detector at particular fixed angles as utilized in the time-resolved experiment. The black line is the integrated intensity distribution of (c) along $q_\text{x}$. Panels (d--f) display the laser-induced shift of the Bragg peaks of sample 1 that is used to determine the layer-specific transient strain according to Eq.~\eqref{eq:II_1_strain_definition}.}
\label{fig:II_pux_data}
\end{figure*} 

Fig.~\ref{fig:II_pux_data}(a) introduces the sample structures and the specular \ie symmetric and coplanar diffraction geometry. A femtosecond x-ray pulse derived from a laser-based plasma x-ray source \cite{schi2013c} is incident under the angle $\alpha_\text{in}$ with respect to the sample surface. The pixels of a position-sensitive detector (PSD) record the x-ray intensity distribution diffracted from the individual layers of the heterostructure. On the PSD we observe three maxima separated within the diffraction plane which correspond to the Bragg peaks of the individual metal layers. The respective diffraction angles $\alpha_\text{out}$ are determined by the reciprocal lattice vector $\bm{G}$ via the Laue condition:
\begin{align}
\bm{q}=\bm{k}_\text{out} - \bm{k}_\text{in} =k \begin{pmatrix} \cos{(\alpha_\text{out})} - \cos{(\alpha_\text{in})} \\ \sin{(\alpha_\text{in})} + \sin{(\alpha_\text{out}) } \end{pmatrix} =\bm{G} \,,
\label{eq:II_0_Laue}
\end{align}
where the scattering vector $\bm{q}$ is determined by the wave vector of the incident ($\bm{k}_\text{in}$) and diffracted ($\bm{k}_\text{out}$) x-rays, which have equal magnitude $k=2 \pi /\lambda$ according to the wavelength $\lambda$ of the elastically scattered x-rays. The magnitude of $\bm{G}$ encodes the out-of-plane lattice constant $d_3$ and the diffraction order $n$ via $|\bm{G}|=2\pi n/ d_3$. Each detector pixel of the PSD probes a small volume around a specific point in reciprocal space and a symmetric scan of $\alpha_\text{in}$ and $\alpha_\text{out}$ maps the reciprocal space along the out-of-plane direction defining the $q_\mathrm{z}$-axis. A representative RSM, \ie the intensity scattered along  $\bm{q}$, of a Pt-Cu-Ni heterostructure is depicted in Fig.~\ref{fig:II_pux_data}(c) and reveals Bragg peaks of each (111)-oriented layer. The width of the Bragg peaks along $q_\text{x}$ and $q_\text{z}$ is given by the crystalline quality of the layers characterized by the mosaicity and the in-plane and out-of-plane coherence length of the crystallites forming the metal layers. A more detailed discussion of the RSM and the transformation of the angle-dependent intensity to reciprocal space $q_\text{x}$-$q_\text{z}$ is provided in appendix~\ref{sec:VI_a_rsm} and \cite{schi2013c}. 

The integration of the RSM along $q_\text{x}$ yields the intensity distribution along the reciprocal coordinate $q_\text{z}$ (Fig.~\ref{fig:II_pux_data}(b)) that encodes the average out-of-plane lattice constant $d_{3}$ of the diffracting layers by the center-position of their Bragg peaks $q_\text{z}=2\pi n/d_{3}$ (see Eq.~\eqref{eq:II_0_Laue}). Following the laser-induced time-dependent shift of the Bragg peaks along $q_\text{z}$ in a pump-probe experiment yields the layer-specific transient strain $\eta(t)$ which represents the relative change of the average out-of-plane lattice constant with respect to its value before excitation ($t<0$):
\begin{align}
	\eta(t) =  \frac{d_{3}(t)-d_{3}(t<0)}{d_{3}(t<0)}= \frac{q_\text{z}(t<0)-q_\text{z}(t)}{q_\text{z}(t)}\,.
	\label{eq:II_1_strain_definition}	
\end{align}  
Instead of scanning the incidence angle $\alpha_\text{in}$ to create an RSM at each pump-probe delay, we often follow the transient Bragg peak position by the time-efficient RSS method \cite{zeus2021} that is further discussed in Section~\ref{sec:VI_b_rss}. The PSD simultaneously probes a subset of the reciprocal space as indicated in Fig.~\ref{fig:II_pux_data}(c) by the blue and yellow lines for two different fixed $\alpha_\text{in}$ corresponding to the Bragg angles of Ni and Pt, respectively. The intensity on the detector exemplarily shown in the inset of panel (a) for an intermediate $\alpha_\text{in}$ probing all three peaks simultaneously with less intensity is integrated along the $q_\text{y}$ coordinate of the detector. For the yellow and blue lines $\alpha_\text{in}$ is chosen to efficiently capture slices through the different maxima of the RSM (Fig.~\ref{fig:II_pux_data}(c)), in order to collect RSS (Fig.~\ref{fig:II_pux_data}(b)) which closely resemble the integrated RSM (black line). We relate the transient shifts of the Bragg peaks on the detector to a transient shifts along $q_\text{z}$ depicted in Fig.~\ref{fig:II_pux_data}(d--f) that yield the layer-specific transient out-of-plane strains according to Eq.~\eqref{eq:II_1_strain_definition}.  

\begin{figure}[t!]
\centering
\includegraphics[width = 1.00\columnwidth]{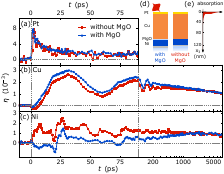}
\caption{\textbf{Comparison of the picosecond strain response in Pt-Cu-(MgO-)Ni heterostructures:} Transient strain in the (a) Pt, (b) Cu and (c) Ni layers of the heterostructures with and without MgO interlayer as depicted in (d). Blue lines are for the heterostructure with an insulating MgO barrier between Cu and Ni. (e) Absorption of the pump-pulses occurs only in the Pt layer and in the first few nm of Cu. The suppression of the electronic heat transport from Pt to Ni by the MgO interlayer changes the strain response of the Ni detection layer from expansion to compression and delays the rise of the quasi-static expansion.}
\label{fig:II_PtCuNiSi_comparison}
\end{figure}

The resulting strain response of the Pt transducer, the Cu propagation layer, and the Ni detection layer are compared in Fig.~\ref{fig:II_PtCuNiSi_comparison}(a--c) for two different heterostructures with and without a $5\,\text{nm}$ thin insulating MgO interlayer in front of the buried Ni detection layer. The near-infrared pump-pulse mainly deposits energy in the $7\,\text{nm}$ thin Pt transducer as illustrated by the absorption profile in Fig.~\ref{fig:II_PtCuNiSi_comparison}(e). In absence of an MgO interlayer, we observe a rapid expansion of both the Pt and the Ni layer upon laser-excitation and the Cu layer is compressed within the first picoseconds. The rapid expansion of the Ni layer compressing the adjacent Cu layer originates from a fast transport of hot electrons from the laser-excited Pt transducer through Cu to the buried Ni layer where they release their energy to phonons. This dominant electronic heat transport to Ni causes the compression of the Cu layer that is heated only on a longer timescale owing to its weak electron-phonon coupling. This surprising observation was referred to as "heat transport without heating" in a previous publication \cite{pude2020b} that provides additional information on the role of layer-specific electron-phonon coupling and the modelling of the strain response.

Here, we highlight the suppression of the crucial electronic energy transport from Pt to Ni, if the additional MgO interlayer is introduced (blue data in Fig.~\ref{fig:II_PtCuNiSi_comparison}). Now the Ni layer is rapidly compressed by the expansion of Cu and compressed even more at $20\,\text{ps}$, when the strain pulse generated in Pt reaches the Ni layer. The expansion of the Ni detection layer only rises on the timescale of hundreds of picoseconds after laser-excitation via phononic heat transport from Cu through the MgO interlayer separating sound and heat in the time domain. The suppressed electronic energy transport to the buried Ni detection layer yields a background-free signal of the strain pulse and an increased expansion of the Cu layer compared to the pure metal heterostructure.

The experiment nicely illustrates PUX: We obtain the strain response by tracking the shift of layer specific diffraction peaks that are well separated due to their material-specific lattice constants.  The total strain response is a superposition of propagating strain pulses and a quasi-static expansion due to heating that dominates after the strain pulses have propagated into the substrate. Since PUX measures the strain amplitude, both the strain pulses and the quasi-static strain from laser-induced temperature changes can be quantitatively evaluated on equal footing. In addition, the transient strain accesses the layer thicknesses via the timing of the expansion and compression pulses. 

\section{\label{sec:III_model_theory} Concepts for modelling the strain response}
In this section, we discuss the fundamental concepts for quantitatively modelling the picosecond strain response, driven by a laser-induced spatio-temporal stress. These dynamics are generally described by the elastic wave equation, and we elaborate on the special case of a laterally homogeneous excitation of continuous thin films and heterostructures investigated in typical picosecond ultrasonic experiments. The spatio-temporal stress is proportional to the local contributions to the energy density, which changes in time according to the heat transport within the sample. The different degrees of freedom such as electrons and phonons contribute differently to the transport and have to be accounted for in nanoscale metals. We advocate the Gr\"uneisen concept to model the effect of the energy transfer between these subsystems on the laser-induced stress. The Gr\"uneisen parameters linearly relate the energy deposited in different degrees of freedom to the respective stress contributions. Their superposition yields the total external stress driving the strain response. Finally, we provide an example of modelling the strain response of a sample by numerically solving the elastic wave equation for an educative case of an inhomogeneously excited transducer on a non-absorbing detection layer.

\subsection{\label{sec:III_a_strain_3D} Poisson stresses in a 3D strain response}
In general, the strain response of an elastic solid to a time- and space-dependent stress is found as a solution of the elastic wave equation, \ie the equation of motion for the displacement field $u_i(\bm{x},t)$ at a specific position $\bm{x}$ in the sample and at a time $t$.\footnote{To enhance the readability of this section the space and time dependencies $(\bm{x},t)$ are not listed further after the introduction of physical quantities.} The index~$i$ enumerates the three spatial dimensions. The mass density $\rho_\text{m}(\bm{x})$ is accelerated by the spatial gradient of a total stress $\sigma^\text{tot}_{ij}(\bm{x},t)$, as described by the elastic wave equation:
\begin{align}
	\begin{split}
		\rho_\text{m}\frac{\partial^2 u_i}{\partial t^2} &= \sum_{j} \frac{\partial }{\partial x_{j}} \sigma_{ij}^\text{tot} \\ 
		&= \sum_{j}  \frac{\partial}{\partial x_{j}} \left( \sum_{k,l}  c_{ijkl}\eta_{kl} - \sigma_{ij}^\text{ext}\right)\,.
	\end{split}
	\label{eq:III_1_general_wave_equation}
\end{align}
The deformation of the solid is described by the strain
\begin{align}
    \eta_{kl}(\bm{x},t) = \frac{\partial u_k(\bm{x},t)}{\partial x_l}\,,
    \label{eq:III_0_strain}
\end{align}
which is determined by the displacement $u_k$. 
The proportionality constants of the elastic stress are the direction-dependent elastic constants $c_{ijkl}(\bm{x})$\footnote{Here, $c_{ijkl}$ does not necessarily denote the elastic constants along the high symmetry axis of the crystal but along the spatial directions given by the sample geometry. If the high symmetry axis do not match the principal axis of the sample a transformation of the elastic tensor into the coordinate system of the sample is required.}. Adding an external stress $\sigma_{ij}^\text{ext}(\bm{x},t)$ drives the atomic motion.

In this publication, we limit our discussion to longitudinal laser-induced stresses $\sigma^\text{ext}_{ii}$ and strains $\eta_{kk}$, \ie volume changing elements of the stress-strain relation. Under this limitation the elastic wave equation \eqref{eq:III_1_general_wave_equation} simplifies to:
\begin{align}
\begin{split}
\rho_\text{m} \frac{\partial^2 u_i}{\partial t^2} &= \frac{\partial}{\partial x_{i}} \left( c_{iiii}\eta_{ii} + \sum_{k \neq i}  c_{iikk}\eta_{kk}- \sigma_{ii}^\text{ext} \right) \\
    &= \frac{\partial }{\partial x_{i}} \left( \sigma^\text{elastic}_{ii}+ \sigma_{ii}^\text{Poi} - \sigma_{ii}^\text{ext} \right) \,.
\end{split}
\label{eq:III_2_wave_equation_without_shear}
\end{align}

Here, the negative sign in front of the external stress is chosen such that a positive longitudinal stress $\sigma^\mathrm{ext}_{ii}$ that acts from within the material leads to an expansion $(\eta_{ii}>0)$ along the direction of the stress. Note that other works \cite{ruel2015, mats2015b} use the opposite sign for $\sigma^\mathrm{ext}_{ii}$, which yield intuitive results when the material is subjected to an external force, for instance, through pressing in a diamond anvil cell where a positive stress leads to compression $(\eta_{ii}<0)$.  If the gradients in the external stress rise faster than the elastic stress which is proportional to the strain propagating at sound velocity, a propagating strain pulse is launched. Its propagation is affected by interfaces between different layers within the sample structure unless they are acoustically impedance matched\footnote{The acoustic impedance $Z$ of a material  is defined by the product of its mass density $\rho_\mathrm{m}$ and the appropriate sound velocity $v_\mathrm{s}= \sqrt{c_{3333}/\rho_\text{m}}$  \ie  $Z= \rho_\mathrm{m} v_\mathrm{s}$. The acoustic reflection coefficient $R$ for a strain pulse traversing the interface from material 1 to material 2 considered in the model, is given by $R = \frac{Z_2-Z_1}{ Z_1 + Z_2}$, which implicitly assumes perfect adhesion \cite{gros2017, roye2000}.}. The strain pulses are partially reflected from the interfaces. When reflection occurs at an interface to a medium with lower acoustic impedances, \eg in particular for reflections at the surface of the sample, the sign of the strain pulse changes. The strain $\eta_{ii}$ induces an elastic stress $\sigma^\text{elastic}_{ii}(\bm{x},t)$ that partially compensates the external stress. In addition, the three-dimensional response of the solid introduces Poisson stress contributions $\sigma_{ii}^\text{Poi}$ that originate from the spatio-temporal strains $\eta_{kk}$ along the perpendicular directions analogously determined by the elastic wave equation along the respective directions. In total, the three-dimensional strain response of the solid to the external stresses $\sigma_{ii}^\text{ext}$ requires a solution of the three coupled differential equations~\eqref{eq:III_1_general_wave_equation}.

\begin{figure}[t!]
\centering
\includegraphics[width =1\columnwidth]{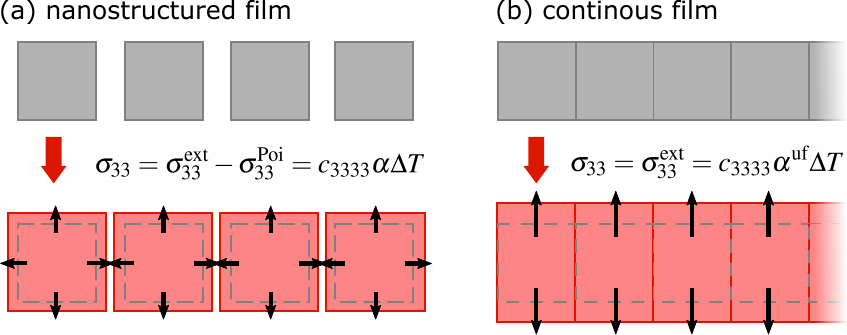}
\caption{\textbf{Morphology-dependent strain response of nanoparticles and thin films:} Sketch of the laser-driven quasi-static expansion for an in-plane nanostructured film (a) and an in-plane homogeneous and continuous thin film (b) of an isotropic material without an attached substrate. The out-of-plane expansion of the continuous film is enhanced by the absence of contractive Poisson stress contributions that would partially compensate the expansive external stress. The arrows indicate the effective driving stress $\sigma_{ii}$.}
\label{fig:III_alpha_ultrafast}
\end{figure}

When the driven strain pulses and their reflections have propagated out of the volume of interest, the elastic and the Poisson stress contributions fully compensate the external stress and the vanishing total stress marks a quasi-static state\footnote{The quasi-static state slowly changes due to diffusion of heat and other slow processes.}. The vanishing time derivative in Eq.~\eqref{eq:III_2_wave_equation_without_shear} then determines the quasi-static strain $\eta_{ii}^\text{qs}$ to:
\begin{align}
	\begin{split}
		\eta_{ii}^\mathrm{qs} &= \frac{\sigma_{ii}^\text{ext}-\sigma_{ii}^\text{poi}}{c_{iiii}} \\
		&= \frac{\sigma_{ii}^\text{ext}}{c_{iiii}}-\sum_{k \neq {i}} \frac{c_{iikk}}{c_{iiii}}\eta_{kk}^
		\text{qs}= \alpha_{i} \Delta T \,.
	\end{split}
	\label{eq:III_3_general_quasi_static_strain}
\end{align}
The anisotropic linear expansion coefficient $\alpha_{i}(\bm{x})$ relates this quasi-static strain to a temperature increase $\Delta T(\bm{x},t)$ as in thermal equilibrium\footnote{Note, this approach is only true for small $\Delta T=T_\text{f}-T_\text{i}$. Otherwise, $\eta_{ii}^\mathrm{qs}=\int_{T_\text{i}}^{T_\text{f}} \alpha_i(T') \text{d}T'$ holds.}. Here, the expansion is not only driven by the externally induced stress $\sigma_{ii}^\text{ext}$ but also reduced by a Poisson contribution that arises from the expansion along the perpendicular directions $\eta_{kk}^\text{qs}$ driven by the external stresses $\sigma_{kk}^\text{ext}$ (see Fig.~\ref{fig:III_alpha_ultrafast}(a)).

The parametrization of $\eta_{kk}^\text{qs}$ in Eq.~\eqref{eq:III_3_general_quasi_static_strain} by their respective linear thermal expansion coefficient $\alpha_{k}$ yields a relation between the external stress and the temperature increase: 
\begin{align}
	\begin{split}
		\sigma_{ii}^\text{ext}(x_i,t) &= c_{iiii} \left( \alpha_{i} + \sum_{i \neq k} \frac{c_{iikk}}{c_{iiii}} \alpha_{k}\right) 	
		\Delta T \\
		&= \Gamma_{i} \rho_{Q}\,.
	\end{split}
	\label{eq:III_4_general_external_stress}	
\end{align}
This temperature increase originates from an optically deposited energy density $\rho_Q$: $\Delta T= \rho_Q/C_V$ determined by the heat capacity per constant volume $C_V$\footnote{Note, this approach is only true for small $\Delta T=T_\text{f}-T_\text{i}$ where $C_V(T)$ only barely changes. Otherwise it should read $\rho_{Q}=\int_{T_\text{i}}^{T_\text{f}} C_V(T') \text{d}T'$. However, the chosen notation simplifies the introduction of the Gr\"uneisen parameter.}. Equation~\eqref{eq:III_4_general_external_stress} introduces the direction-dependent Gr\"uneisen parameter $\Gamma_{i}(\bm{x})$:
\begin{align}
	\Gamma_{i} &=  \frac{c_{iiii}}{C_V} \underbrace{\left( \alpha_{i} + \sum_{i \neq k} \frac{c_{iikk}}{c_{iiii}} \alpha_{k} \right)}_{\alpha_{i}^\text{uf}} \,,
	\label{eq:III_5_definition_Gr\"uneisen_constant}	
\end{align}
that linearly relates this deposited energy density to the induced external stress $\sigma_{ii}^\text{ext}$. It describes how efficiently energy density generates stress and is determined by the direction-dependent $\alpha_{i}^\text{uf}$. This is the expansion coefficient along direction $i$ assuming that the lattice is clamped along other spatial directions, \eg for the ultrafast excitation of a homogeneous thin film (see next section).  
 
The advantage of using the Gr\"uneisen concept is that the ratios of the intrinsically temperature-dependent quantities $\alpha_{i}(T)$, $\alpha_{i}^\text{uf}(T)$ and $C_V(T)$ in Eq.~(\ref{eq:III_5_definition_Gr\"uneisen_constant}) can be combined to temperature-independent unitless parameters $\Gamma_{i}$. Equation~\eqref{eq:III_4_general_external_stress} states that the energy density $\rho_Q$ is proportional to the stresses $\sigma_{ii}^\text{ext}$ which themselves linearly relate to the quasi-static strain $\eta_{ii}^\text{qs}$. Moreover, in materials with several subsystems hosting the energy density, we find a simple recipe for modeling the out-of-equilibrium expansion response by adding the stresses for all subsystems, even if their energy content may correspond to different equilibrium subsystem temperatures \cite{nie2006,nico2011,heni2016,mald2017} (see Section~\ref{sec:III_d_Gr\"uneisen_stress} for further details).

\subsection{\label{sec:III_b_strain_1D} Constraints of the thin film geometry}
In most picosecond ultrasound experiments the footprint of the excitation laser pulse (sub-mm) is much larger than both the thickness of the transducer (sub-$\text{\textmu}$m) and the footprint of the probe pulse, which results in a laterally homogeneous excitation of the probed sample volume.

Therefore, on the timescale ($t_\text{1D} \ll d_\text{L}/v_\text{s}$) given by the size of the pump-laser footprint $d_\text{L}$ and the sound velocity $v_\text{s}(\bm{x})$, the in-plane stresses are balanced and only the spatial derivative $\frac{\partial}{\partial x_{3}} \sigma_{33}^\text{ext}$ remains. Under this condition, the system of coupled Eqs.~\eqref{eq:III_2_wave_equation_without_shear} simplifies to a one-dimensional equation for the strain along the out-of-plane direction ($x_3$) of the thin film:

\begin{align}
\label{eq:III_6_1d_wave_equation_strain}
\rho_\text{m}\frac{\partial^2 u_3}{\partial t^2} &= \frac{\partial}{\partial x_3} \left( c_{3333} \eta_{33} 
- \sigma^\text{ext}_{33} \right)\, .
\end{align}

Figure~\ref{fig:III_alpha_ultrafast} compares the three-dimensional response of a thin film with in-plane nanostructure to the purely one-dimensional expansion of a continuous thin film. This anisotropic picosecond strain  of continuous thin films occurs even in otherwise isotropic solids. It is driven by gradients in the external stress along the out-of-plane direction which typically appear at the sample surface or layer interfaces or at the slopes of the excitation profile that is slowly changing via heat transport. The arising out-of-plane strain $\eta_{33}$ upon laser excitation  partially compensates the external stress until the total stress vanishes when the strain pulses have propagated into the substrate. The absence of in-plane expansion for $t\leq t_\text{1D}$ suppresses any Poisson stress contributions. Therefore, the remaining quasi-static expansion $\eta_{33}^\text{qs}$ is directly related to the external stress $\sigma_{33}^\text{ext}$ which simplifies Eq.~\eqref{eq:III_3_general_quasi_static_strain} to:

\begin{align}
	\eta_{33}^\mathrm{qs} = \frac{\sigma^\mathrm{ext}_{33}}{c_{3333}} = \frac{\Gamma_{3}\rho_Q}{c_{3333}} = \alpha_{3}^\text{uf} \Delta T \,.
	\label{eq:III_8_1d_quasi_static_strain}
\end{align}
Using the concept of a Gr\"uneisen parameter we express the external stress by $\sigma^\mathrm{ext}_{33}=\Gamma_{3}\rho_Q$. With the definition of the Gr\"uneisen parameter in Eq.~\eqref{eq:III_5_definition_Gr\"uneisen_constant} the quasi-static expansion in the thin film geometry is related to the ultrafast expansion coefficient $\alpha_{3}^\text{uf}$, which differs from the corresponding equilibrium expansion coefficient $\alpha_{3}$ that is used for three-dimensional expansion (Eq.~\eqref{eq:III_3_general_quasi_static_strain}):
\begin{equation}
\alpha_{3}^\text{uf} = \alpha_{3} \left( 1 + \sum_{k \neq 3} \frac{c_{33kk} \alpha_{k}}{c_{3333} \alpha_{3}} \right)\,.
\label{eq:III_9_alpha_ultrafast}
\end{equation}
In case of isotropic solids this expression simplifies due to identical values of the off-diagonal elements of the elastic tensor and isotropic expansion coefficients $\alpha_{i}=\alpha$. For a metal with a typical Poisson constant of $\nu=c_{1133}/(c_{3333}+c_{1133})\approx 1/3$, the ultrafast expansion coefficient 
\begin{equation}
\alpha^\text{uf} = \alpha \left( 1 + 2 \cdot \frac{c_{1133}}{c_{3333}} \right) \approx 2\alpha\,.
\label{eq:III_10_alpha_ultrafast_isotropic_solid}
\end{equation}
is approximately twice times larger than the equilibrium constant, because the Poisson stresses are absent in  typical picosecond ultrasonic experiments on homogeneous thin films.
Therefore, the Poisson effect requires considering the morphology of the thin film (Fig.~\ref{fig:III_alpha_ultrafast}), because it influences the dimensionality of the strain response and hence the amplitude of the strain response to the laser-induced temperature increase.

The one-dimensional nature of the picosecond strain response of continuous thin films was already considered implicitly by the seminal work by Thomsen \etal \cite{thom1986}. Their formulation of the one-dimensional wave equation~\eqref{eq:III_13_wave_equation_thomson} can be transformed to the much simpler Gr\"uneisen formulation (Eq.~\eqref{eq:III_11_wave_equation_Gr\"uneisen}) by considering the bulk modulus $B=(c_{3333} +2c_{3311})/3$ and the Poisson factor $\nu$:
\begin{align}
\label{eq:III_13_wave_equation_thomson}
\rho_\text{m}\frac{\partial^2 u_3}{\partial t^2} &= \frac{\partial}{\partial x_3} \left( 3\frac{1-\nu}{1+\nu}B\eta_{33}-3B\alpha \Delta T \right)\\
\label{eq:III_12_wave_equation_thomson_comparison} 
&= \frac{\partial}{\partial x_3} \left( c_{3333} \eta_{33}  - c_{3333} \alpha^\text{uf} \Delta T \right) \\
\label{eq:III_11_wave_equation_Gr\"uneisen}
&= \frac{\partial}{\partial x_3} ( \underbrace{c_{3333} \eta_{33}}_{\sigma^\text{elastic}_{33}}  - \underbrace{\Gamma\rho_Q}_{\sigma^\text{ext}_{33}} )\,. 
\end{align}
Equation~\eqref{eq:III_12_wave_equation_thomson_comparison} illustrates that the modified thermal expansion coefficient $\alpha^\text{uf}$ has to be used to quantify the laser-induced stress in the absence of an in-plane expansion. This formulation is useful as it connects $\alpha^\text{uf}$ to the quasi-static strain $\eta_{33}^\text{qs}$ (see Eq.~\eqref{eq:III_8_1d_quasi_static_strain}) that exists in homogeneous thin films on timescales where strain pulses have propagated out of the investigated region of interest, but in plane motion is still negligible due to the large, homogeneously exited area of the thin film. When $t>t_\text{1D}$ the film laterally relaxes and induces the Poisson stresses that re-establish the thermal expansion coefficients used in Eq.~\eqref{eq:III_3_general_quasi_static_strain}. 

Finally, the simple formulation of the wave equation given in Eq.~\eqref{eq:III_11_wave_equation_Gr\"uneisen} highlights the direct access to the spatio-temporal energy density by measuring the transient strain response using the Gr\"uneisen concept. This perspective is particularly useful in the context of materials, where the excitation of several degrees of freedom (\eg electrons, phonons and spins) simultaneously contribute to the stress with different time-dependencies as discussed in Section~\ref{sec:III_d_Gr\"uneisen_stress}.

\subsection{\label{sec:III_c_heat_transport} Energy transfer processes and diffusive two temperature models }
The elastic wave equation~\eqref{eq:III_12_wave_equation_thomson_comparison} relates the picosecond strain response of a homogeneous thin film to a laser-induced stress that is characterized by a temperature increase $\Delta T=T_\text{f}-T_\text{i}$. This simple formulation contains the strong assumption that the temperatures of the different degrees of freedom - in particular electrons and phonons - is the same. 

Under this assumption the shape of both the driven picosecond strain pulses and the spatio-temporal distribution of the remaining quasi-static expansion $\eta_{33}^\text{qs}$ is determined by the spatio-temporal profile of the temperature $T(x_3,t)$ that is a solution to the one-dimensional heat equation:  
\begin{equation}
    C_V(T)\frac{\partial T}{\partial t}=\frac{\partial}{\partial x_3} \left( \kappa(T) \frac{\partial T}{\partial x_3} \right) + S \,,
   \label{III_14_1D_heat_diffusion_1TM}
\end{equation}
with the thermal conductivity $\kappa=\kappa(T(x_3,t))$ that inherits its depth-dependence from the temperature profile and differs for different materials. In addition, the description of heat transport across interfaces may require considering interface resistances that account for different dispersion relations of the involved quasi-particles \cite{redd2005,oomm2022,herz2022}. The absorption of energy from the incoming photons is treated by the source term $S(x_3,t)$ which is in the most simple case described by Lambert-Beer's law. Especially in heterostructures and films thinner than the optical skin depth it is often necessary to consider internal optical reflections, which can be accounted for using transfer matrix approaches \cite{yeh2005,leg2013}. The one-dimensional approach is again only valid for thin, lateral homogeneously excited films where in-plane thermal transport within the probed volume can be neglected.

However, for most of the materials assuming a single temperature, \ie quasi-instantaneous equilibration of the electrons, phonons or any other energy reservoirs in the solid is a strong oversimplification. In typical metals such as Au, Cu and Pt the laser excitation leads to a sudden increase of the energy density in the electron system, which is subsequently transferred to phonons within few picoseconds. This coupling of the subsystems on a timescale comparable to the relaxation of the lattice may be crucial for the induced stress and the driven picosecond strain pulses as experimentally demonstrated for Al \cite{nie2006}, Au \cite{nico2011} and Ni \cite{wang2008}. In contrast, the strain response of materials that exhibit a very strong electron-phonon coupling such as SrRuO$_3$ are in some scenarios sufficiently well described by a single temperature \cite{schi2014b,korf2008}.

In addition to the energy distribution among different degrees of freedom, a one-temperature model also oversimplifies spatial heat transport within metal heterostructures because it disregards non-equilibrium transport phenomena like ballistic \cite{bror1987,hohl1997b} and super-diffusive \cite{batt2012,mali2018} electron transport. Already the modeling of the diffusive transport within the Pt-Cu-Ni metal stack discussed in Section~\ref{sec:II_PtCuNi_example} requires a two-temperature model (2TM) that captures the electronic thermal conductivity $\kappa^\text{el}\propto T^\text{el}/T^\text{ph}$ \cite{pude2020b, hohl2000} enhanced by a long lasting non-equilibrium of electrons and phonons ($T^\text{el}(x_3,t) \gg T^\text{ph}(x_3,t)$) due to weak electron-phonon coupling in Cu. In general, the propagation of quasiparticle excitations following the non-equilibrium after optical excitation can be discussed in a quantitative and state-resolved way by Boltzmann-transport equations \cite{nenn2018,wais2021,chen2001b}. However, in most cases the conceptually simpler diffusive 2TM suffices:
\begin{align}
\begin{split}
    C^\text{el}(T^\text{el}) \frac{\partial T^\text{el}}{\partial t} &= \frac{\partial}{\partial x_3} \left( \kappa^\text{el}(T^\text{el},T^\text{ph}) \frac{\partial T^\text{el}}{\partial x_3} \right)\\
    &~ - g^\text{el-ph}\left(T^\text{el}-T^\text{ph}\right) + S\,,\\
   C^\text{ph}(T^\text{ph}) \frac{\partial T^\text{ph}}{\partial t} &= \frac{\partial}{\partial x_3} \left( \kappa^\text{ph}(T^\text{ph}) \frac{\partial T^\text{ph}}{\partial x_3} \right)\\
   &~ + g^\text{el-ph}\left(T^\text{el}-T^\text{ph}\right)\,.
\end{split}   
\label{eq:III_heat_diffusion_2TM}
\end{align}

Such a diffusive 2TM not only includes the coupling between the two subsystems (here by the electron-phonon coupling constant $g^\text{el-ph}$) but also the individual diffusion of electrons and phonons. Transport across interfaces can be treated via the spatial dependence of the thermo-physical parameters. By modeling the thermal conductivity of electrons $\kappa^\text{el}$ as a parameter depending on both the electron and phonon temperature, we can even rationalize the picosecond ultrasonic response of thin metal nanolayers, where ballistic or superdiffusive transport occurs. However, the analysis of our PUX experiments has up to now not depended on finer details of the electron transport since very rapid processes may be masked by the comparatively slow rise of the strain limited by the sound velocity. Therefore, studying details of non-equilibrium electron transport would require ultrathin detection layers.

In materials with magnetic order, the excitation of the magnetic degrees of freedom has to be treated in addition to the electron and phonon subsystems. Studies of the electron-phonon coupling in ferromagnetic (FM) transition metal elements (Ni, Fe, Co) in the high excitation fluence regime observe a distinct fluence and temperature dependence of the electron-phonon coupling timescale \cite{wang2010, zahn2021, zahn2022}. In order to explain this observation the authors explicitly treated the excitation of magnetic degrees of freedom via electrons by extending the 2TM. However, the variety of different demagnetization behaviors \cite{koop2010,  batt2010, roth2012,frie2015} is related to different timescales of energy transfer to magnetic degrees of freedom.  Therefore, modeling the spatio-temporal excitation of quasi-particles requires the explicit treatment of magnetic excitations in three-temperature, N-temperature or even more complex models \cite{shok2017,sieg2019,frie2020}.

\subsection{\label{sec:III_d_Gr\"uneisen_stress} Subsystem-specific stresses and Gr\"uneisen parameters}
The prevailing paradigms in picosecond ultrasonics are thermoelastic stresses that drive the observed picosecond strain response according to the elastic wave equation~\eqref{eq:III_11_wave_equation_Gr\"uneisen}. Here, the introduced Gr\"uneisen parameter $\Gamma_{3}$ linearly relates the laser-induced stress $\sigma_{33}^\text{ext}(x_3,t)$ to the spatio-temporal energy density $\rho_Q(x_3,t)$.

This energy density initially deposited to electronic excitations is subsequently distributed within the sample structure and also locally transferred to other degrees of freedom as already discussed in Sec.~\ref{sec:III_c_heat_transport}. This heat transport and subsystem couplings determine the spatio-temporal energy densities $\rho_Q^r(x_3,t)$ stored in each subsystem $r$ that add up to the total deposited energy density $\rho_Q(x_3,t)$:
\begin{align}
\label{eq:III_subsystem_energy_density}
\rho_Q(x_3,t) = \sum_r \rho_Q^r(x_3,t)\,. 
\end{align}
Within the Gr\"uneisen approach the energy density deposited in the respective degrees of freedom is linearly related to a stress contribution $\sigma_{33}^r=\Gamma_{3}^r\rho_Q^r$ by subsystem-specific Gr\"uneisen parameters $\Gamma_{3}^r(x_3)$ \cite{nico2011,ruel2015,heni2016,pude2019,repp2020}. In case of different subsystem-specific Gr\"uneisen parameters $\Gamma_{3}^r$ the total stress in Eq.~(\ref{eq:III_11_wave_equation_Gr\"uneisen}) has to be adapted in order to individually treat the subsystem contributions to the laser-induced stress $\sigma_{33}^\text{ext}$:
\begin{align}
\label{eq:III_subsystem_stress_Gruneisen}
\sigma_{33}^\text{ext} = \sum_r \sigma_{33}^r(x_3,t)=\sum_r \Gamma_{3}^r \rho_Q^r(x_3,t)\,,
\end{align}
\ie their superposition gives the total laser-induced stress driving the picosecond strain response.

\begin{figure}[t!]
\centering
\includegraphics[width =1\columnwidth]{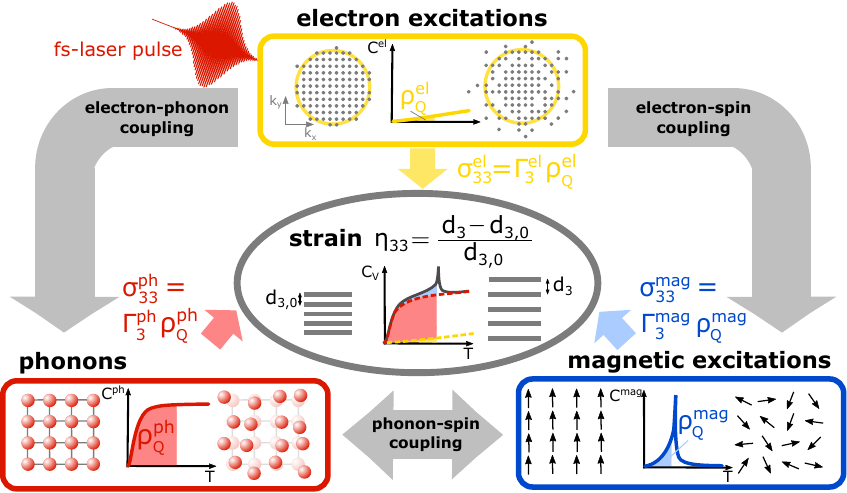}
\caption{\textbf{Gr\"uneisen concept for laser-induced stress on the lattice:} This viewgraph depicts the treatment of magnetic excitations in addition to the optically excited electrons and phonons for the laser-induced stress using the Gr\"uneisen concept. The total equilibrium heat capacity is the superposition of contributions from all subsystems. The energy density stored in their excitations generates stresses on the lattice according to the subsystem-specific Gr\"uneisen parameters $\Gamma_3^r$. Finally, the strain response is driven by the time- and space-dependent superposition of all subsystem-stress contributions that exclusively depend on the energy transfer into each subsystem.}
\label{fig:III_triangle}
\end{figure}

It depends on the properties of the material under investigation, which degrees of freedom have to be considered separately to account for the total spatio-temporal external stress. In general, the individual treatment of different degrees of freedom is only necessary for describing the picosecond strain response if the subsystem-specific Gr\"uneisen parameters differ. Conversely, the strain measurement only provides access to ultrafast microscopic processes if the involved quasi-particle excitations contribute differently to the total external stress. The separation of subsystem contributions to the strain response of the atomic lattice is schematically visualized in Fig.~\ref{fig:III_triangle} for the case of a material where electronic excitations, phonons and magnetic excitations contribute.

The quasi-instantaneous electronic stress contribution is captured by an electronic Gr\"uneisen parameter $\Gamma^\text{el}= \frac{\partial \ln{\gamma_S}}{\partial \ln{(V)}}$ that can be derived as first-order approximation from the electronic density of states at the Fermi level that determines the Sommerfeld constant $\gamma_S$ \cite{barr1980}. The value depends on details of the band structure and for the idealized case of free electrons in a parabolic band with spherical symmetry in the Sommerfeld approximation one obtains $\Gamma^\text{el}= 2/3$ \cite{barr1980}. The subsequent transfer of energy to phonons gives rise to a phonon stress that can be parameterized to first order by a macroscopic Gr\"uneisen constant $\Gamma^\text{ph}$ assuming a similar phonon population as in thermal equilibrium. However, mode-specific electron-phonon coupling can give rise to long lasting non-equilibria of the phonon system itself \cite{heni2016,mald2017} which becomes highly relevant for the modeling of picosecond acoustics in case of strongly mode-specific Gr\"uneisen parameters $\Gamma^\text{ph}_l = \frac{\partial \ln{(\hbar \omega_l)}}{\partial \ln{ (V)}}$. Here the change of the respective phonon energy $\hbar \omega_l$ with volume $V$ parametrizes the efficiency of a class of phonon modes $l$ to generate stress. In various semiconducting materials such as Tellurides the transverse phonon modes exhibit negative Gr\"uneisen parameters in contrast to the typically positive Gr\"uneisen parameter of longitudinal phonon modes \cite{whit1993,barr1980}. In thermal equilibrium this results in NTE at low temperatures since their low-frequency transverse acoustic modes are already excited at lower temperatures, as opposed to the longitudinal acoustic modes with higher frequency. 
In case of strong non-equilibria between phonon modes with strongly mode-specific Gr\"uneisen parameters upon laser-excitation, the individual treatment of the different modes is necessary. However, in metals the application of a macroscopic phononic Gr\"uneisen parameter is typically sufficient to quantify the transient phonon stress \cite{nie2006, nico2011, wang2008, pude2020b, repp2016b}.

In laser-excited magnetically ordered metals, the laser-induced spin disorder provides an additional magnetic stress contribution in addition to the electron and phonon stresses. For a Heisenberg exchange interaction the respective Gr\"uneisen parameter can be expressed as $\Gamma^\text{mag} = \frac{\partial \ln{(J)}}{\partial \ln{(V)}}$ resulting from the dependence of the exchange constant $J(V)$ on the volume of the unit cell \cite{whit1962,pytt1965,argy1967}. However, so far only few experiments \cite{reid2018,pude2019,repp2020,repp2020b,matt2021} have investigated the magnetic stress that adds to the electron and phonon stress contribution as discussed in Sec.~\ref{sec:IV_a_dy}.

Figure~\ref{fig:III_triangle} graphically represents the main idea of the Gr\"uneisen approach treating these three stresses $\sigma^r_{3}=\Gamma^r_{3} \rho_Q^r$ with $r=\{\text{el,ph,mag}\}$ that all act on the strain $\eta_{33}$. The central idea is, that by Hooke's law -- and more generally by Eq.~\eqref{eq:III_8_1d_quasi_static_strain} -- the strain is a linear measure of each of the stress contributions. In the Gr\"uneisen model for several subsystems within the same material each of the stresses is proportional to the respective energy density $\rho_Q^r$ which can be expressed via the specific heat contributions $C^r$ in thermal equilibrium. Introducing subsystem-specific Gr\"uneisen parameters is useful as the thermal expansion coefficients and the heat capacities often share the same temperature-dependence which originates from the same occupation probability of the underlying quantum states \cite{ashc2012}. In essence, the macroscopic out-of-plane Gr\"uneisen parameters $\Gamma^r_{3}$ encode, how efficiently energy densities $\rho_Q^r(x_3,t)$ in each of the subsystems generate out-of-plane stress $\sigma^r_{33}(x_3,t)$. Their extraction is exemplarily demonstrated for the rare-earth element Dy in Section~\ref{sec:IV_a_dy} and requires the separation of the specific heat and thermal expansion into the subsystem contributions \cite{nix1941,nick1971,barr2005,whit1993}

In total, the Gr\"uneisen approach provides a separation of the laser-induced stress into distinct degrees of freedom, which is linear in the contributing energy densities, and can be used even if these subsystems are out-of-equilibrium with each other for many picoseconds \cite{repp2016,mald2017,heni2016, mald2020, ritz2020}. In such situations, the stress and strain only depend on the energy density transferred between the subsystems under the boundary condition of energy conservation. Therefore, the subsystem-specific Gr\"uneisen parameters determining the transient stress contributions enable the description of expansion and contraction in thermal equilibrium and after photoexcitation on an equal footing.

\subsection{\label{sec:III_e_numerical_simulation} Numerical modelling of the picosecond strain response}
\begin{figure*}[tbh!]
\centering
\includegraphics[width =1\textwidth]{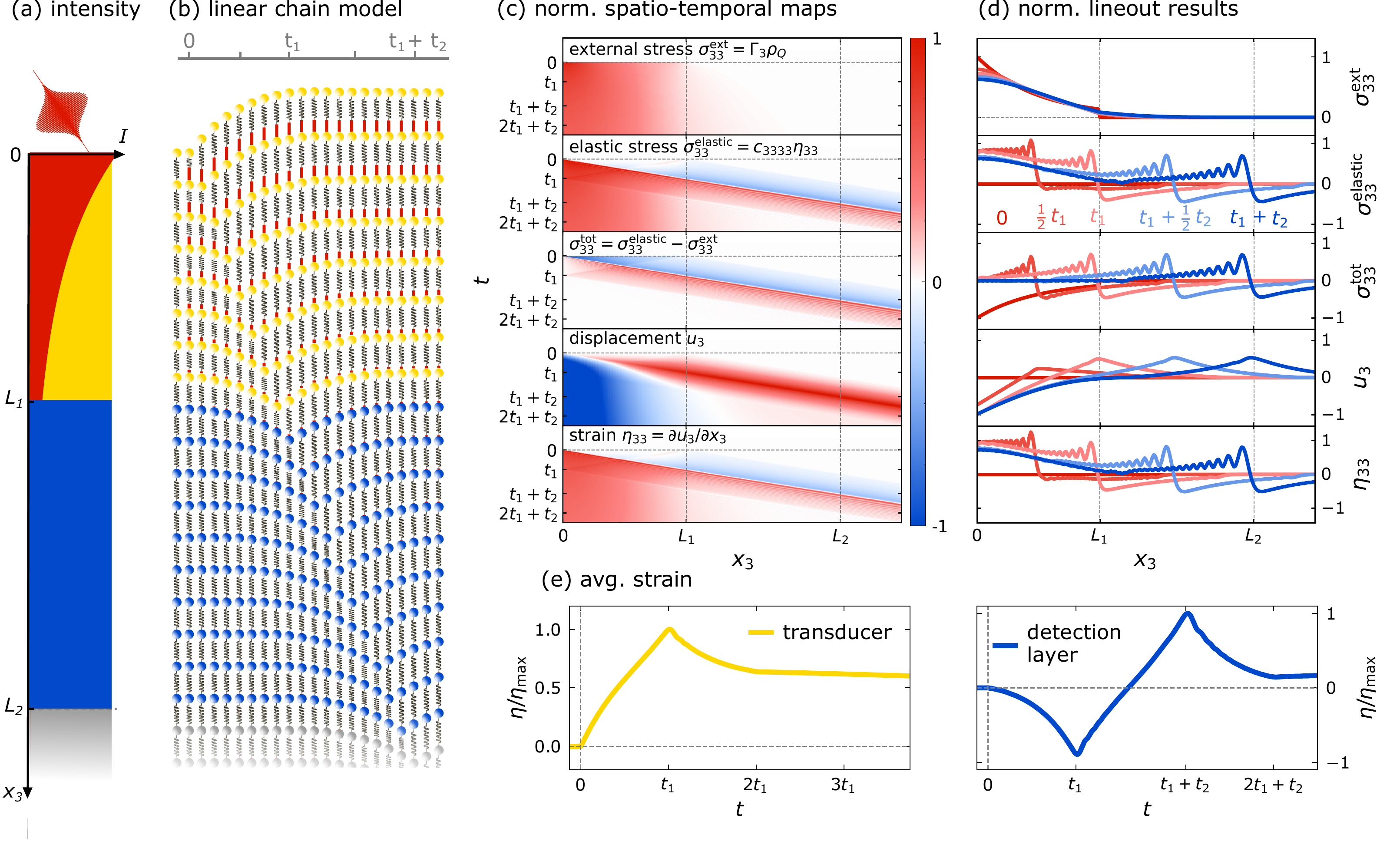}
\caption{\textbf{Visualization of the contributions to the elastic response of a laser-excited heterostructure:}  The transducer layer (thickness $L_1$) on top of a non-absorbing detection layer (thickness $L_2 > L_1$)  exhibits an absorption profile for the optical excitation that is indicated as red area on top of the sample structure (a). (b) LCM representation of the elastic response wherein spheres represent a local mass element, springs encode the elastic coupling and incompressible spacer sticks represent the laser-induced external stress. The motion of the masses is greatly enlarged, and thermal fluctuations are omitted for better visibility. The quasi-static strain is reached when the springs have relaxed to their equilibrium length and a new inter-atomic distance is attained. (c) Spatio-temporal maps of the external stress $\sigma^\text{ext}_{33}$, the elastic stress $\sigma^\text{elastic}_{33}$, the total stress $\sigma^\text{tot}_{33}$, the atomic displacement $u_{3}$, and its spatial derivative, the strain $\eta_{3}$. (d) Spatial profiles of the same quantities at selected times. (e) shows the spatially averaged strain $\eta$ of the individual layers of the bilayer heterostructure, which can be compared with the center-of-mass evolution of the Bragg peaks extracted from a PUX experiment.}
\label{fig:III_toolbox_simulation}
\end{figure*}
Modelling the time-dependent strain response allows us to obtain the spatio-temporal stress profile $\sigma_{33}^\text{ext}(x_3,t)$ that occurs as source term in the one-dimensional elastic wave equation~\eqref{eq:III_6_1d_wave_equation_strain}. This provides insights into energy transfer processes as the energy density distribution between the subsystems $\rho_Q^r(x_3,t)$ determines the external stress contributions $\sigma^{r}_{33}(x_3,t)$ as introduced in Section~\ref{sec:III_d_Gr\"uneisen_stress}. Different approaches for solving  Eq.~\eqref{eq:III_11_wave_equation_Gr\"uneisen} for a given $\sigma_{33}^\text{ext}(x_3,t)$ exist. Analytical solutions for the strain field can be constructed for time-independent stress profiles as shown in \cite{thom1986,boja2015,mats2015b}. Numerical approaches \cite{schi2014, schi2021} may be easier to implement, when the time dependence of the stresses by sub-system couplings, thermal diffusion and interface effects need to be accounted for. A natural spatial grid in a numerical simulation is provided by the atomic layers, or unit-cells which also represent the smallest physically meaningful discretization of the strain response. Linear chain models (LCMs) of masses and springs provide an intuitive approach for the numerical calculation of the strain field $\eta_{33}(x_3,t)$ with unit-cell precision \cite{herz2012b}. Publicly available implementations of a LCM are, for example, provided by the \textsc{udkm1Dsim} \textsc{Matlab} \cite{schi2014} and Python \cite{schi2021} code-libraries, which include modules for modelling the pump-laser absorption profile using a transfer matrix model, heat-transport via diffusive N-temperature models, the strain response and dynamical x-ray scattering, which can also include magnetic scattering.

In the following, we display and discuss the modeled laser-driven strain response for a generic sample structure that consists of an opaque transducer with thickness $L_1$ and a transparent detection layer of thickness $L_2$ grown on a semi-infinite, transparent substrate. Fig.~\ref{fig:III_toolbox_simulation} illustrates the sample structure and the space- and time-dependent results for the terms in the elastic wave that is solved numerically using in the \textsc{udkm1Dsim} code \cite{schi2014, schi2021}. The strain response can be visualized by a LCM as shown in Fig.~\ref{fig:III_toolbox_simulation}(b), which serves as a mechanistic analog for the time-dependent elastic response of the bilayer structure sketched in Fig.~\ref{fig:III_toolbox_simulation}(a). The time-dependent position of the masses in the LCM visualizes the displacement $u_3$ averaged over a $5\,\text{nm}$ length-fraction of the sample, where the elongation of the adjacent spring encodes the corresponding local elastic stress $\sigma^\text{elastic}_{33}$. The thickness $L_1$ of the metallic transducer is chosen to exceed the optical penetration depth leading to an inhomogeneous energy deposition shown in Fig.~\ref{fig:III_toolbox_simulation}(a). For simplicity, we assume an instantaneous rise of the laser-induced external stress $\sigma^\text{ext}_{33}$ that is indicated by the appearance of incompressible red spacer sticks compressing the springs of the linear chain directly after excitation at $t = 0$ (see Fig.~\ref{fig:III_toolbox_simulation}(b)). Unbalanced gradients in the external stress drive the elastic response, which consists of an expansion wave starting at the sample surface and a compression within the inhomogeneously excited transducer. The resulting strain pulse propagates at the speed of sound from the transducer through the detection layer into the substrate. Reflections at the interfaces are omitted for clarity by choosing perfect acoustic impedance matching of all constituents. When the strain pulse has passed, the material reaches its new equilibrium position where the residual quasi-static expansion is represented by the spacer sticks' length that varies due to heat diffusion within the heterostructure. Note that the LCM representation of the strain response in Fig.~\ref{fig:III_toolbox_simulation}(b) omits thermal fluctuations and strongly exaggerates the strain response.

The color maps and selected profiles in Fig.~\ref{fig:III_toolbox_simulation}(c) and (d) provide a numerical representation of the elastic stress, the total stress, the driven displacement $u_3$ and the corresponding strain $\eta_{33}$ that appear in the elastic wave equation~\eqref{eq:III_6_1d_wave_equation_strain}. Figures~\ref{fig:III_toolbox_simulation}(c) and (d) illustrate that the finite total stress $\sigma_{33}^\text{tot}$ arising from the unbalanced external stress $\sigma_{33}^\text{ext}$ drives a displacement of the lattice. The rising strain response $\eta_{33}$ induces an elastic stress $\sigma_{33}^\text{elastic}$ that lowers the total stress by partially compensating the external stress within the transducer. After the strain pulse has propagated into the substrate the elastic stress arising from the induced quasi-static expansion compensates the external stress which results in a vanishing total stress, which marks the quasi-static state. The resulting strain is given by the derivative of the displacement and displays both the quasi-static expansion and the strain pulse propagating through the heterostructure\footnote{The pronounced high frequency oscillations in panel (d) of Fig.~\ref{fig:III_toolbox_simulation} originate from surface vibrations, which occur due to the discretization of the strain response, which can be relevant for atomically flat samples \cite{chan1997,meln2003}. They are not present in analytic solutions of the continuum elastic wave equation.}.

Fig.~\ref{fig:III_toolbox_simulation}(e) displays the average strain of the transducer and the detection layer that can be inferred from the simulated strain field $\eta_{33}(x_3,t)$ and compared to the strain extracted from the Bragg peak shift in a PUX experiment. While the average strain of the laser-excited transducer contains contributions from the strain pulse and the quasi-static expansion, the strain response of the non-absorbing detection layer is dominated by the propagating strain pulse. The timing of the inflection points in the average strain response depends on the layer thicknesses, whereas the shape of the rising and falling edge in the strain response is related to the stress-profile as discussed in \cite{schi2014b,repp2020}. At the delay $t_1=L_1/v_\text{s}$ the strain pulse has propagated through the transducer and the compressive part has fully entered the detection layer. This causes a maximum expansion of the transducer due to the superposition of the quasi-static expansion and the expansive part of the strain pulse as well as a maximum compression of the detection layer. Subsequently, also the expansive part enters the detection layer, which reduces the expansion of the transducer until the strain pulse has completely left the transducer at $2t_1$. The entrance of the expansive part and the propagation of the compressive part from the detection layer into the substrate both result in an overall expansion of the detection layer. As the compressive part of the strain pulse has completely entered the substrate at $t_1+t_2$ the detection layer reaches its maximum expansion which subsequently decreases as the expansive part of the strain pulse exits towards the substrate. Finally, for  $t\geq 2t_1+t_2$ only the quasi-static expansion remains in the detection layer, which varies due to thermal transport.

Modeling the laser-driven strain response via the elastic wave equation is independent of experimental probing technique \eg optical probe pulses or x-ray diffraction. A direct comparison between experimental data with the model, however, requires a weighting of the modeled spatio-temporal strain map $\eta_{33}(x_3,t)$ by a sensitivity function that is specific for the detection method. The large penetration depth of hard x-rays on the order of microns enables the homogeneous probing of strain in nanostructures. Therefore, the center of mass evolution of the Bragg observed in PUX experiments is, in many cases, well-approximated by the average strain in the probed material. This no longer applies in case of variations of the crystallinity within a layer which modifies the sensitivity and thus require a weighting the modelled strain map. Furthermore, strongly inhomogeneously strained layers exhibit peak broadening or even splitting effects \cite{schi2014b,high2007} that complicate the relation between Bragg peak shift and average strain and require explicitly modelling the Bragg peak evolution via kinematical or dynamical x-ray diffraction \cite{schi2014,schi2021}. Furthermore the occurrence of laser-induced structural phase transitions may complicate the extraction of strain from the diffraction peak evolution  \cite{mari2012,mari2014, wall2018}. The presented numerical modelling approach not only helps to rationalize experimental data, but also aids in the design of the sample structure as the amplitude, shape and timing of the detected strain signal can be predicted even for complex heterostructures \cite{pude2020b,zeus2019, matt2022, repp2020}.

\section{\label{sec:IV_examples}Use cases for PUX}
This section is dedicated to the presentation of PUX experiments that illustrate the theoretical concepts discussed in the previous section by utilizing the capabilities of hard x-ray diffraction as a quantitative, material specific probing technique. 

In Section~\ref{sec:IV_a_tbfe} we exemplify how the large x-ray penetration depth extends the sensitivity of classical picosecond ultrasonics beyond the near-surface region of a metallic transducer. A quantitative comparison of the strain response of a nanogranular and a continuous FePt film in \ref{sec:IV_b_fept} highlights the importance of the Poisson stresses and geometrical constraints discussed in \ref{sec:III_a_strain_3D} and \ref{sec:III_b_strain_1D}.  Signatures of the energy transfer processes discussed in \ref{sec:III_c_heat_transport} are demonstrated in \ref{sec:IV_c_auni} using PUX on a nanoscopic Au/Ni heterostructure with and without an insulating MgO interlayer.
Section~\ref{sec:IV_a_dy}, demonstrates the utility of sub-system specific Gr\"uneisen parameters introduced in \ref{sec:III_d_Gr\"uneisen_stress} for a Dy transducer that exhibits giant magnetic stress contributions which result in a contraction upon laser-excitation.

\subsection{\label{sec:IV_a_tbfe}Sensing shape and timing of strain pulses in buried layers}
Observing the timing, shape and amplitude of picosecond strain pulses is central to picosecond ultrasonics experiments. Here, we showcase the ability of PUX to track the propagation of strain pulses within an opaque heterostructure consisting of a thick transducer on a thin detection layer \cite{zeus2019}. We illustrate that the presence of a transparent capping layer on top of the transducer leads to the emission of unipolar strain pulses towards the sample surface and a pronounced asymmetry of the bipolar strain-pulse that propagates towards the substrate. 

\begin{figure}[t!]
\centering
\includegraphics[width =1\columnwidth]{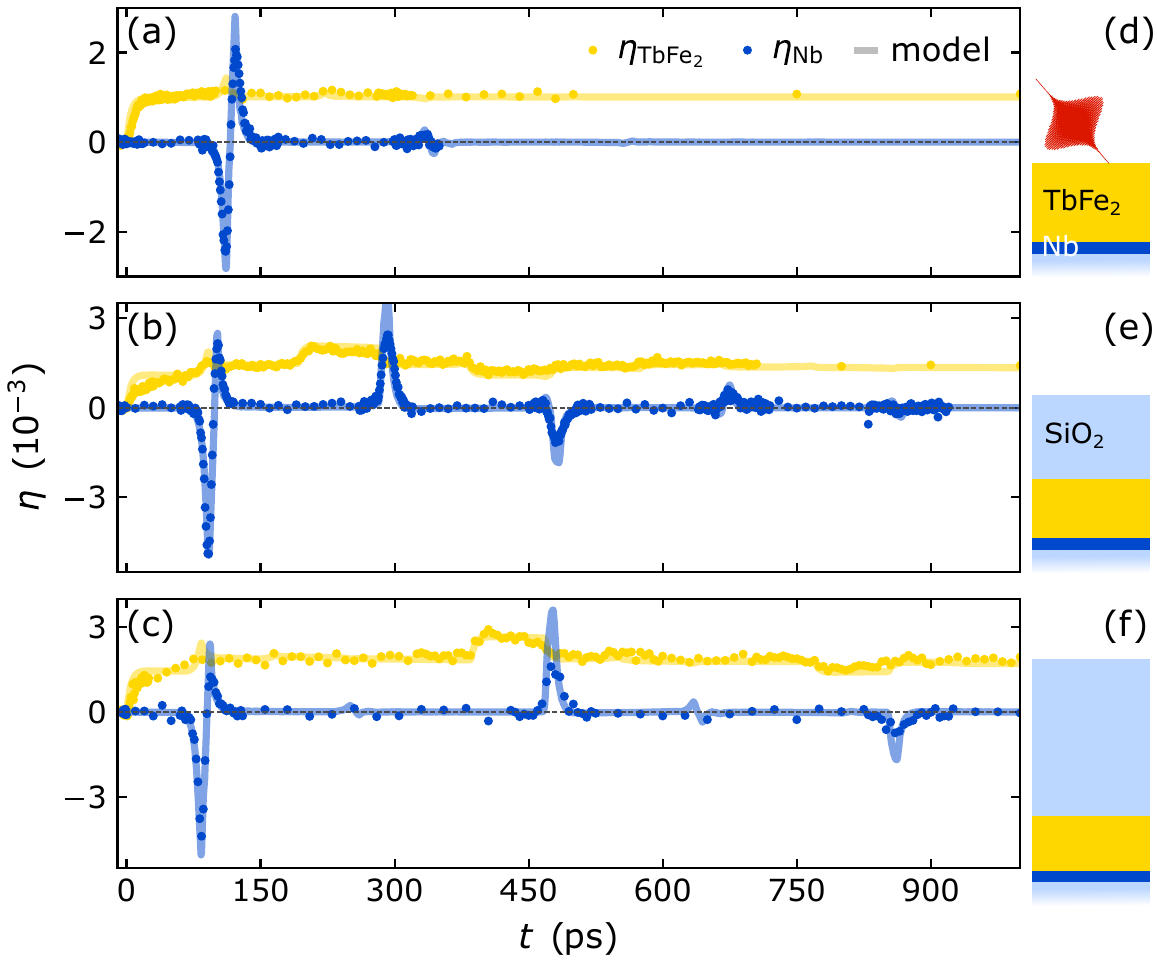}
\caption{\textbf{Sensing propagating strain waves within an opaque heterostructure with a variable capping:} Transient strain of the metallic TbFe$_2$ transducer (yellow) and Nb detection layer (blue) for an uncapped structure (a),  a structure with a transparent $550\,\text{nm}$ SiO$_2$ capping (b) and a $1100\,\text{nm}$ SiO$_\mathrm{2}$ capping (c). Solid lines represent the modeled strain response using the \textsc{udkm1Dsim} toolbox \cite{schi2014} using parameters given in \cite{zeus2019}. Panels (d) to (f) sketch the corresponding sample structures that are laser-excited from the TbFe$_\mathrm{2}$ side. A comparison of the strain response shows that the SiO$_\mathrm{2}$ capping layer separates the bipolar strain pulse emitted by the uncapped transducer (a) into an asymmetric bipolar strain pulse and a unipolar strain pulse (b, c) whose delay with respect to the bipolar pulse is set by the thickness of the capping layer and its sound velocity.}
\label{fig:IV_tbfe_data}
\end{figure}

In particular, we discuss the strain response of heterostructures that consist of a few hundred nm thick TbFe$_\mathrm{2}$ transducer on top of a $50\,\text{nm}$ thin Nb detection layer grown on a Al$_\mathrm{2}$O$_\mathrm{3}$ substrate, which are capped with an amorphous, transparent SiO$_\mathrm{2}$ layer with a variable thickness ranging from $0$ to $1100\,\text{nm}$ as sketched in Fig.~\ref{fig:IV_tbfe_data}(d--f). The samples are excited by femtosecond laser pulses with an optical penetration depth of $30\,\text{nm}$ in the TbFe$_\mathrm{2}$ layer and the resulting strain responses observed via PUX are depicted in Fig.~\ref{fig:IV_tbfe_data}(a--c). Further details on the sample growth and experimental parameters are given in \cite{zeus2019}. 

The femtosecond laser pump-pulse deposits energy into the near surface region inducing an expansion of the $450\,\text{nm}$-thick TbFe$_\mathrm{2}$ transducer of the uncapped structure, that launches a bipolar strain pulse with a leading compression and a trailing expansion propagating through the metal heterostructure into the substrate as discussed in Sec.~\ref{sec:III_e_numerical_simulation}. Fig.~\ref{fig:IV_tbfe_data}(a) displays the corresponding rise of the average strain of the TbFe$_\mathrm{2}$ layer and the bipolar shape of the strain response of the Nb detection layer that are extracted from the diffraction peak shifts. While the strain pulse travels at the speed of sound, the deposited energy density causing a quasi-static expansion of the transducer diffuses into the Nb detection layer on a nanosecond timescale. The different propagation speeds of sound and heat thus yield a background-free signature of the strain pulse in the buried Nb detection layer. The average strain in Nb is determined by the integral over the part of the strain pulse within the Nb layer. When the compressive part enters the Nb layer at $100\,\text{ps}$ it exhibits a negative strain until $125\,\text{ps}$ when the trailing tensile part has entered the layer and the compressive part has progressed towards the substrate. 
Subsequently, the tensile part dominates and the resulting positive strain recedes as the strain pulse propagates into the substrate. The average strain of the Nb layer detected via PUX is thus determined by the spatial shape of the bipolar strain pulse convoluted with the propagation of the strain pulse through the layer.

Adding an additional transparent SiO$_\mathrm{2}$ layer on top of a $350\,\text{nm}$ TbFe$_\mathrm{2}$ transducer modifies both the shape and timing of the emitted strain pulses as shown in Fig.~\ref{fig:IV_tbfe_data}(b) and (c). As before, the expansion of the TbFe$_\mathrm{2}$ launches a bipolar strain pulse propagating towards the Nb detection layer. In addition, a unipolar compression pulse is launched into the SiO$_\mathrm{2}$ capping. Consequently, the amplitude of the tensile part of the initial bipolar strain pulse detected in the buried Nb layer is reduced due to conservation of elastic energy. Subsequently, we observes a train of unipolar strain pulses with alternating sign and a period that is given by the propagation time of the strain pulse back and forth through the capping layer. The presence of the unipolar strain pulse inside the thick TbFe$_\mathrm{2}$ layer is heralded by small plateau-like signatures in the measured TbFe$_\mathrm{2}$ strain, which is rather insensitive to the shape of the strain pulses in comparison to the short, pronounced peaks that occur in the strain response of the thin Nb detection layer. It is interesting to note that if we extrapolated the results in Fig.~\ref{fig:IV_tbfe_data}(b) and (c) with a cap layer thickness of $550\,\text{nm}$ and $1100\,\text{nm}$, respectively, to a cap layer thickness of zero, the positive unipolar strain pulse would superimpose onto the asymmetric bipolar pulse in such a way that the symmetric bipolar pulse from Fig.~\ref{fig:IV_tbfe_data}(a) is recovered. One can therefore rationalize the occurrence of the tensile part of the bipolar strain pulse shape by an instantaneous and complete reflection of a compression wave driven at the transducer-air interface.
\begin{figure}[b!]
\centering
\includegraphics[width =1\columnwidth]{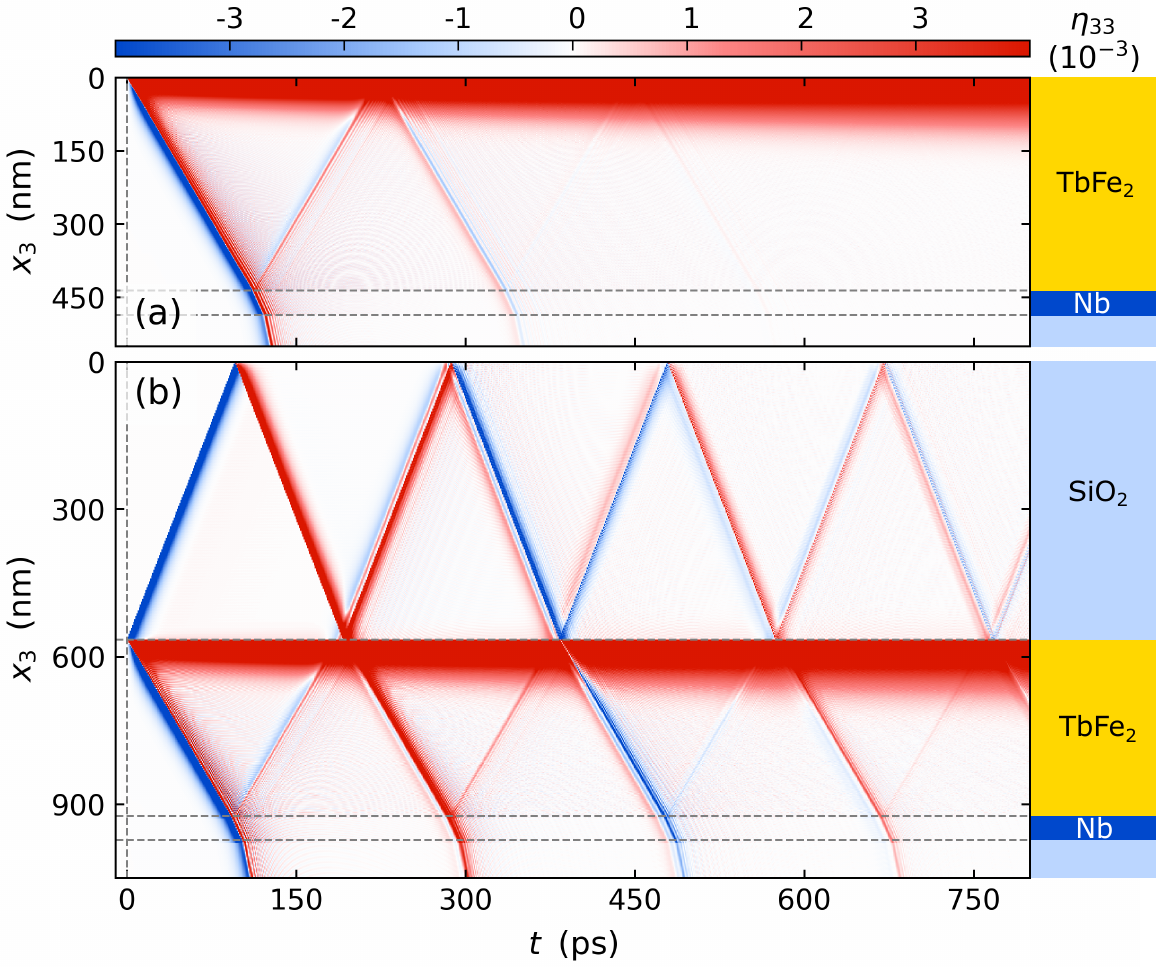}
\caption{\textbf{Modeled strain map:} Spatio-temporal strain of the sample structure without SiO$_\mathrm{2}$ capping (a) and with $550\,\text{nm}$ capping (b) simulated using the modular \textsc{udkm1Dsim} toolbox \cite{schi2014}. The spatio-temporal strain displays the propagation of the strain pulses through the heterostructure and the distribution of heat via heat diffusion indicated by the slowly growing expanded part of the TbFe$_\mathrm{2}$ transducer.}
\label{fig:IV_tbfe_simulation}
\end{figure}

We modelled the strain response of the TbFe$_\mathrm{2}$ transducer and the Nb detection layer for all heterostructures using the approach introduced in Section~\ref{sec:III_e_numerical_simulation} with parameters given in \cite{zeus2019}. The modeled average layer strains shown as solid lines in Fig.~\ref{fig:IV_tbfe_data} match the experimental data indicated by symbols. Moreover, the modelling yields detailed spatio-temporal maps of the strain inside the heterostructures without and with SiO$_\mathrm{2}$ capping layer which are depicted in Fig.~\ref{fig:IV_tbfe_simulation}(a) and (b), respectively. The modeled strain maps provide a detailed depiction of the strain profile within the structure at any given time, which extends the insights from the average layer strain obtained via PUX experiments. Fig.~\ref{fig:IV_tbfe_simulation}(b) shows that the unipolar strain pulse launched into the capping layer inverts its sign upon reflection at the sample surface and that a considerable fraction of the returning wave is reflected at the SiO$_\mathrm{2}$-TbFe$_\mathrm{2}$-interface due to an acoustic impedance mismatch between these two materials.  This rationalizes the multiple echoes of unipolar pulse that consecutively traverse the structure as revealed in Fig.~\ref{fig:IV_tbfe_data}(b) and (c). Modeling the strain allows us to precisely calibrate the layer thicknesses from the timing of the detected strain pulses. However, the modeled strain map does not only help to rationalize the strain observed in a PUX experiment but can furthermore guide and support the interpretation of strain signatures in all-optical data, \eg from time-resolved magneto-optics and reflectivity \cite{peze2016, zeus2019, parp2021}.  

Overall, we find that the observation of the strain pulses in a thin crystalline detection layer is advantageous compared to the analysis of the strain response of a thick transducer layer. The separation of the detection layer from the laser-excitation region separates the strain signatures of sound and heat in the time domain which can be exploited for the detection of strain pulses with an unconventional shape, as shown in Section~\ref{sec:IV_a_dy}. The current example shows that in principle only the detection layer needs to be crystalline to observe the strain pulses, which extends the applicability of PUX to a large class of heterostructures and transducer materials.

\subsection{\label{sec:IV_b_fept}Quantifying a morphology-dependent strain response} 
Here, we compare the qualitative and quantitative picosecond strain response of thin films for various in-plane expansion constraints that are shown to affect the out-of-plane strain response. Different sample morphologies change the nature of the ultrafast strain response from one-dimensional in the case of a homogeneous film to three-dimensional in the case of nanograins even if they are attached to a substrate. Describing the ultrafast strain response of granular morphologies requires a model for the three-dimensional elastic response in accordance with Eq.~\eqref{eq:III_2_wave_equation_without_shear} which includes Poisson stress contributions that couple in-plane and out-of-plane motion. 

\begin{figure}[tbh!]
\centering
\includegraphics[width = 1\columnwidth]{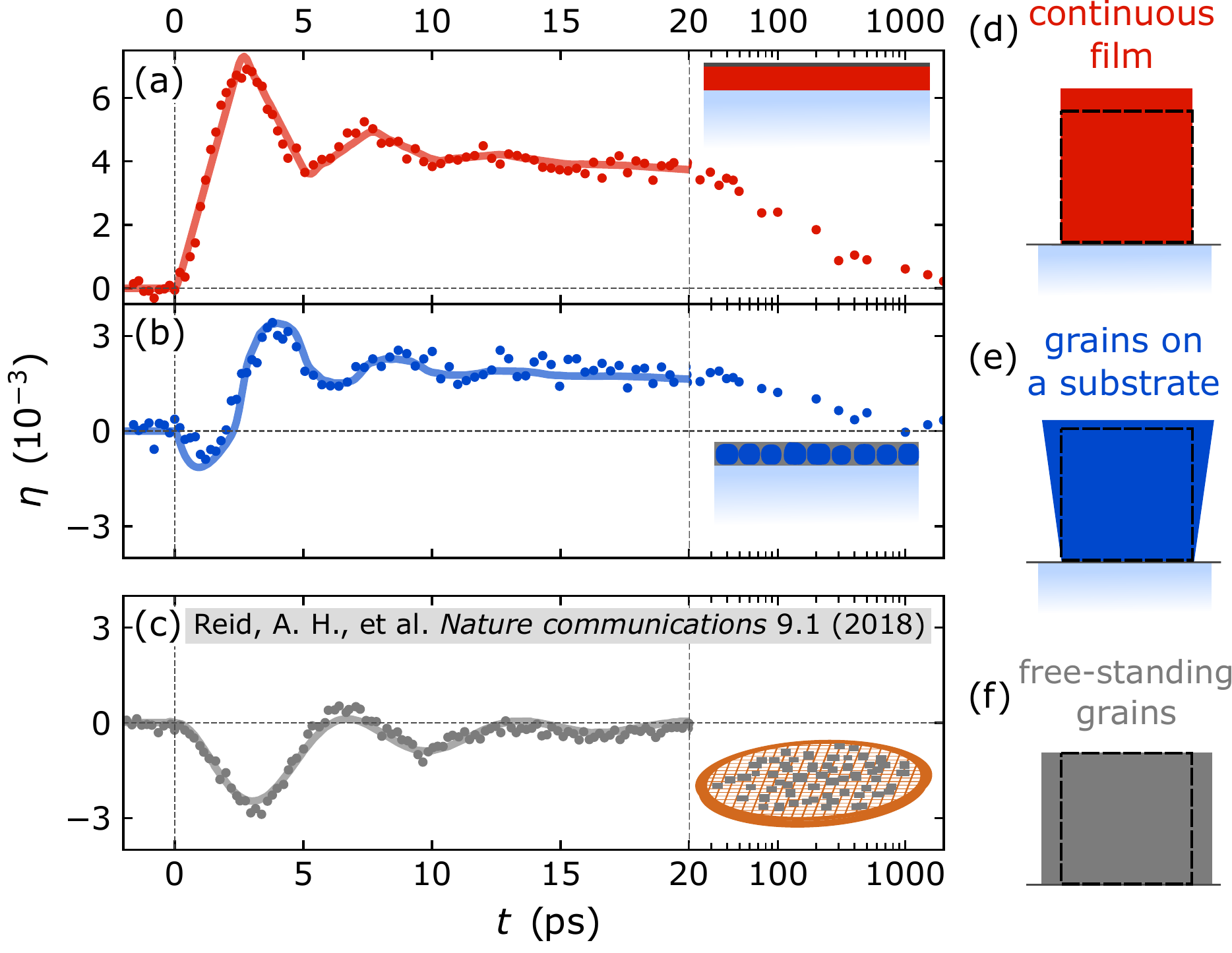}
\caption{\textbf{Morphology-dependent strain response of FePt $10\,\text{nm}$ specimen:} Comparison of the laser-induced out-of-plane strain-response of $\text{L1}_0$-phase FePt for three different in-plane boundary conditions: (a) continuous thin film (b) FePt nanograins on an MgO substrate (c) free-standing FePt grains on a TEM grid. The data in panel (a) and (b) are from UXRD measurements \cite{repp2018} at an incident laser fluence of $\approx 6\,\mathrm{{mJ}/{cm^2}}$. The data in panel (c) are reproduced from the publication by Reid \etal \cite{reid2018} in order to highlight the qualitative differences in the FePt strain response.  Reid \etal determined the strain response for a comparable laser-fluence of $5\,\mathrm{{mJ}/{cm^2}}$ using time-resolved electron diffraction. Solid lines in in (a--c) are derived from modeling approaches discussed in the corresponding publications \cite{repp2018, reid2018}  and schematic insets  depict the sample morphologies. Panels (d--f) illustrate the constrained expansion with respect to the initial condition indicated as black dashed line.}
\label{fig:IV_comparison_fept} 
\end{figure}

Specifically, we discuss the picosecond strain response of three $10\,\text{nm}$ thin crystalline $\text{L1}_0$-phase FePt layers having their tetragonal c-axis oriented out-of-plane. Interestingly, the $\text{L1}_0$-phase of FePt exhibits a vanishing expansion along its c-axis ($\alpha_3 \approx  0$  \cite{tsun2004,repp2020b}) under near-equilibrium heating conditions and a distinct expansion along the in-plane directions ($\alpha_1 = \alpha_2  \approx  9 \cdot 10^{-6}$)  \cite{rasm2005,repp2020b}. Fig.~\ref{fig:IV_comparison_fept} contrasts the laser-induced strain response along the tetragonal c-axis for a continuous and a granular thin film on an MgO substrate as measured via UXRD \cite{repp2018} and a quasi-free-standing granular film transferred to a transmission electron microscopy (TEM) grid measured via ultrafast electron diffraction \cite{reid2018}. The compiled experiments used comparable laser excitation conditions with a fluence of $\approx 6\,\mathrm{mJ/cm^2}$. Both substrate-supported FePt samples, \ie the continuous and the granular thin film, respond on average by an out-of-plane expansion upon laser-excitation despite the Invar-like behavior in near-equilibrium heating conditions. However, the free-standing grains on the TEM grid exhibit an initial out-of-plane contraction while the lattice expands in-plane (not shown here)  \cite{reid2018}. After the coherent strain pulse oscillations have ceased, one observes a nearly vanishing out-of-plane expansion, in agreement with the Invar-like behavior that is expected from quasi-static heating experiments.   

The schematic depictions (Fig.~\ref{fig:IV_comparison_fept}(d--f)) adjacent to the data illustrate the hypothesis for the data interpretation that rationalizes the observed behaviors \cite{repp2018, repp2020b}. They sketch the equilibrium film dimensions as a dashed black rectangle and the change in dimensions upon laser excitation in color. In case of the free-standing grains the modeling of the picosecond strain response requires the three-dimensional elastic wave equation Eq.~\eqref{eq:III_2_wave_equation_without_shear} that includes a coupling of the in-plane and out-of-plane strain response. The non-vanishing laser-driven in-plane strains introduce a time-dependent out-of-plane Poisson stress contribution in addition to the laser-induced out-of-plane stresses from electrons, phonons and spin-excitations \cite{repp2020b, reid2018}. The rising in-plane strain induces a contractive out-of-plane Poisson stress that dominates the strain response of the free-standing grains. When the coherent motion has ceased at $\approx 20\,\text{ps}$, the remaining quasi-static expansion is given by the near equilibrium thermal expansion coefficient $\alpha_{3}\approx 0$ according to Eq.~\eqref{eq:III_3_general_quasi_static_strain} as indicated by the out-of-plane Invar behavior and an in-plane expansion. 

In contrast, the strictly one-dimensional picosecond strain response of the continuous film along the out-of-plane direction can be described by the one-dimensional elastic wave equation~\eqref{eq:III_6_1d_wave_equation_strain} derived in Section~\ref{sec:III_b_strain_1D}. Because the homogeneous laser excitation of a thin film lacks in-plane stress gradients, there is no in-plane motion on ultrashort timescales. The absence of the Poisson stress contributions modifies the out-of-plane thermal expansion coefficient on ultrafast timescales according to Eq.~\eqref{eq:III_9_alpha_ultrafast}, which in the case of FePt simplifies to: 
\begin{align}
\label{eq:IV_1_FePt }
\alpha_{3}^\text{uf} &= \alpha_{3}  +  2 \frac{c_{3311}}{c_{3333}}\alpha_{1} = \alpha_{3} + 2 \frac{\nu}{1-\nu} \alpha_{1}\,. 
\end{align}
Assuming a Poisson ratio $\nu = 1/3$, as is common for metals one finds that $\alpha_{3}^\text{uf}$ is equal to $\alpha_{1} $ and thus positive despite the vanishing $\alpha_{3}$. According to Eq.~\eqref{eq:III_5_definition_Gr\"uneisen_constant} this is related to a positive out-of-plane Gr\"uneisen parameter that translates deposited energy density to a positive out-of-plane stress $\sigma_{33}^\text{ext}$ that is fully compensated by the Poisson stress in thermal equilibrium.

The nanograins attached to a substrate represent an intermediate case between the freestanding grains and the continuous film morphology. Accordingly, they exhibit an intermediate behavior with a short-lived contraction that is followed by a similar, albeit smaller out-of-plane thermal expansion compared to the continuous FePt film. The clamping to the substrate partially hinders the in-plane expansion of the FePt grains as sketched in Fig.~\ref{fig:IV_comparison_fept}(e). The nature of the strain response for this specimen is intrinsically three-dimensional and further complicated by the additional boundary conditions introduced by the clamping to the substrate, by a carbon filling in between the FePt grains and the grain size distribution. Finite-element modeling is however able to reproduce the observed short-lived contraction followed by an out-of-plane expansion for these boundary conditions \cite{repp2020b}. The partially allowed in-plane expansion induces a Poisson stress contribution that reduces the observed out-of-plane expansion in the quasi-static limit for $t \geq 20\,\mathrm{ps}$ compared to the continuous thin film but does not lead to the Invar-like behavior of the free-standing grains.

This use case illustrates the crucial role of the constraints for the in-plane expansion on the out-of-plane strain response. Utilizing the picosecond strain response for a quantitative determination of the laser-induced energy density or even temperature change therefore requires a careful consideration of the in-plane boundary conditions in the sample. Application of the near-equilibrium linear thermal expansion coefficient in the common thin film geometry overestimates the laser-induced temperature change, due to the absence of the Poisson stress contributions on ultrafast timescales. Nanoscale granularity drastically influences the amplitude of the out-of-plane strain, not only for samples purposely grown as lateral nanostructures but also for thin granular films or islands that unintentionally form during sample growth. However, if the Poisson effect is taken into account, time-resolved strain measurements via PUX may serve for tracking ultrafast energy transfer processes and enable thermometry applications in nanoscale structures.

\subsection{\label{sec:IV_c_auni}Identifying non-equilibrium heat transport between ultrathin layers}
Here, we showcase the ability of PUX to observe the non-equilibrium energy transfer within nanoscale heterostructures down to few-unit-cell thickness. Time-resolved, quantitative strain measurements extend the applicability of thermal expansion as a proxy for energy content and temperature changes from quasi-equilibrium to the picosecond timescale. The shape of the picosecond strain pulses provides supporting insights into the spatio-temporal stress profile that is determined by the energy density distribution and heat transfer within the structure which we quantify by modelling the strain response.

In the following, we discuss the ultrafast energy transport within a metallic bilayer of $5\,\text{nm}$ Au and $13\,\text{nm}$ Ni in the framework of a diffusive 2 TM (Eq.~\eqref{eq:III_heat_diffusion_2TM}) that we previously applied to capture the energy transfer and resulting strain response in similar Au-Ni \cite{pude2018,herz2022} and Au-Fe bilayers \cite{matt2022} across a large range of layer thicknesses. The picosecond strain response of both Au and Ni serves as a reference for the heat transfer that is dominated by electronic transport within the first hundreds of femtoseconds. In order to experimentally verify this crucial electron energy transport we compare the strain response of the Au-Ni bilayer to the strain response of a Au-MgO-Ni trilayer where the $8\,\text{nm}$ thin insulating MgO interlayer inhibits electronic energy transfer among the metal layers.
\begin{figure}[t!]
\centering
\includegraphics[width = 1\columnwidth]{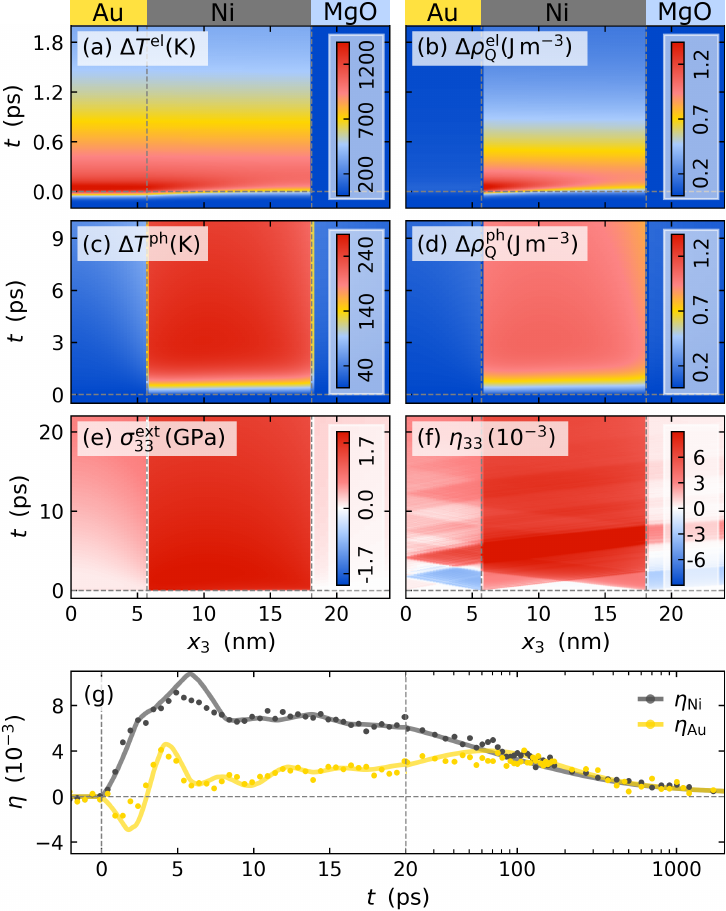}
\caption{\label{fig:IV_auni_sim} \textbf{Subsystem-specific results from modelling the strain response of a thin Au-Ni bilayer excited by 400\,nm pump pulses using a diffusive 2TM:} (a) spatio-temporal  electron temperature increase $\Delta T^\text{el}$, (b) energy density increase in the electron system $\Delta \rho_Q^\text{el}$, (c) phonon temperature increase $\Delta T^\text{ph}$ and (d) phonon energy density increase $\Delta \rho_Q^\text{el}$. Panel (e) displays the resulting laser-induced stress that drives the strain response  depicted in panel (f). (g) Layer-selective spatial averaging yields the average strain of Au and Ni (solid lines) in reasonable qualitative and quantitative agreement with the experimental strain response (dots). The modelling illustrates the ultrafast electronic energy transfer, that is evident from the expansion of the Ni which compresses the Au, that in turn only heats up only within tens of ps as discussed in \cite{pude2018,herz2022}.}
\end{figure}

Fig.~\ref{fig:IV_auni_sim}(g) displays the picosecond strain response of the Au (yellow dots) and the Ni (grey dots) layer of the Au-Ni bilayer to a $100\,\text{fs}$ laser pulse with a central wavelength of $400\,\text{nm}$ observed via PUX. The modeling process yields spatio-temporal electron and phonon temperature maps and the corresponding energy densities that are depicted in Fig.~\ref{fig:IV_auni_sim}(a--d). The electron temperature map in Fig.~\ref{fig:IV_auni_sim}(a) displays an equilibration of the electron temperature across the metal stack within the first $200\,\text{fs}$ after laser excitation. The flat 3d-bands of Ni close to the Fermi level provide a large density of states, which results in a large electronic specific heat and also a large electron-phonon coupling strength compared to Au \cite{lin2008}. The thermalization of the Au and Ni electrons thus rapidly transfers most of the energy to the Ni electrons as visualized by the electronic energy density map in Fig.~\ref{fig:IV_auni_sim}(b). Subsequently, the energy is transferred to phonons in Ni via electron-phonon coupling within the first $1\,\text{ps}$ to $2\,\text{ps}$, which lowers the overall electron temperature and the energy density stored in electron excitations. The phonon temperatures of both layers equilibrate only on the timescale of tens of picoseconds as discussed in \cite{pude2018,herz2022} and illustrated by Fig.~\ref{fig:IV_auni_sim}(c). 

The rapidly rising energy density in the electrons and the phonons of Ni induces a rapidly rising expansive stress via the respective positive Gr\"uneisen parameters (see Section~\ref{sec:III_d_Gr\"uneisen_stress}) \cite{nix1941,wang2008} which remains unbalanced at the Au-Ni interface and drives a rapid expansion of the Ni layer that compresses the adjacent Au layer.
The compression pulse propagates through the Au layer to the sample surface where it is turned into an expansion pulse upon reflection. This propagation of the strain pulse through the heterostructure superimposed with the quasi-static expansion of the layers is displayed in the strain map in Fig.~\ref{fig:IV_auni_sim}(f) that is obtained by solving the elastic wave equation~\eqref{eq:III_6_1d_wave_equation_strain} for the modelled external stress $\sigma_{33}^\text{ext}(x_3,t)$. Averaging the spatio-temporal strain over the Au and the Ni layer, respectively, enables the comparison of the model with the experimental data for the average strains of the layers shown in Fig.~\ref{fig:IV_auni_sim}(g). The bipolar feature in the Au strain in combination with the rapidly rising expansion of Ni indicates the initial compression of the Au layer due to the dominant energy transfer into Ni. In addition, the slowly rising expansion of Au on tens of picoseconds indicates the surprisingly slow equilibration of both metal layers due to a backward transfer of energy from Ni into the Au phonon system via conduction electrons and weak electron-phonon coupling in Au and via phonon transport across the Ni-Au interface. For heterostructures with Au thinner than $\approx 10\,\text{nm}$ one finds that the energy transfer between the two metals occurs predominantly via phonons \cite{herz2022}. The Au layer reaches its maximum quasi-static strain and thus temperature at $\approx 100\,\text{ps}$ after which the entire bilayer cools towards the MgO substrate via diffusive phonon transport, which results in the overall decrease of the strain in Fig.~\ref{fig:IV_auni_sim}(g).

\begin{figure}[tbp!]
\centering
\includegraphics[width = \columnwidth]{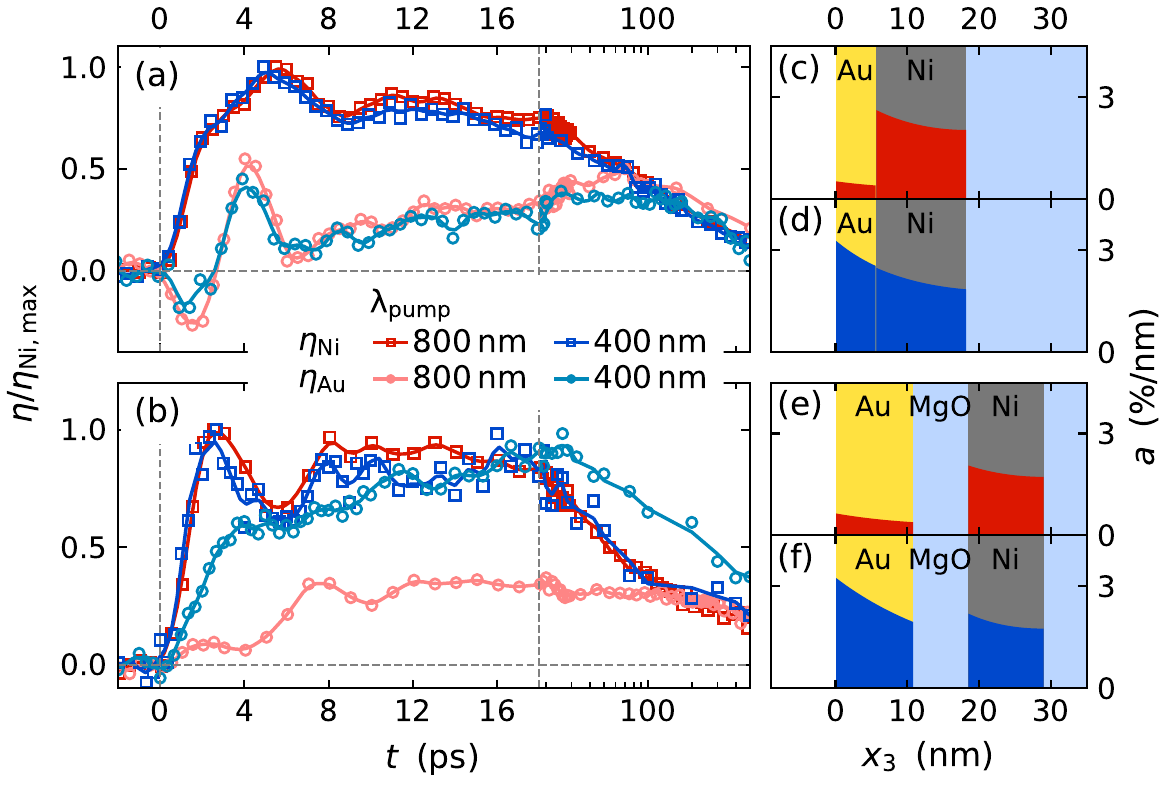}
\caption{\label{fig:IV_auni} \textbf{Strain response for wavelength-dependent energy deposition profiles in bi- and trilayer Au-Ni samples:} Comparison between the strain response of a sample without insulating barrier (a) and with insulating MgO barrier (b) subjected to a laser pulses with $800\,\text{nm}$ (red lines) and $400\,\text{nm}$ (blue lines) central wavelength. Solid lines are spline interpolations to the data that serve as guide to the eye. Panels (c--f) represent the modeled absorbance per length $a$ in the bi- and trilayer structure from a transfer matrix model that uses the refractive indices for similarly thin Au films \cite{yaku2019} for $800$ (red areas) and $400\,\text{nm}$ (blue areas) excitation.}
\end{figure}

It is instructive to consider further excitation scenarios and sample structures that support and extend our findings. Fig.~\ref{fig:IV_auni}(a) compares the strain response of the Au and the Ni layer to an ultrashort laser pulse excitation with a central wavelength of $400\,\text{nm}$ and $800\,\text{nm}$. Fig.~\ref{fig:IV_auni}(c) and (d) display the corresponding spatial absorption profiles calculated from literature values for the optical constants using a transfer matrix model \cite{khor2014,esch2014,pude2018}. While the absorption in Ni is essentially independent of the excitation wavelength, the Au absorption is heavily increased at $400\,\text{nm}$. However, the strain response in the bilayer does not depend on where the optically deposited energy is absorbed except for an overall scaling factor, because the heat energy in the electron system is rapidly distributed throughout the metallic heterostructure. The strong electron-phonon coupling in Ni localizes the energy in Ni phonons that dominate the overall strain response. The spatial profile of the source term $S(x_3,t)$ in the diffusive 2TM~\eqref{eq:III_heat_diffusion_2TM} is not highly relevant. 

This changes if we insert an insulating MgO layer between Au and Ni that inhibits the ultrafast electronic energy transfer between the layers (Fig.~\ref{fig:IV_auni}(b)). Now the Au and Ni layers both expand individually due to the optical excitation of electrons and subsequent local electron-phonon coupling. The absorbed energy density in the Au and Ni layers (Fig.~\ref{fig:IV_auni}(e) and (f)) determines the  stress profile which sets shape and amplitudes of the picosecond strain pulses and the quasi-static expansion of the individual layers within the first picoseconds. Phonon heat transport subsequently equilibrates the temperature in the trilayer within tens of picoseconds and cools it towards the substrate on a timescale of hundreds of picoseconds. In the trilayer, both the coherent phonon excitation (picosecond strain pulses) and the incoherent phonon population (thermal energy) depend drastically on the excitation wavelength as seen in Fig.~\ref{fig:IV_auni}(b). According to the increased absorption of the Au layer for 400\,nm excitation, Au shows a strongly enhanced quasi-static expansion around $10\,\text{ps}$ in comparison to $800\,\text{nm}$ excitation. The strain oscillation signature of the coherently driven picosecond strain pulse observed at $4\,\text{ps}$ in Au, which is attributed to the compression pulse launched by the Ni expansion also heralds the varied energy distribution. In comparison to the bilayer the compression signature is much less pronounced in the trilayer where the electron energy transfer from Au is prohibited. Moreover for $400\,\text{nm}$ excitation the laser-induced stress of the Au and Ni layers is almost equal as indicated by the similar quasi-static strain around $10\,\text{ps}$ to $20\,\text{ps}$. Accordingly, the amplitudes of the strain pulses launched by Au and Ni become comparable and the oscillatory signatures in the early Au strain response essentially cancel.

This use case thus displays various signatures of energy transfer in nanoscale heterostructures and illustrates that the insertion of an insulating interlayer suppresses the electronic transport that is key for the surprisingly slow thermal equilibration in similar structures \cite{pude2018,matt2022,herz2022}. Ultrafast electronic energy transport within metallic heterostructures has been reported and utilized previously in various all-optical experiments \cite{wang2012,choi2014,shin2020}. However, the layer dimensions often need to be larger than the optical skin depth in the involved metals in order to unequivocally attribute the experimental signals to the layers. The layer-specificity of PUX overcomes this limitation and thus enables such investigations on few-nm thin films.

\subsection{\label{sec:IV_a_dy} Detecting ultrafast negative thermal expansion}
In this section, we discuss the strain response of a magnetostrictive transducer which exhibits NTE caused by a contractive stress originating from magnetic excitations. Depending on the temperature, \ie the initial magnetic state, we observe a laser-induced expansion or contraction of the transducer, which we identify via Bragg peak shifts. The strain response of a buried detection layer reveals that parts of the inhomogeneously excited transducer expand while others contract upon laser excitation. We apply the Gr\"uneisen concept to individually treat the stress contributions of electrons, phonons and magnetic excitations via subsystem-specific Gr\"uneisen parameters extracted from near-equilibrium thermal expansion experiments. Based on the strain response of the transducer and the detection layer, which encodes the shape of the strain pulse, we extract the spatio-temporal subsystem-specific stress contributions. The proportionality of energy and stress in the Gr\"uneisen concept thus provides insight into energy transfer processes between the subsystems.

\begin{figure}[tb!]
\centering
\includegraphics[width = 1\columnwidth]{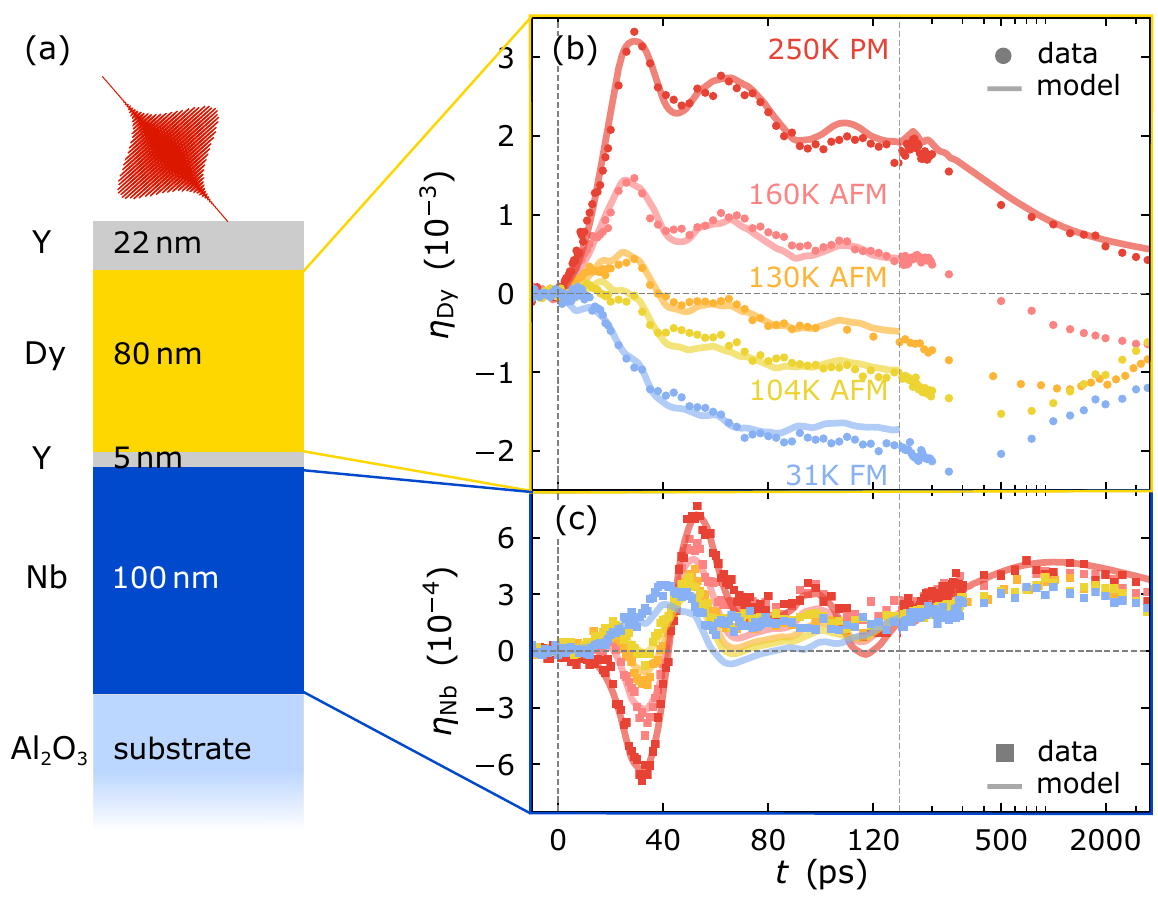}
\caption{\textbf{Temperature-dependent picosecond strain driven by the laser-excited Dy transducer:} The investigated heterostructure sketched in (a) contains an opaque Dy transducer and a buried Nb detection layer. Panels (b) and (c) display the picosecond strain response of Dy and Nb at sample temperatures above and below $T_\text{N}=180\,\text{K}$ for a fixed laser fluence of $7.2\,\mathrm{mJ/cm^2}$. Lowering the temperature below $T_\text{N}$ changes the response of Dy from expansive to contractive and modifies the shape of the emitted picosecond strain pulse that is detected in the Nb layer. Solid lines depict the modelling results as reported in a previous publication \cite{repp2020}.}
\label{fig:IV_dy_transient_strain}
\end{figure}

Specifically, we discuss the picosecond strain response of a heterostructure that consists of a $80\,\text{nm}$ Dy transducer embedded between an Y capping and buffer layer on top of a $100\,\text{nm}$ buried Nb detection layer (Fig.~\ref{fig:IV_dy_transient_strain}(a)). Below its Néel temperature ($T_\text{N}=180\,\text{K}$) the rare-earth Dy transducer hosts a helical antiferromagnetic (AFM) order of its large localized 4f-magnetic moments along the c-axis of its hexagonal unit cell which is oriented along the out-of-plane direction. At its Curie-temperature ($T_\text{C}=60\,\text{K}$) the Dy layer undergoes a first-order phase transition below which the magnetic order becomes FM \cite{dume1996}. 

Figures~\ref{fig:IV_dy_transient_strain}(b) and (c) display the transient average strain of the Dy transducer and Nb detection layer, respectively, extracted from the shift of their material-specific Bragg peaks for a fixed laser fluence of $7.2\,\mathrm{mJ/cm^2}$  \cite{repp2020}. Measurements at selected temperatures above and below $T_\text{N}=180\,\text{K}$ compare the response for initial states with and without magnetic order, which directly shows that the laser-induced disorder of the spin system provides a contractive magnetic stress. The Dy transducer discussed in this section serves as a representative of the class of heavy rare-earth elements. We obtain similar findings for Gd \cite{koc2017} and Ho \cite{pude2019} in their respective FM and AFM phases.

In the paramagnetic (PM) phase at $250\,\text{K}$, we observe the conventional strain response of a laser-excited metallic transducer and a buried detection layer (see Sec.~\ref{sec:III_e_numerical_simulation}). The optical excitation of electrons in Dy within the optical penetration depth of $\approx 25\,\text{nm}$ and the subsequent energy transfer to phonons via electron-phonon coupling induces a rapidly rising expansive electron and phonon stress. This drives a bipolar strain pulse with a leading compression propagating from the Y-Dy interface through the Dy transducer into the Nb detection layer. In addition, the laser-induced stress gives rise to an expansion of the Dy layer reaching its maximum at $\approx 29\,\text{ps}$ at the time when the compressive part has completely left the Dy layer (Fig.~\ref{fig:IV_dy_transient_strain}(b)). Partial back reflections at interfaces cause the damped oscillations in the strain response of Dy that superimpose with the quasi-static expansion, which decays on a nanosecond timescale due to thermal transport into the buried Nb layer.

The entrance of the leading compression into the unexcited Nb layer leads to a negative average strain in Nb, while the exit of the compression and the entrance of the tensile part of the strain pulse cause an expansion (Fig.~\ref{fig:IV_dy_transient_strain}(c)). In total, the bipolar strain response of the Nb layer within the first picoseconds maps out the spatial stress profile within the inhomogeneously excited Dy transducer, which determines the spatial profile of the propagating strain pulse according to Eq.~\eqref{eq:III_6_1d_wave_equation_strain}. Finally, the quasi-static expansion of Nb due to heating rises on hundreds of picoseconds via the slow diffusive transport of thermal energy from Dy into Nb shown in Fig.~\ref{fig:IV_dy_transient_strain}(c).

The excitation of Dy in its AFM or FM state below $T_\text{N}=180\,\text{K}$ changes the strain response by the presence of an additional contractive stress originating from the excitation of the spin system that adds to the expansive electron-phonon stress. Already slightly below the Néel temperature at $160\,\text{K}$ we observe a reduced expansion of Dy that continuously changes to a pronounced contraction at $31\,\text{K}$ upon further decreasing the initial sample temperature. At the same time, the strain response of Nb changes from a conventional bipolar shape at $250\,\text{K}$ to a unipolar expansion at $31\,\text{K}$. At intermediate temperatures we observe an unconventional composite shape of the strain response consisting of a leading expansion and a bipolar contribution that indicates a complex space and time dependence of the total external stress within the inhomogeneously excited Dy transducer. With decreasing temperature, the leading expansion continuously becomes more pronounced and the amplitude of the bipolar component decreases.
\begin{figure}[tb!]
\centering
\includegraphics[width = 1.00\columnwidth]{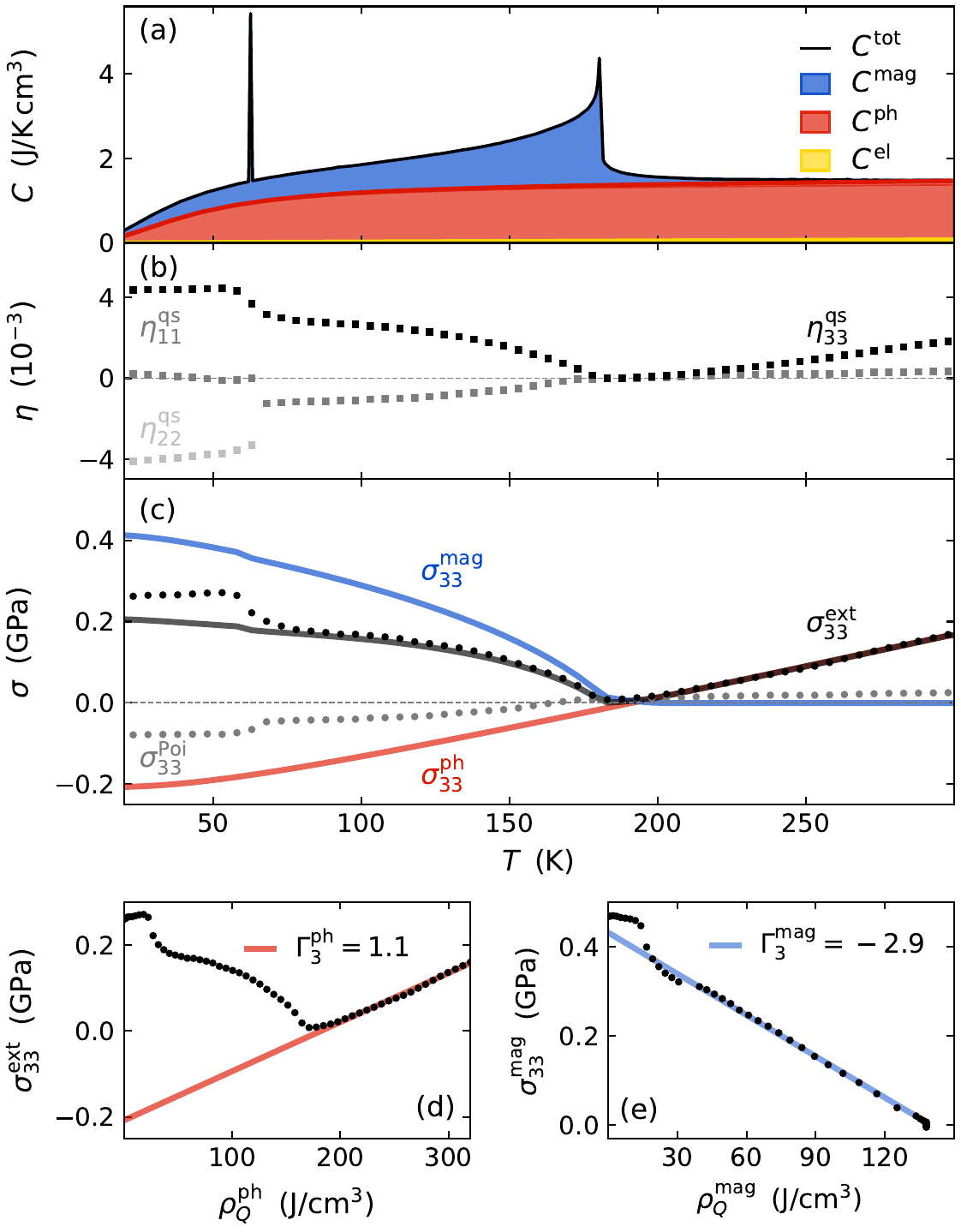}
\caption{\textbf{Separation of the magnetic and phonon stresses in Dy thin films using the Gr\"uneisen concept:} (a) Specific heat of Dy \cite{pech1996} separated into a magnetic-, phonon- and a very small electron contribution \cite{repp2020, repp2016}. (b) Thermal expansion of Dy along the hexagonal c- and a-axis. (c) The associated Poisson stress $\sigma_{33}^\text{Poi}$ (grey dots) and external stress $\sigma_{33}^\text{ext}$ (black dots) along the out-of-plane separated into the magnetic $\sigma_{33}^\text{mag}$ ( solid blue line) and phononic $\sigma_{33}^\text{ph}$ ( solid red line) contribution by the respective Gr\"uneisen parameters determined from the linear dependence of the stress on the energy densities $\rho^Q_\text{mag}$ and $\sigma^Q_\text{ph}$ in panels (d) and (e). The zero stress level is arbitrarily chosen at $T_\text{N}=180\,\text{K}$.}
\label{fig:IV_dy_Gr\"uneisen}
\end{figure}

The solid lines in Figs.~\ref{fig:IV_dy_transient_strain}(b) and (c) represent the modeled strain response from a previous work \cite{repp2020} that reproduces the observed temperature dependence and yields subsystems specific stress contributions. This provides insights into the energy transfer processes among the different degrees of freedom. The modeling approach utilizes the concept of subsystem-specific Gr\"uneisen parameters to individually treat the stress contributions of electrons, phonons and magnetic excitations as sketched in Figure~\ref{fig:III_triangle}. Fig.~\ref{fig:IV_dy_Gr\"uneisen} exemplifies the determination of the subsystem-specific Gr\"uneisen parameters along the out-of-plane direction of the Dy film from heat capacity data and the anisotropic thermal expansion in equilibrium. In Fig.~\ref{fig:IV_dy_Gr\"uneisen}(b) the temperature-dependent out-of-plane lattice strain of the Dy layer along its hexagonal c-axis $\eta_{33}^\text{qs}$ illustrates an anomalous expansion below $180\,\text{K}$ that is concomitant to the onset of magnetic order and results in a NTE during heating. This behavior is universal for the class of the heavy rare-earth elements from Gd to Er which exhibit a pronounced NTE below their respective magnetic ordering temperatures as summarized in the works of Darnell \etal \cite{darn1963,darn1963b,darn1963c}.

Utilizing the anisotropic expansion of Dy $\eta_{kk}^\text{qs}(T)$ ($k=\{ 1,2,3\}$) shown in Fig.~\ref{fig:IV_dy_Gr\"uneisen}(b), we extract the external stress $\sigma_{33}^\text{ext}(T)$ along the out-of-plane direction taking the Poisson stress contribution $\sigma_{33}^\text{Poi}(T)$ into account that arises from the three-dimensional expansion in thermal equilibrium according to Eq.~\eqref{eq:III_3_general_quasi_static_strain}. The hexagonal symmetry of Dy above $T_\text{C}$ simplifies Eq.~\eqref{eq:III_3_general_quasi_static_strain} by $c_{1133}=c_{2233}$ and $\eta_{11}^\text{qs}(T)=\eta_{22}^\text{qs}(T)$ to:
\begin{align}
\begin{split}
\sigma_{33}^\text{ext}(T)&=c_{3333}\eta_{33}^\text{qs}(T) + \sigma_{33}^\text{Poi}(T)\\
&=c_{3333}\eta_{33}^\text{qs}(T) + 2 c_{1133}\eta_{11}^\text{qs}(T) \,,
\end{split}
\label{eq:IV_1_dy_external_stress}  
\end{align}
with the Poisson stress contribution arising from the expansion within the hexagonal plane $\eta_{11}^\text{qs}(T)$ \footnote{We assume the same in-plane expansion as for bulk Dy taken from literature \cite{bula1996, cher2008}. However, we shift the strain signature of the first-order phase transition in temperature to account for the lower Curie-temperature of our thin film sample in comparison to bulk Dy due to the Y layers, which stabilize the spin helix \ie the AFM order \cite{dume1996}.}. Figure~\ref{fig:IV_dy_Gr\"uneisen}(c) displays the resulting out-of-plane Poisson (grey dots) and external stress that is dominated by phonon excitations well above $T_\text{N}=180\,\text{K}$ due to a vanishing magnetic and a negligible electron heat capacity contribution shown in Fig.~\ref{fig:IV_dy_Gr\"uneisen}(a). The linear dependence of the external stress on the energy density in phonon excitations is given by the integral of the heat capacity contribution $\rho_Q^\text{ph}=\int C^\text{ph}\text{d}T$. This yields the phononic Gr\"uneisen parameter $\Gamma_{3}^\text{ph}=1.1$ (Fig.~\ref{fig:IV_dy_Gr\"uneisen}(d)) that determines the phonon stress contribution over the entire temperature range. The difference of the total external stress and the phonon contribution is given by the magnetic stress contribution that vanishes above $T_\text{N}$ since no additional energy density can be deposited in magnetic excitations, as indicated by the finite integral of the magnetic heat capacity contribution. Although $\sigma_{33}^\text{mag}$ is a strongly nonlinear function of the temperature, it depends linearly on the energy density in magnetic excitations $\rho_Q^\text{mag}=\int C^\text{mag}\text{d}T$ (Fig.~\ref{fig:IV_dy_Gr\"uneisen}(e)). The resulting large negative magnetic Gr\"uneisen parameter $\Gamma_{3}^\text{mag}=-2.9$ rationalizes that the contractive magnetic stress dominates the smaller phonon driven expansion below $T_\text{N}$. In Fig.~\ref{fig:IV_dy_Gr\"uneisen}(c) the superposition of the temperature-dependent magnetic (blue solid line) and phononic stress contribution (red solid line) modelled within the Gr\"uneisen approach (solid black line) matches the temperature-dependent total external stress and  thus describes the expansion of Dy along the out-of-plane direction in thermal equilibrium.
\begin{figure}[tb!]
\centering
\includegraphics[width = 1\columnwidth]{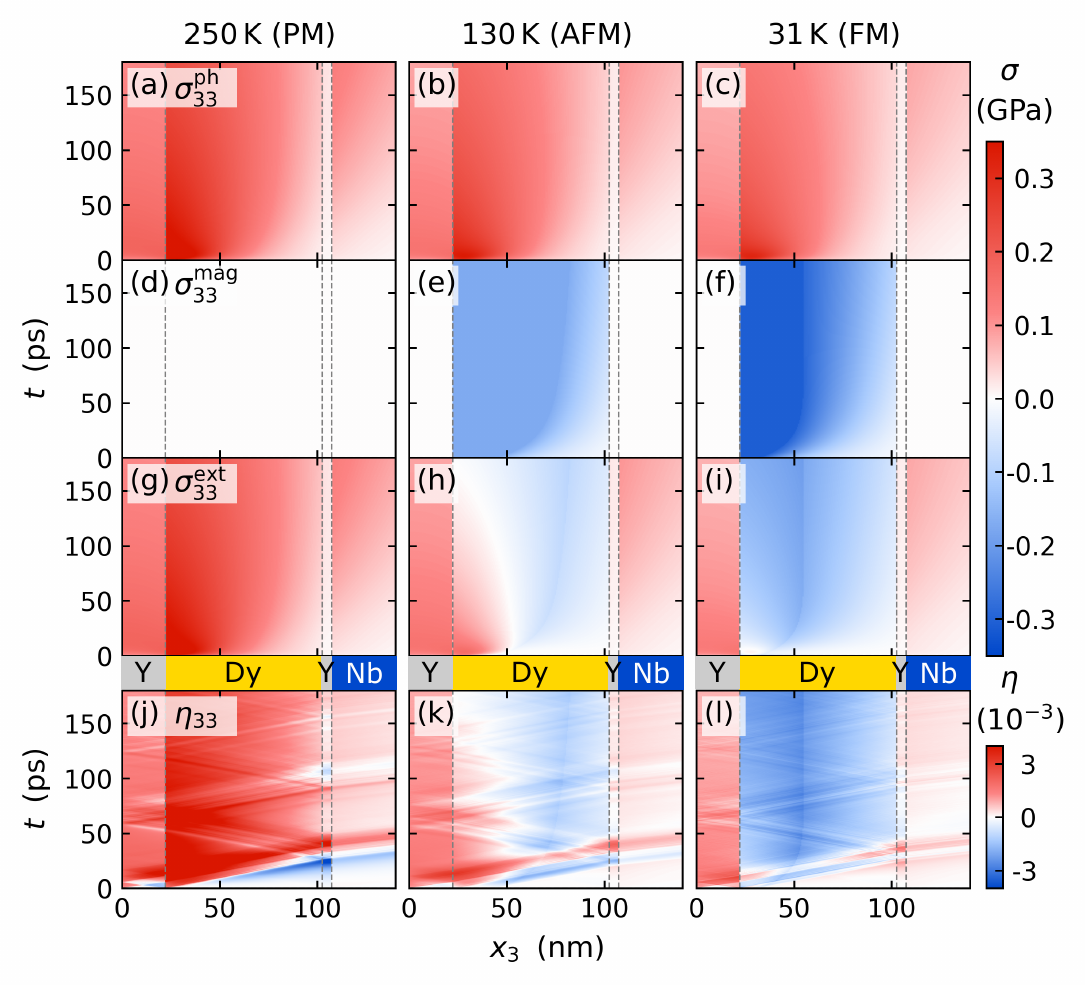}
\caption{\textbf{Model of the external stress separated into subsystem contributions:} Extracted spatio-temporal electron-phonon stress contribution (a--c), magnetic stress contribution (d--f)  and total external stress (g--i) for three representative sample temperatures. The laser-induced magnetic stress contributions at $130\,\text{K}$ and $31\,\text{K}$ display a saturation of the magnetic stress in the strongly excited near-surface region and an increase of the maximum magnetic stress for lower temperatures. The  spatio-temporal strain (j--l) driven by the total external stress shows that at $130\,\text{K}$ the saturated magnetic stress results in a non-monotonous total stress that launches a composite strain pulse into Nb due to an expansion of Dy at the front and a contraction at its backside.}
\label{fig:IV_dy_stress_maps}
\end{figure}

Finally, the extracted Gr\"uneisen parameters are used to model the laser-induced strain response of Dy and Nb shown in Fig.~\ref{fig:IV_dy_transient_strain}(b) and (c) \cite{repp2020}. The modelled strain response (solid lines) excellently reproduces the experimental data and thus reveals the spatio-temporal laser-induced stresses separated into an expansive electron-phonon and a contractive magnetic stress contribution shown in Fig.~\ref{fig:IV_dy_stress_maps}. Within the Gr\"uneisen approach introduced in Section~\ref{sec:III_d_Gr\"uneisen_stress} the total laser-induced stress is given by the superposition of the subsystem contributions that are determined by their Gr\"uneisen parameters and the deposited energy densities as stated in Eq.~\eqref{eq:III_subsystem_stress_Gruneisen}. This approach fulfills energy conservation as the local sum of the transient subsystem energy density contributions corresponds to the total laser-deposited energy density (Eq.~\eqref{eq:III_subsystem_energy_density}) which we calibrate in the PM phase in the absence of magnetic excitations. In addition, we determine the finite rise time of the electron-phonon stress from the amplitude and shape of the strain pulse detected by the Nb strain response, which we model by assuming a smaller Gr\"uneisen parameter for the electrons than for the phonons and introduce an electron-phonon coupling constant that corresponds to a timescale of $2\,\text{ps}$ for the used excitation. However, we limit our discussion in the following to the combined expansive electron-phonon stress in contrast to the contractive magnetic stress since the phonon contribution dominates by far after electron-phonon equilibration.

The magnetic degrees of freedom are locally excited by energy transfer from the electron and phonon subsystem on two timescales with a sub-picosecond contribution and a slower $15\,\text{ps}$ timescale resulting in a reduced expansive phonon stress contribution below $T_\text{N}$ as shown in Fig.~\ref{fig:IV_dy_stress_maps}(b) and (c). The maximum amount of energy density transferred to magnetic excitations is given by the finite integral of the magnetic heat capacity (Fig.~\ref{fig:IV_dy_Gr\"uneisen}(a)) starting at the initial sample temperature $T_\text{i}$. The Gr\"uneisen concept linearly relates this energy density to a maximum magnetic stress that is reached in the strongly excited near-surface region of the Dy transducer (see Fig.~\ref{fig:IV_dy_stress_maps}(e) and (f)). In contrast, the phonon system can take up very large amounts of energy up to the melting point. For large pump fluences, the expansive electron-phonon stress in the near-surface region therefore dominates the saturated magnetic stress (red color in Fig.~\ref{fig:IV_dy_stress_maps}(h)). In contrast, the contractive magnetic stress (blue) dominates in the Nb-near region of the transducer, because the weak excitation does not saturate the spin excitation. Hence the energy is shared among the spin- and phonon system almost equally according to Fig.~\ref{fig:IV_dy_Gr\"uneisen}(a) and the contractive spin stress prevails because of its large negative Gr\"uneisen constant.

This complex interplay of a saturable magnetic stress and a non-saturable phonon stress results in a spatially non-monotonic total stress within the inhomogeneously excited Dy transducer at $130\,\text{K}$ and $31\,\text{K}$ as displayed in Figs.~\ref{fig:IV_dy_stress_maps}(h) and (i), respectively. The corresponding strain maps, which are generated by solving the elastic wave equation~\eqref{eq:III_11_wave_equation_Gr\"uneisen}, are depicted in Fig.~\ref{fig:IV_dy_stress_maps}(k) and (l). They display an expansion of Dy at the frontside launching a bipolar strain pulse into Nb. The weaker and more slowly rising contraction at the backside drives a unipolar expansion pulse into the adjacent Nb layer. Their superposition yields the composite shape of the average strain response of the Nb detection layer as revealed by PUX in Fig.~\ref{fig:IV_dy_transient_strain}(c) for intermediate temperatures. With decreasing temperature and increasing maximum magnetic stress the expansive electron-phonon stress and the amplitude of the bipolar strain pulse are less pronounced, while the amplitude of the unipolar expansive strain pulse driven by the contraction in Dy is enhanced. Furthermore, the total stress averaged over the Dy transducer becomes more contractive, which results in the increasing contraction of Dy that is even enhanced within tens of picoseconds due to the second slowly rising magnetic stress contribution (cf.  Fig.~\ref{fig:IV_dy_transient_strain}(b)). In summary, the Gr\"uneisen model reveals that the saturation of the magnetic stress due to the finite amount of energy that can be deposited in magnetic excitations is crucial to explain the temperature dependence of the strain response of the Dy and the Nb layer. Utilizing the sensitivity of the Nb strain to the stress profile within Dy we were also able to extract the spatial-profile of the magnetic stress and thus gain insights into the spatio-temporal excitations of the AFM order in Dy from our PUX experiment.

In summary, this use case demonstrates the unambiguous detection of NTE on picosecond timescales via PUX and exemplifies the extraction of the subsystem-specific stress contributions utilizing subsystem-specific Gr\"uneisen constants extracted from thermal expansion in equilibrium. In heavy rare-earth elements NTE arises from  magnetostrictive stresses. However, several other mechanisms for NTE and Invar behavior exist \eg dominant transverse phonon contributions at low temperature and repulsive deformation potential contributions \cite{barr2005, dove2016}. For each of these mechanisms it is interesting to investigate, if the NTE is also active in non-equilibrium states after ultrashort laser pulse excitation. An ultrafast NTE or Invar behavior may lead to novel transducer applications that may allow for tunable waveforms for picosecond ultrasonics as discussed in \cite{matt2023}.

\section{\label{sec:V_conclusion}Conclusion and Outlook}
We have presented an overview of the capabilities and concepts of PUX, where hard x-ray probe pulses detect laser-induced transient material strain via Bragg peak shifts.
Layered heterostructures provide excellent use cases, because the strain propagation and reflections connect the signals from different layers, which -- thanks to the large penetration depth of x-rays -- are accessible even beneath thick metal layers. PUX experiments often provide an intuitive picture of both the strain pulses and the quasistatic strain from the local increase in the energy density as these effects can be separated in the time-domain due to their different propagation speeds. 

Motivated by the quantitative access to the strain response, we have revisited continuum elasticity theory, to connect the established formulation of picosecond ultrasonics in homogeneously excited continuous thin films via the thermal expansion coefficient to a formulation that incorporates NTE and is able to reveal the Poisson effect on ultrafast timescales.

We have shown that PUX experiments may serve as an ultrafast probe of the local energy content in nanoscopic layers. The observed strain is driven by a stress, which is in many scenarios directly proportional to the energy densities in different subsystems. The proportionality constants are the sub-system specific Gr\"uneisen parameters for electrons, phonons and magnetic excitations which enable a simplified expression for the driving stress in the inhomogeneous elastic wave equation.

The selected examples of PUX experiments illustrate scenarios where x-ray probing is advantageous compared to established and versatile all-optical experiments. Strain pulse detection in buried layers that allows us to separate propagating strain pulses and quasi-static strain in the time-domain is one of the key advantages of PUX experiments as demonstrated for a thin Nb layer buried below an opaque TbFe$_\text{2}$ transducer.  We discussed the case of granular vs. continuous FePt films to illustrate the morphology dependence of the strain response and expansion amplitude. For Au-Ni heterostructures that are thinner than the optical penetration depth of visible probe pulses we showcased that the material-specific Bragg peaks in PUX experiments enable the detection of energy transfer, where all-optical techniques would convolute the contributions from both layers. Finally, the NTE on ultrafast timescales was exemplified by a Dy transducer layer whose ultrafast giant magnetostriction is capable of launching unconventional picosecond strain pulses into a buried Nb layer. 

The presented experiments work in the symmetric diffraction of rather dim and poorly collimated x-rays from a laser-driven plasma source \cite{schi2013c}. Already this technologically simple approach has many more applications beyond the current manuscript. Moreover, PUX can be extended to asymmetric diffraction geometries that are sensitive to in-plane strain and to shear waves \cite{juve2020}. Optical phonons, \ie coherent atomic motions within the crystallographic unit cell, are accessible via the modulations of the x-ray intensity through the structure factor \cite{soko2003, john2009b, kozi2017}. The excitation can be tailored by changing the wavelength and duration of the pump-pulse, by multipulse-excitation sequences \cite{boja2015,repp2020b} or by combining optical excitation with externally applied magnetic or electric fields \cite{matt2023}. Time-resolved studies of phase transitions that are accompanied by lattice strain \cite{mari2012,jong2013, mogu2020} or non-thermal stresses that arise in laser-excited piezo- and ferroelectric materials \cite{schi2013,chen2016} represent another intriguing field for the application of PUX. 

Large scale facilities from synchrotron radiation sources to FELs provide many additional opportunities beyond the basic PUX experiments discussed here. The excellent collimation and energy resolution of the x-ray beam supports grazing-incidence measurements \cite{mari2012,john2010}, which allow for tuning the penetration depth of x-rays in the bulk and provide a higher resolution of the transient shape changes of the diffraction peaks in reciprocal space. The enormous photon flux at FELs enables the investigation of even thinner layers, pushing the spatial strain-pulse dimensions further towards the single layer limit.  Inelastic scattering provides alternative access to phonon dynamics via thermal diffuse scattering \cite{trig2010}, Fourier-Transform inelastic scattering \cite{trig2013}, TDBS in the x-ray range \cite{boja2013,heni2016,shay2022} and time-resolved x-ray reflectivity on non-crystalline specimen \cite{nusk2011, jal2017}. The coherence of \nth{4} generation synchrotrons and FELs even allows for advanced strain-field imaging techniques \cite{robi2009, newt2014, hols2022}.

\renewcommand\thesection{\Alph{section}}
\setcounter{section}{0}
\numberwithin{equation}{section}
\renewcommand{\theequation}{\thesection.\arabic{equation}}
\numberwithin{figure}{section}
\renewcommand{\thefigure}{\thesection.\arabic{figure}}
\numberwithin{table}{section}
\renewcommand{\thetable}{\thesection.\arabic{table}}

\section{\label{sec:VI_xray_diffraction}Appendix: Detecting Bragg peak shifts in reciprocal space}
Here, we outline details of the x-ray diffraction process that is central to PUX experiments. We discuss both RSM and RSS schemes that utilize a PSD for the detection of Bragg peak shifts that encode the laser-induced strain response of the sample.

In this article, we limit the discussion to the  symmetric and coplanar scattering geometry, where the sample normal is parallel to the scattering plane, generated by the incident and the diffracted x-ray beam. This specular Bragg diffraction geometry specifically accesses the out-of-plane lattice constant.  
A quantitative analysis of the three-dimensional strain response as discussed in Section~\ref{sec:IV_b_fept} requires access to  asymmetric Bragg reflections. This is also true for the detection of shear waves via PUX \cite{juve2020,dorn2019}. More details on the theory of x-ray diffraction can be found in dedicated books \cite{als2011,warr1990,auth2008}, and treatises that specialize on x-ray diffraction from thin films \cite{noya1995,holy1999} or time-resolved x-ray diffraction \cite{barg2005,schi2013c,hell2007}. 

\subsection{\label{sec:VI_a_rsm}Reciprocal space mapping}
As typical for picosecond ultrasonics experiments, we employ a pump-probe geometry where the probe pulses are focused sufficiently well that they are scattered from a region of the sample that is homogeneously laser-excited within the sample plane. The laser-induced strain response then exclusively develops along the out-of-plane direction of the sample as explained in Section~\ref{sec:III_b_strain_1D}. This results in a change of the out-of-plane lattice constant $d_{3}$, \ie a change of the spacing of the lattice planes parallel to the sample surface that can be probed via symmetric x-ray diffraction.

\begin{figure}[tb]
\centering
\includegraphics[width = \columnwidth]{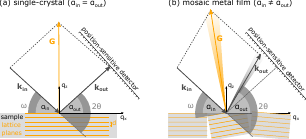}
\caption{\textbf{Sketch of the diffraction geometry:} (a) symmetric diffraction geometry where the angle of incidence relative to the sample surface plane $\alpha_\text{in}$ is equal to the reflection angle $\alpha_\text{out}$ which in turn are equal to the angles in respect to the lattice planes $\omega$ and $\theta$, respectively. In this case the scattering vector is aligned along the out-of plane reciprocal space direction $q_\text{z}$. (b) Sketch of an asymmetric scattering event where reflection occurs from a scattering vector that has a component along the in-plane reciprocal space coordinate $q_\text{x}$. The asymmetry can equally well result from the tilting of nanoscale crystallites in respect to the surface and from the associated nanoscale inhomogeneity within the film plane (finite coherence length). The diffraction processes in both panels are elastic so that $|\bm{k}_\text{in}|= |\bm{k}_\text{out}|$, but the scattered photons are detected on different pixels of the PSD. Sketch adapted from the book of Hol\'y et al \cite{holy1999}.}
\label{fig:VI_1_diffraction_geometry} 
\end{figure}

Figure~\ref{fig:VI_1_diffraction_geometry}(a) introduces the corresponding experimental geometry: The incident angles of the x-ray beam with respect to the sample surface and to the lattice planes are $\alpha_\text{in}$ and $\omega$, respectively. The diffracted intensity is recorded on the PSD under the angles $\alpha_\text{out}$ and $\theta$ relative to the sample surface and the lattice planes, respectively. In case of symmetric diffraction from lattice planes parallel to the sample surface, \ie $\alpha_\text{in}=\omega$ and $\alpha_\text{out}=\theta$, Bragg's law
\begin{align}
n \lambda = 2d_{3} \sin(\theta_\text{B})
\label{eq:VI_1_Braggs_law}  
\end{align}
relates the out-of-plane lattice constant $d_{3}$ to the Bragg angle $\theta_\text{B}$ for a fixed wavelength of the x-rays $\lambda$ and a given diffraction order $n$. The Bragg angle $\theta_\text{B}$ represents the center and the maximum of the Bragg peak intensity reflected by a set of parallel lattice planes that are exposed to a monochromatic x-ray probe beam when scanning the angle $\theta$ relative to the lattice planes.
Note that irrespective of the detector position, Bragg diffraction only occurs if the incoming x-rays are incident at $\theta_\text{B}$ with respect to the lattice planes, \ie $\omega=\theta_\text{B}$. Within this simplified picture it is required to carry out $\theta-2\theta$ scans that symmetrically vary both $\alpha_\text{in}$ and $\alpha_\text{out}$ to identify the changing lattice constant $d_{3}$ from the corresponding change of the Bragg angle $\theta_\text{B}$.

This basic description of x-ray diffraction via Bragg's law has to convoluted with the finite divergence of the x-ray sources. However, investigating thin films with mosaicity or finite crystalline coherence length requires more work. Figure~\ref{fig:VI_1_diffraction_geometry}(b) illustrates the diffraction from a mosaic film that consists of small crystallites tilted with respect to the sample surface in comparison to the idealized situation depicted in Fig.~\ref{fig:VI_1_diffraction_geometry}(a). In case b) the diffraction under the condition of $\alpha_\text{in} \neq \omega$ and $\alpha_\text{out} \neq \theta$ the description by the Laue condition
\begin{equation}
\bm{k}_\text{out} - \bm{k}_\text{in} = \bm{G} \,
\label{eq:VI_2_laue_condition}
\end{equation}
is better suited. Here, \nolinebreak{$\bm{k}_\text{in}$ and $\bm{k}_\text{out}$} are the wave-vectors of the incident and elastically scattered x-rays ($|\bm{k}_\text{in}| = |\bm{k}_\text{out}|= k = {2\pi}/{\lambda}$) which define the scattering vector $\bm{q}=\bm{k}_\text{out} - \bm{k}_\text{in}$.
In reciprocal space, Bragg peaks occur at scattering vectors $\bm{q}$ being equal to a reciprocal lattice vector $\bm{G}$ as stated by the Laue condition Eq.~\eqref{eq:VI_2_laue_condition}. As the reciprocal lattice vectors $|\bm{G}| = {2\pi n}/{d_3}$ are inversely related to the lattice plane spacing $d_3$, the determination of the position of a Bragg peak in reciprocal space yields the lattice strain.

Figure~\ref{fig:VI_1_diffraction_geometry}(b) displays the application of this description to diffraction from a tilted crystallite with lattice planes slightly tilted with respect to the surface of the sample. Even under this condition the diffraction from the lattice planes only occurs at Bragg condition when $\omega=\theta=\theta_B$. However, the corresponding reciprocal lattice vector $\bm{G}$ does not point along the out-of-plane direction anymore but is tilted relative to the reciprocal space coordinates $q_\text{x}$ and $q_\text{z}$ that are defined with respect to the sample surface, where $q_\text{z}$ denotes the out-of-plane direction. Under this condition the maximum of the Bragg peak occurs at $\alpha_\text{in} \neq \alpha_\text{out}$. In addition, a finite nanoscopic in-plane size of the crystallites relaxes the $q_\text{x}=0$ condition by broadening of the diffraction condition along $q_\text{x}$, even in case of lattice planes parallel to the surface. This leads to diffraction intensity even if the Bragg condition is not fulfilled as illustrated in Fig.~\ref{fig:VI_1_diffraction_geometry}(b) by the diffuse spread of the reciprocal lattice vector $\bm{G}$. In total, mosaicity leads to a broadening of the Bragg peak along the in-plane direction $q_\text{x}$ and similarly along $q_\text{y}$. Hence, diffraction can even occur if both the incoming angle $\alpha_\text{in}$ and the scattering angle $\alpha_\text{out}$ deviate from $\theta_B$. The PSD simultaneously captures a range of $\alpha_\text{out}$ and additionally probes diffraction intensity perpendicular to the diffraction plane, \ie along $q_\text{y}$. However, we restrict our discussion to the diffraction plane characterized by $q_\text{x}$ and $q_\text{z}$ and integrate the recorded intensity along $q_\text{y}$. This assumes equivalent in-plane directions of the film and allows us to treat the PSD effectively as line detector. The transformation from angular $\alpha_\text{in}$-$\alpha_\text{out}$-space to reciprocal $q_\text{x}$-$q_\text{z}$-space by geometric considerations is then given by \cite{holy1999,schi2013c}:
\begin{equation}
\bm{q} = \begin{pmatrix} q_\text{x} \\ q_\text{z}  \end{pmatrix} =k \begin{pmatrix} \cos{(\alpha_\text{out})} - \cos{(\alpha_\text{in})} \\ \sin{(\alpha_\text{in})} + \sin{(\alpha_\text{out}) } \end{pmatrix}\,.
\label{eq:VI_3_transformation}
\end{equation}
According to this transformation the PSD captures a range of $\alpha_\text{out}$ and therefore probes a subset of the reciprocal $q_\text{x}$-$q_\text{z}$-space for a fixed $\alpha_\text{in}$. The individual treatment of $\alpha_\text{in}$ and $\alpha_\text{out}$ is a generalization of Bragg's law~\eqref{eq:VI_1_Braggs_law} that is recovered from the Laue condition~\eqref{eq:VI_2_laue_condition} for $\alpha_\text{in}=\alpha_\text{out}=\theta$ implying $q_\text{x}=0$ as in the symmetric case in Fig.~\ref{fig:VI_1_diffraction_geometry}(a). The variation of $\alpha_\text{in}$ during a symmetric $\theta-2\theta$ scan with a line detector yields the diffraction intensity along several lines within the $q_\text{x}$-$q_\text{z}$-space with an offset along $q_\text{z}$. This set of reciprocal space slices yields the RSM in the vicinity of the Bragg peak in the form of diffraction intensity versus $q_\text{x}$ and $q_\text{z}$ as exemplified by Fig.~\ref{fig:II_pux_data}(c).

While the width of the Bragg peak along $q_\text{x}$ is a measure of the in-plane coherence length and  mosaicity of the layer, its width along $q_\text{z}$ encodes the layer thickness or the out-of-plane coherence length. The central observable of PUX is the $q_\text{z}$-position of the Bragg peak which encodes the average out-of-plane lattice constant of the corresponding crystalline layer. Therefore, a heterostructure consisting of different materials with distinct out-of-plane lattice constants results in layer-specific Bragg peaks separated in reciprocal space along $q_\text{z}$ which enables probing the layer-specific laser-induced strain response as demonstrated in Section~\ref{sec:II_PtCuNi_example}. The angular divergence and the spectral width of the x-rays impinging on the sample, \ie the instrument function, broaden the recorded diffraction peaks \cite{zeus2021}.

Time-resolved RSM measurements follow the shift of the material-specific Bragg peaks along $q_\text{z}$ by recording the diffraction intensity in reciprocal space via $\omega$ scans or $\theta-2\theta$ scans for each pump-probe delay \cite{schi2013c}. The transient position of the center-of-mass of the Bragg peak yields the laser-induced out-of-plane strain by the relative change of the $q_\text{z}$ coordinate according to Eq.~\eqref{eq:II_1_strain_definition}.

Most PUX experiments focus on the average strain of each layer within the heterostructure. However, strain gradients within a material can be observed as peak broadening. Especially the case of inhomogeneous energy deposition within the transducer region or localized, high amplitude strain pulses can result in a Bragg peak splitting for the inhomogeneously strained layer where one maximum corresponds to the compressed part and one maximum for the expanded part of the layer \cite{schi2014b}. In extreme cases dynamical x-ray diffraction analysis is required to reconstruct the underlying strain profile from the observed Bragg peak deformations. Such capabilities are provided by modules of the \textsc{udkm1Dsim} \textsc{Matlab} \cite{schi2014} and Python \cite{schi2021} code libraries.

\subsection{\label{sec:VI_b_rss}Reciprocal space slicing}
Recording RSMs around multiple Bragg peaks for each time delay between laser pump pulse and x-ray probe pulse contains the full information about changes of the crystalline structure. To date, most PUX experiments only measure the out-of-plane expansion encoded by Bragg peak shifts along $q_\text{z}$, because this is the natural strain response on a picosecond timescale in the standard experimental setting utilizing a thin film excited by a pump laser profile with much larger lateral extent than the film thickness (see Section~\ref{sec:III_b_strain_1D}). In these situations we can often apply the RSS technique, which considerably speeds up the measurement, because only one or very few slices of reciprocal space and hence angular positions $\alpha_\text{in}$ are required. This means the measurement is done at an angle that offers a large number of diffracted x-ray photons, whereas RSMs contain many positions in the reciprocal space with a low number of diffracted photons. Details of this technique and its application to different types of x-ray sources have been reported previously \cite{zeus2021}.

\begin{figure}[t!]
\centering
\includegraphics[width = 1.00\columnwidth]{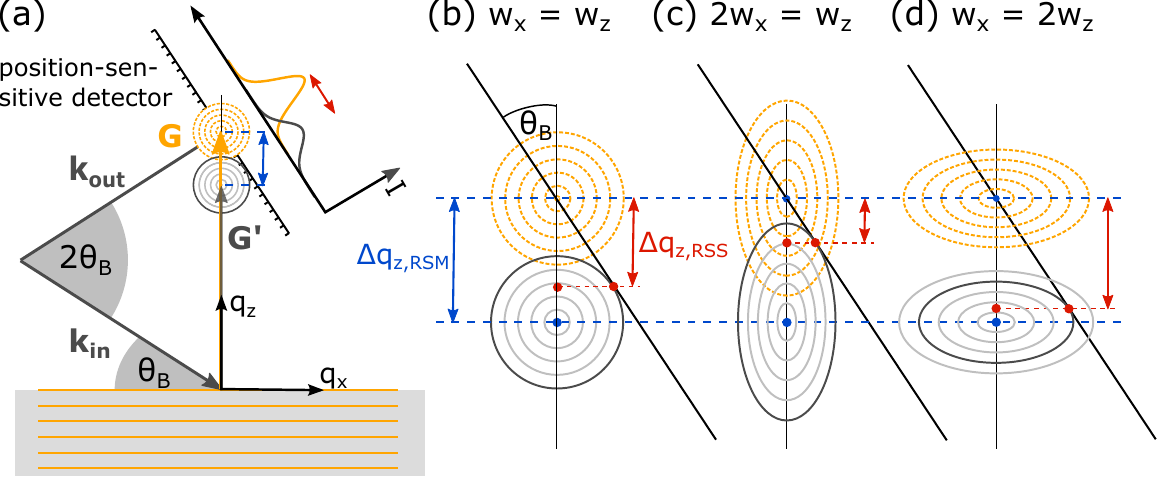}
\caption{\textbf{Sketch of the RSS scheme:} (a) RSS utilizes a PSD that probes multiple diffraction angles $\alpha_\mathrm{out}$ without any mechanical movements of the goniometer. The detector is approximated as a straight line in reciprocal space with $\alpha_\mathrm{in}$ and $\alpha_\mathrm{out}$ chosen such that the maximum intensity occurs on the center pixel of the detector. The peak shifts by $\Delta q_\mathrm{z,RSM}$ as a result of out-of-plane strain. The projection $\Delta q_\mathrm{z,RSS}$ of this shift in the reciprocal space slice defined by the PSD yields the measured strain. Panels (b--d) display the relation of both shifts for different shapes of the Bragg peak in reciprocal space, where both shifts are nearly identical for a Bragg peak that is elongated along $q_\text{x}$.}
\label{fig:VI_2_rss}
\end{figure}

Fig.~\ref{fig:VI_2_rss} displays the central idea of the RSS technique. Due to the finite coherence length of nanoscopic thin films both along $q_\mathrm{z}$ (thickness) and $q_\mathrm{x}$ (mosaicity) asymmetrically scattered x-rays contribute to the scattered intensity distribution around the reciprocal lattice vector $\bm{G}$. Eventually, this results in a characteristic extent of the Bragg peak as indicated by the yellow contour lines around the equilibrium reciprocal lattice vector $\bm{G}$ in Fig.~\ref{fig:VI_2_rss}(a).
In the case of a purely one-dimensional strain response of the film along the out-of-plane direction, as in typical picosecond ultrasonic experiments, the Bragg peak exclusively shifts along $q_\text{z}$ as represented by the grey contour lines around $\bm{G'}$. The pixels of the PSD simultaneously record the x-ray intensities at various diffraction angles $\alpha_\text{out}$ for a fixed $\alpha_\text{in}$.
The PSD positioned at a fixed angle thus records a specific lineout of the Bragg peak's intensity profile as indicated by the line with ticks. The PSD position is initially chosen to slice the reciprocal space through the center of the Bragg peak which maximizes the recorded diffraction intensity profile (orange profile). This exactly corresponds to the symmetrical Bragg condition. After the shift of the Bragg peak, the fixed PSD intersects the Bragg peak's intensity distribution in its wings. Hence, the recorded intensity profile shifts on the detector and is reduced in intensity (grey peak). The red arrow denotes the shift of the Bragg peak projection observed on the detector which is closely related to the actual shift of the Bragg peak (contour lines) along $q_\text{z}$ in reciprocal space.

Figures~\ref{fig:VI_2_rss}(b--d) show a schematic zoom into the intersection of the detector slice (black line) and shifting Bragg peaks with different widths along $q_\text{x}$ and $q_\text{z}$. The ratio of the shift of the recorded intensity profile on the detector $\Delta q_\text{z,RSS}$ and the shift of the Bragg peak in reciprocal space $\Delta q_\text{z,RSM}$  depend on the shape of the Bragg peak in reciprocal space. These shifts are related by geometric arguments. The corresponding strain $\eta_\mathrm{RSM}$ is given by the projection of the RSM:
 \begin{align}
 \begin{split}
\eta_\mathrm{RSM} &=-\frac{\Delta q_\text{z,RSM}}{q_\text{z,0}} \\&=-\frac{\Delta q_\text{z,RSS}}{q_\text{z,0}} \underbrace{\left(1+\left(\frac{w_\text{z}}{w_\text{x}}\right)^2\tan^2{(\theta_\text{B})} \right)}_{:=s} \,,
\label{eq:VI_correction_factor}
\end{split}
\end{align}

and can be related to the shift  $-(\Delta q_\text{z,RSM})/q_\text{z,0}$ of the RSS on the PSD via a scaling factor $s$. Here, $w_\text{z}$ and $w_\text{x}$ denote the width of the Bragg peak in reciprocal space along the $q_\text{z}$ and $q_\text{x}$ coordinate, respectively, which can be characterized for a given sample by standard RSM techniques. Under symmetrical scattering conditions the angle $\theta_\text{B}$ in Fig.~\ref{fig:VI_2_rss}(b) corresponds to the chosen fixed incident angle of the x-ray beam $\alpha_\text{in}=\theta_\text{B}$ that corresponds to the Bragg angle of the film yielding maximum intensity on the detector by slicing the Bragg peak through its center. Figures~\ref{fig:VI_2_rss}(b--d) illustrate the dependence of the correction factor $s$ on the width of the Bragg peak in reciprocal space. The observed shift on the detector and in reciprocal space nearly coincide for a peak that is broad along $q_\text{x}$ (Fig.~\ref{fig:VI_2_rss}(d)), resulting in a correction factor $s\approx 1$. However, $s$ can become rather large for a Bragg peak that is narrow along $q_\text{x}$ and broad along $q_\text{z}$ (Fig.~\ref{fig:VI_2_rss}(c)). This situation is met \eg for very thin films with excellent in-plane epitaxy, \ie with negligible mosaicity. Although the strain may considerably shift the Bragg peak along $q_\text{z}$, the shift of the peak on the detector can become very small in such cases, thus limiting the sensitivity and accuracy of the RSS technique.

In general, samples exhibiting Bragg peaks elongated along $q_\text{x}$ and small Bragg angles $\theta_\text{B}$ are well suited for the RSS probing scheme that is able to reduce the measurement time by an order of magnitude by avoiding $\theta-2\theta$ scans or full RSMs at each pump probe delay. The specimen discussed in the current manuscript all exhibit Bragg peaks that are broad along $q_\text{x}$ and the shifts were recorded via RSS. In contrast, some specimen do require ultrafast RSM experiments: if the Bragg peak becomes broad along $q_\text{z}$ due to very thin films or if the Bragg peak is in close vicinity of a substrate peak  that provides a strong background due to its crystal truncation rod \cite{matt2021, rong2023}. Furthermore, the analysis of transient peak intensities in terms of the Debye-Waller effect requires RSM measurements since the peak shift induces an additional decrease of the peak intensity within the RSS scheme.

\section*{Acknowledgments and funding sources}
We gratefully acknowledge the DFG for financial support via No.\ BA 2281/11-1 and Project-No.\ 328545488 – TRR 227, project A10.

\bibliographystyle{elsarticle-num}
\bibliography{references.bib}

\begin{thebibliography}{100}
\expandafter\ifx\csname url\endcsname\relax
  \def\url#1{\texttt{#1}}\fi
\expandafter\ifx\csname urlprefix\endcsname\relax\def\urlprefix{URL }\fi
\expandafter\ifx\csname href\endcsname\relax
  \def\href#1#2{#2} \def\path#1{#1}\fi

\bibitem{grah1989}
H.~T. Grahn, H.~J. Maris, J.~Tauc, {Picosecond ultrasonics}, {IEEE Journal of
  Quantum Electronics} 25~(12) (1989) 2562--2569.
\newblock \href {https://doi.org/10.1109/3.40643} {\path{doi:10.1109/3.40643}}.

\bibitem{mari1998}
H.~Maris, \href{http://www.jstor.org/stable/26057627}{{Picosecond
  Ultrasonics}}, {Scientific American} 278~(1) (1998) 86--89.
\newline\urlprefix\url{http://www.jstor.org/stable/26057627}

\bibitem{tas1994}
Tas, Maris, {Electron diffusion in metals studied by picosecond ultrasonics},
  {Physical review. B, Condensed matter} 49~(21) (1994) 15046--15054.
\newblock \href {https://doi.org/10.1103/PhysRevB.49.15046}
  {\path{doi:10.1103/PhysRevB.49.15046}}.

\bibitem{tas1998}
G.~Tas, J.~J. Loomis, H.~J. Maris, A.~A. Bailes, L.~E. Seiberling, {Picosecond
  ultrasonics study of the modification of interfacial bonding by ion
  implantation}, {Applied Physics Letters} 72~(18) (1998) 2235--2237.
\newblock \href {https://doi.org/10.1063/1.121276}
  {\path{doi:10.1063/1.121276}}.

\bibitem{anto2006}
G.~A. Antonelli, B.~Perrin, B.~C. Daly, D.~G. Cahill, {Characterization of
  Mechanical and Thermal Properties Using Ultrafast Optical Metrology}, {MRS
  Bulletin} 31~(8) (2006) 607--613.
\newblock \href {https://doi.org/10.1557/mrs2006.157}
  {\path{doi:10.1557/mrs2006.157}}.

\bibitem{mats2015b}
O.~Matsuda, M.~C. Larciprete, R.~{Li Voti}, O.~B. Wright, {Fundamentals of
  picosecond laser ultrasonics}, {Ultrasonics} 56 (2015) 3--20.
\newblock \href {https://doi.org/10.1016/J.ULTRAS.2014.06.005}
  {\path{doi:10.1016/J.ULTRAS.2014.06.005}}.

\bibitem{thom1986}
Thomsen, Grahn, Maris, Tauc, {Surface generation and detection of phonons by
  picosecond light pulses}, {Physical review. B, Condensed matter} 34~(6)
  (1986) 4129--4138.
\newblock \href {https://doi.org/10.1103/physrevb.34.4129}
  {\path{doi:10.1103/physrevb.34.4129}}.

\bibitem{ruel2015}
P.~Ruello, V.~E. Gusev, {Physical mechanisms of coherent acoustic phonons
  generation by ultrafast laser action}, {Ultrasonics} 56 (2015) 21--35.
\newblock \href {https://doi.org/10.1016/J.ULTRAS.2014.06.004}
  {\path{doi:10.1016/J.ULTRAS.2014.06.004}}.

\bibitem{wrig1991b}
O.~B. Wright, T.~Hyoguchi, {Ultrafast vibration and laser acoustics in thin
  transparent films}, {Optics letters} 16~(19) (1991) 1529--1531.
\newblock \href {https://doi.org/10.1364/ol.16.001529}
  {\path{doi:10.1364/ol.16.001529}}.

\bibitem{lejm2014}
M.~Lejman, V.~Shalagatskyi, O.~Kovalenko, T.~Pezeril, V.~V. Temnov, P.~Ruello,
  {Ultrafast optical detection of coherent acoustic phonons emission driven by
  superdiffusive hot electrons}, {Journal of the Optical Society of America B}
  31~(2) (2014) 282.
\newblock \href {https://doi.org/10.1364/JOSAB.31.000282}
  {\path{doi:10.1364/JOSAB.31.000282}}.

\bibitem{guse2014}
V.~E. Gusev, {Detection of nonlinear picosecond acoustic pulses by
  time-resolved Brillouin scattering}, {Journal of Applied Physics} 116~(6)
  (2014) 064907.
\newblock \href {https://doi.org/10.1063/1.4893183}
  {\path{doi:10.1063/1.4893183}}.

\bibitem{boja2013}
A.~Bojahr, M.~Herzog, S.~Mitzscherling, L.~Maerten, D.~Schick, J.~Goldshteyn,
  W.~Leitenberger, R.~Shayduk, P.~Gaal, M.~Bargheer, {Brillouin scattering of
  visible and hard X-ray photons from optically synthesized phonon
  wavepackets}, {Optics express} 21~(18) (2013) 21188--21197.
\newblock \href {https://doi.org/10.1364/OE.21.021188}
  {\path{doi:10.1364/OE.21.021188}}.

\bibitem{guse2018}
V.~E. Gusev, P.~Ruello, {Advances in applications of time-domain Brillouin
  scattering for nanoscale imaging}, {Applied Physics Reviews} 5~(3) (2018)
  031101.
\newblock \href {https://doi.org/10.1063/1.5017241}
  {\path{doi:10.1063/1.5017241}}.

\bibitem{wrig1995b}
O.~B. Wright, {Laser picosecond acoustics in double-layer transparent films},
  {Optics letters} 20~(6) (1995) 632--634.
\newblock \href {https://doi.org/10.1364/ol.20.000632}
  {\path{doi:10.1364/ol.20.000632}}.

\bibitem{peze2016}
T.~Pezeril, \href{http://dx.doi.org/10.1016/j .optlastec.2016.03.019}{{Laser
  generation and detection of ultrafast shear acoustic waves in solids and
  liquids}}, {Optics and Laser Technology} 83 (2016) 177--188.
\newblock \href {https://doi.org/10.1016/j. optlastec.2016.03.019}
  {\path{doi:10.1016/j. optlastec.2016.03.019}}.
\newline\urlprefix\url{http://dx.doi.org/10.1016/j .optlastec.2016.03.019}

\bibitem{van2010}
{van Capel, P. J. S.}, J.~I. Dijkhuis, {Time-resolved interferometric detection
  of ultrashort strain solitons in sapphire}, {Physical Review B} 81~(14)
  (2010) 422.
\newblock \href {https://doi.org/10.1103/PhysRevB.81.144106}
  {\path{doi:10.1103/PhysRevB.81.144106}}.

\bibitem{daly2004}
B.~C. Daly, N.~C.~R. Holme, T.~Buma, C.~Branciard, T.~B. Norris, D.~M. Tennant,
  J.~A. Taylor, J.~E. Bower, S.~Pau, {Imaging nanostructures with coherent
  phonon pulses}, {Applied Physics Letters} 84~(25) (2004) 5180--5182.
\newblock \href {https://doi.org/10.1063/1.1764599}
  {\path{doi:10.1063/1.1764599}}.

\bibitem{pere2020}
F.~P{\'e}rez-Cota, R.~Fuentes-Dom{\'i}nguez, S.~{La Cavera}, W.~Hardiman,
  M.~Yao, K.~Setchfield, E.~Moradi, S.~Naznin, A.~Wright, K.~F. Webb, A.~Huett,
  C.~Friel, V.~Sottile, H.~M. Elsheikha, R.~J. Smith, M.~Clark, {Picosecond
  ultrasonics for elasticity-based imaging and characterization of biological
  cells}, {Journal of Applied Physics} 128~(16) (2020) 160902.
\newblock \href {https://doi.org/10.1063/5.0023744}
  {\path{doi:10.1063/5.0023744}}.

\bibitem{barg2006}
M.~Bargheer, N.~Zhavoronkov, M.~Woerner, T.~Elsaesser, {Recent progress in
  ultrafast X-ray diffraction}, {Chemphyschem : a European journal of chemical
  physics and physical chemistry} 7~(4) (2006) 783--792.
\newblock \href {https://doi.org/10.1002/cphc.200500591}
  {\path{doi:10.1002/cphc.200500591}}.

\bibitem{lind2017}
A.~M. Lindenberg, S.~L. Johnson, D.~A. Reis, {Visualization of Atomic-Scale
  Motions in Materials via Femtosecond X-Ray Scattering Techniques}, {Annual
  Review of Materials Research} 47~(1) (2017) 425--449.
\newblock \href {https://doi.org/10.1146/annurev-matsci-070616-124152}
  {\path{doi:10.1146/annurev-matsci-070616-124152}}.

\bibitem{trig2021}
M.~Trigo, M.~P.~M. Dean, D.~A. Reis,
  \href{http://arxiv.org/pdf/2108.05456v1}{{Ultrafast x-ray probes of dynamics
  in solids}}, {arXiv preprint} (2021).
\newline\urlprefix\url{http://arxiv.org/pdf/2108.05456v1}

\bibitem{lind2000}
Lindenberg, Kang, Johnson, Missalla, Heimann, Chang, Larsson, Bucksbaum,
  Kapteyn, Padmore, Lee, Wark, Falcone, {Time-resolved X-Ray diffraction from
  coherent phonons during a laser-induced phase transition}, {Physical review
  letters} 84~(1) (2000) 111--114.
\newblock \href {https://doi.org/10.1103/PhysRevLett.84.111}
  {\path{doi:10.1103/PhysRevLett.84.111}}.

\bibitem{lars2002}
J.~Larsson, A.~Allen, P.~H. Bucksbaum, R.~W. Falcone, A.~Lindenberg, G.~Naylor,
  T.~Missalla, D.~A. Reis, K.~Scheidt, A.~Sj{\"o}gren, P.~Sondhauss, M.~Wulff,
  J.~S. Wark, {Picosecond X-ray diffraction studies of laser-excited acoustic
  phonons in InSb}, {Applied Physics A} 75~(4) (2002) 467--478.
\newblock \href {https://doi.org/10.1007/s003390201421}
  {\path{doi:10.1007/s003390201421}}.

\bibitem{haya2006}
Y.~Hayashi, Y.~Tanaka, T.~Kirimura, N.~Tsukuda, E.~Kuramoto, T.~Ishikawa,
  {Acoustic pulse echoes probed with time-resolved X-ray triple-crystal
  diffractometry}, {Physical review letters} 96~(11) (2006) 115505.
\newblock \href {https://doi.org/10.1103/PhysRevLett.96.115505}
  {\path{doi:10.1103/PhysRevLett.96.115505}}.

\bibitem{heni2016}
T.~Henighan, M.~Trigo, S.~Bonetti, P.~Granitzka, D.~Higley, Z.~Chen, M.~P.
  Jiang, R.~Kukreja, A.~Gray, A.~H. Reid, E.~Jal, M.~C. Hoffmann, M.~Kozina,
  S.~Song, M.~Chollet, D.~Zhu, P.~F. Xu, J.~Jeong, K.~Carva, P.~Maldonado,
  P.~M. Oppeneer, M.~G. Samant, S.~S.~P. Parkin, D.~A. Reis, H.~A. D{\"u}rr,
  {Generation mechanism of terahertz coherent acoustic phonons in Fe},
  {Physical Review B} 93~(22) (2016) 220301.
\newblock \href {https://doi.org/10.1103/PhysRevB.93.220301}
  {\path{doi:10.1103/PhysRevB.93.220301}}.

\bibitem{dorn2019}
C.~Dornes, Y.~Acremann, M.~Savoini, M.~Kubli, M.~J. Neugebauer, E.~Abreu,
  L.~Huber, G.~Lantz, C.~A.~F. Vaz, H.~Lemke, E.~M. Bothschafter, M.~Porer,
  V.~Esposito, L.~Rettig, M.~Buzzi, A.~Alberca, Y.~W. Windsor, P.~Beaud,
  U.~Staub, D.~Zhu, S.~Song, J.~M. Glownia, S.~L. Johnson, {The ultrafast
  Einstein-de Haas effect}, {Nature} 565~(7738) (2019) 209--212.
\newblock \href {https://doi.org/10.1038/s41586-018-0822-7}
  {\path{doi:10.1038/s41586-018-0822-7}}.

\bibitem{shay2022}
R.~Shayduk, J.~Hallmann, A.~Rodriguez-Fernandez, M.~Scholz, W.~Lu,
  U.~B{\"o}senberg, J.~M{\"o}ller, A.~Zozulya, M.~Jiang, U.~Wegner, R.-C.
  Secareanu, G.~Palmer, M.~Emons, M.~Lederer, S.~Volkov, I.~Lindfors-Vrejoiu,
  D.~Schick, M.~Herzog, M.~Bargheer, A.~Madsen, {Femtosecond x-ray diffraction
  study of multi-THz coherent phonons in SrTiO 3}, {Applied Physics Letters}
  120~(20) (2022) 202203.
\newblock \href {https://doi.org/10.1063/5.0083256}
  {\path{doi:10.1063/5.0083256}}.

\bibitem{trig2010}
M.~Trigo, J.~Chen, V.~H. Vishwanath, Y.~M. Sheu, T.~Graber, R.~Henning, D.~A.
  Reis, {Imaging nonequilibrium atomic vibrations with x-ray diffuse
  scattering}, {Physical review. B, Condensed matter and materials physics}
  82~(23) (2010) 235205--235209.
\newblock \href {https://doi.org/10.1103/PhysRevB.82.235205}
  {\path{doi:10.1103/PhysRevB.82.235205}}.

\bibitem{trig2013}
M.~Trigo, M.~Fuchs, J.~Chen, M.~P. Jiang, M.~Cammarata, S.~Fahy, D.~M. Fritz,
  K.~Gaffney, S.~Ghimire, A.~Higginbotham, S.~L. Johnson, M.~E. Kozina,
  J.~Larsson, H.~Lemke, A.~M. Lindenberg, G.~Ndabashimiye, F.~Quirin,
  K.~Sokolowski-Tinten, C.~Uher, G.~Wang, J.~S. Wark, D.~Zhu, D.~A. Reis,
  {Fourier-transform inelastic X-ray scattering from time- and
  momentum-dependent phonon--phonon correlations}, {Nature Physics} 9~(12)
  (2013) 790--794.
\newblock \href {https://doi.org/10.1038/nphys2788}
  {\path{doi:10.1038/nphys2788}}.

\bibitem{zhu2015}
D.~Zhu, A.~Robert, T.~Henighan, H.~T. Lemke, M.~Chollet, J.~M. Glownia, D.~A.
  Reis, M.~Trigo, {Phonon spectroscopy with sub-meV resolution by femtosecond
  x-ray diffuse scattering}, {Physical Review B} 92~(5) (2015).
\newblock \href {https://doi.org/10.1103/PhysRevB.92.054303}
  {\path{doi:10.1103/PhysRevB.92.054303}}.

\bibitem{wall2018}
S.~Wall, S.~Yang, L.~Vidas, M.~Chollet, J.~M. Glownia, M.~Kozina, T.~Katayama,
  T.~Henighan, M.~Jiang, T.~A. Miller, D.~A. Reis, L.~A. Boatner, O.~Delaire,
  M.~Trigo, {Ultrafast disordering of vanadium dimers in photoexcited VO2},
  {Science (New York, N.Y.)} 362~(6414) (2018) 572--576.
\newblock \href {https://doi.org/10.1126/science.aau3873}
  {\path{doi:10.1126/science.aau3873}}.

\bibitem{rose1999}
C.~Rose-Petruck, R.~Jimenez, T.~Guo, A.~Cavalleri, C.~W. Siders, F.~Rksi, J.~A.
  Squier, B.~C. Walker, K.~R. Wilson, C.~P.~J. Barty,
  {Picosecond--milli{\aa}ngstr{\"o}m lattice dynamics measured by ultrafast
  X-ray diffraction}, {Nature} 398~(6725) (1999) 310--312.
\newblock \href {https://doi.org/10.1038/18631} {\path{doi:10.1038/18631}}.

\bibitem{barg2004}
M.~Bargheer, N.~Zhavoronkov, Y.~Gritsai, J.~C. Woo, D.~S. Kim, M.~Woerner,
  T.~Elsaesser, {Coherent atomic motions in a nanostructure studied by
  femtosecond X-ray diffraction}, {Science (New York, N.Y.)} 306~(5702) (2004)
  1771--1773.
\newblock \href {https://doi.org/10.1126/science.1104739}
  {\path{doi:10.1126/science.1104739}}.

\bibitem{nico2011}
M.~Nicoul, U.~Shymanovich, A.~Tarasevitch, D.~von~der Linde,
  K.~Sokolowski-Tinten, {Picosecond acoustic response of a laser-heated
  gold-film studied with time-resolved x-ray diffraction}, {Applied Physics
  Letters} 98~(19) (2011) 191902.
\newblock \href {https://doi.org/10.1063/1.3584864}
  {\path{doi:10.1063/1.3584864}}.

\bibitem{quir2012}
F.~Quirin, M.~Vattilana, U.~Shymanovich, A.-E. El-Kamhawy, A.~Tarasevitch,
  J.~Hohlfeld, D.~von~der Linde, K.~Sokolowski-Tinten, {Structural dynamics in
  FeRh during a laser-induced metamagnetic phase transition}, {Physical Review
  B} 85~(2) (2012) 498.
\newblock \href {https://doi.org/10.1103/PhysRevB.85.020103}
  {\path{doi:10.1103/PhysRevB.85.020103}}.

\bibitem{zeus2019}
S.~P. Zeuschner, T.~Parpiiev, T.~Pezeril, A.~Hillion, K.~Dumesnil, A.~Anane,
  J.~Pudell, L.~Willig, M.~R{\"o}ssle, M.~Herzog, A.~von Reppert, M.~Bargheer,
  {Tracking picosecond strain pulses in heterostructures that exhibit giant
  magnetostriction}, {Structural dynamics (Melville, N.Y.)} 6~(2) (2019)
  024302.
\newblock \href {https://doi.org/10.1063/1.5084140}
  {\path{doi:10.1063/1.5084140}}.

\bibitem{bost2016}
C.~Bostedt, S.~Boutet, D.~M. Fritz, Z.~Huang, H.~J. Lee, H.~T. Lemke,
  A.~Robert, W.~F. Schlotter, J.~J. Turner, G.~J. Williams, {Linac Coherent
  Light Source: The first five years}, {Reviews of Modern Physics} 88~(1)
  (2016) 207.
\newblock \href {https://doi.org/10.1103/RevModPhys.88.015007}
  {\path{doi:10.1103/RevModPhys.88.015007}}.

\bibitem{abel2017}
R.~Abela, P.~Beaud, J.~A. {van Bokhoven}, M.~Chergui, T.~Feurer, J.~Haase,
  G.~Ingold, S.~L. Johnson, G.~Knopp, H.~Lemke, C.~J. Milne, B.~Pedrini,
  P.~Radi, G.~Schertler, J.~Standfuss, U.~Staub, L.~Patthey, {Perspective:
  Opportunities for ultrafast science at SwissFEL}, {Structural dynamics
  (Melville, N.Y.)} 4~(6) (2017) 061602.
\newblock \href {https://doi.org/10.1063/1.4997222}
  {\path{doi:10.1063/1.4997222}}.

\bibitem{scho2019}
R.~Schoenlein, T.~Elsaesser, K.~Holldack, Z.~Huang, H.~Kapteyn, M.~Murnane,
  M.~Woerner, {Recent advances in ultrafast X-ray sources}, {Philosophical
  Transactions of the Royal Society A: Mathematical, Physical and Engineering
  Sciences} 377~(2145) (2019) 20180384.
\newblock \href {https://doi.org/10.1098/rsta.2018.0384}
  {\path{doi:10.1098/rsta.2018.0384}}.

\bibitem{scho2000}
Schoenlein, Chattopadhyay, Chong, Glover, Heimann, Shank, Zholents, Zolotorev,
  {Generation of femtosecond pulses of synchrotron radiation}, {Science (New
  York, N.Y.)} 287~(5461) (2000) 2237--2240.
\newblock \href {https://doi.org/10.1126/science.287.5461.2237}
  {\path{doi:10.1126/science.287.5461.2237}}.

\bibitem{beau2007}
P.~Beaud, S.~L. Johnson, A.~Streun, R.~Abela, D.~Abramsohn, D.~Grolimund,
  F.~Krasniqi, T.~Schmidt, V.~Schlott, G.~Ingold, {Spatiotemporal stability of
  a femtosecond hard-x-ray undulator source studied by control of coherent
  optical phonons}, {Physical review letters} 99~(17) (2007) 174801.
\newblock \href {https://doi.org/10.1103/PhysRevLett.99.174801}
  {\path{doi:10.1103/PhysRevLett.99.174801}}.

\bibitem{enqu2010}
H.~Enquist, H.~Navirian, R.~N{\"u}ske, C.~von {Korff Schmising},
  A.~Jurgilaitis, M.~Herzog, M.~Bargheer, P.~Sondhauss, J.~Larsson,
  {Subpicosecond hard x-ray streak camera using single-photon counting},
  {Optics letters} 35~(19) (2010) 3219--3221.
\newblock \href {https://doi.org/10.1364/OL.35.003219}
  {\path{doi:10.1364/OL.35.003219}}.

\bibitem{jank2013}
A.~Jankowiak, G.~W{\"u}stefeld, {Low-$\alpha$ Operation of BESSY II and Future
  Plans for an Alternating Bunch Length Scheme BESSY VSR}, {Synchrotron
  Radiation News} 26~(3) (2013) 22--24.
\newblock \href {https://doi.org/10.1080/08940886.2013.791212}
  {\path{doi:10.1080/08940886.2013.791212}}.

\bibitem{ross2021}
M.~R{\"o}ssle, W.~Leitenberger, M.~Reinhardt, A.~Ko{\c{c}}, J.~Pudell,
  C.~Kwamen, M.~Bargheer, {The time-resolved hard X-ray diffraction endstation
  KMC-3 XPP at BESSY II}, {Journal of Synchrotron Radiation} 28~(3) (2021).
\newblock \href {https://doi.org/10.1107/S1600577521002484}
  {\path{doi:10.1107/S1600577521002484}}.

\bibitem{gaal2014}
P.~Gaal, D.~Schick, M.~Herzog, A.~Bojahr, R.~Shayduk, J.~Goldshteyn,
  W.~Leitenberger, I.~Vrejoiu, D.~Khakhulin, M.~Wulff, M.~Bargheer, {Ultrafast
  switching of hard X-rays}, {Journal of synchrotron radiation} 21~(Pt 2)
  (2014) 380--385.
\newblock \href {https://doi.org/10.1107/S1600577513031949}
  {\path{doi:10.1107/S1600577513031949}}.

\bibitem{sand2019}
M.~Sander, R.~Bauer, V.~Kabanova, M.~Levantino, M.~Wulff, D.~Pfuetzenreuter,
  J.~Schwarzkopf, P.~Gaal, {Demonstration of a picosecond Bragg switch for hard
  X-rays in a synchrotron-based pump-probe experiment}, {Journal of synchrotron
  radiation} 26~(Pt 4) (2019) 1253--1259.
\newblock \href {https://doi.org/10.1107/S1600577519005356}
  {\path{doi:10.1107/S1600577519005356}}.

\bibitem{gros2017}
M.~Grossmann, M.~Schubert, C.~He, D.~Brick, E.~Scheer, M.~Hettich, V.~Gusev,
  T.~Dekorsy, {Characterization of thin-film adhesion and phonon lifetimes in
  Al/Si membranes by picosecond ultrasonics}, {New Journal of Physics} 19~(5)
  (2017) 053019.
\newblock \href {https://doi.org/10.1088/1367-2630/ aa6d05}
  {\path{doi:10.1088/1367-2630/ aa6d05}}.

\bibitem{thom1986b}
C.~Thomsen, H.~T. Grahn, H.~J. Maris, J.~Tauc, {Picosecond interferometric
  technique for study of phonons in the brillouin frequency range}, {Optics
  Communications} 60~(1-2) (1986) 55--58.
\newblock \href {https://doi.org/10.1016/0030-4018(86)90116-1}
  {\path{doi:10.1016/0030-4018(86)90116-1}}.

\bibitem{lin1991}
H.~N. Lin, R.~J. Stoner, H.~J. Maris, J.~Tauc, {Phonon attenuation and velocity
  measurements in transparent materials by picosecond acoustic interferometry},
  {Journal of Applied Physics} 69~(7) (1991) 3816--3822.
\newblock \href {https://doi.org/10.1063/1.348958}
  {\path{doi:10.1063/1.348958}}.

\bibitem{berg2016}
N.~Bergeard, M.~Hehn, S.~Mangin, G.~Lengaigne, F.~Montaigne, M.~L.~M. Lalieu,
  B.~Koopmans, G.~Malinowski, {Hot-Electron-Induced Ultrafast Demagnetization
  in Co/Pt Multilayers}, {Physical review letters} 117~(14) (2016) 147203.
\newblock \href {https://doi.org/10.1103/PhysRevLett.117.147203}
  {\path{doi:10.1103/PhysRevLett.117.147203}}.

\bibitem{fert2017}
T.~Fert{\'e}, N.~Bergeard, G.~Malinowski, R.~Abrudan, T.~Kachel, K.~Holldack,
  M.~Hehn, C.~Boeglin, {Ultrafast hot-electron induced quenching of Tb 4f
  magnetic order}, {Physical Review B} 96~(14) (2017).
\newblock \href {https://doi.org/10.1103/PhysRevB.96.144427}
  {\path{doi:10.1103/PhysRevB.96.144427}}.

\bibitem{xu2017}
Y.~Xu, M.~Deb, G.~Malinowski, M.~Hehn, W.~Zhao, S.~Mangin, {Ultrafast
  Magnetization Manipulation Using Single Femtosecond Light and Hot-Electron
  Pulses}, {Adv. Mater.} 29~(42) (2017).
\newblock \href {https://doi.org/10.1002/adma.201703474}
  {\path{doi:10.1002/adma.201703474}}.

\bibitem{berg2020}
N.~Bergeard, M.~Hehn, K.~Carva, P.~Bal{\'a}{\v{z}}, S.~Mangin, G.~Malinowski,
  {Tailoring femtosecond hot-electron pulses for ultrafast spin manipulation},
  {Applied Physics Letters} 117~(22) (2020) 222408.
\newblock \href {https://doi.org/10.1063/5.0018502}
  {\path{doi:10.1063/5.0018502}}.

\bibitem{schi2013c}
D.~Schick, R.~Shayduk, A.~Bojahr, M.~Herzog, C.~von {Korff Schmising}, P.~Gaal,
  M.~Bargheer, {Ultrafast reciprocal-space mapping with a convergent beam},
  {Journal of Applied Crystallography} 46~(5) (2013) 1372--1377.
\newblock \href {https://doi.org/10.1107/S0021889813020013}
  {\path{doi:10.1107/S0021889813020013}}.

\bibitem{zeus2021}
S.~P. Zeuschner, M.~Mattern, J.-E. Pudell, A.~von Reppert, M.~R{\"o}ssle,
  W.~Leitenberger, J.~Schwarzkopf, J.~E. Boschker, M.~Herzog, M.~Bargheer,
  {Reciprocal space slicing: A time-efficient approach to femtosecond x-ray
  diffraction}, {Structural dynamics (Melville, N.Y.)} 8~(1) (2021) 014302.
\newblock \href {https://doi.org/10.1063/4.0000040}
  {\path{doi:10.1063/4.0000040}}.

\bibitem{pude2020b}
J.-E. Pudell, M.~Mattern, M.~Hehn, G.~Malinowski, M.~Herzog, M.~Bargheer, {Heat
  Transport without Heating?---An Ultrafast X--Ray Perspective into a Metal
  Heterostructure}, {Advanced Functional Materials} 30~(46) (2020) 2004555.
\newblock \href {https://doi.org/10.1002/adfm.202004555}
  {\path{doi:10.1002/adfm.202004555}}.

\bibitem{roye2000}
D.~Royer, E.~Dieulesaint, {Elastic waves in solids: Including nonlinear
  dynamics}, 3rd Edition, {Advanced texts in physics}, Springer, Berlin, 2000.

\bibitem{nie2006}
S.~Nie, X.~Wang, H.~Park, R.~Clinite, J.~Cao, {Measurement of the electronic
  Gr{\"u}neisen constant using femtosecond electron diffraction}, {Physical
  Review Letters} 96~(2) (2006) 025901.
\newblock \href {https://doi.org/10.1103/physrevlett.96.025901}
  {\path{doi:10.1103/physrevlett.96.025901}}.

\bibitem{mald2017}
P.~Maldonado, K.~Carva, M.~Flammer, P.~M. Oppeneer, {Theory of
  out-of-equilibrium ultrafast relaxation dynamics in metals}, {Physical Review
  B} 96~(17) (2017) 173.
\newblock \href {https://doi.org/10.1103/PhysRevB.96.174439}
  {\path{doi:10.1103/PhysRevB.96.174439}}.

\bibitem{redd2005}
P.~Reddy, K.~Castelino, A.~Majumdar, {Diffuse mismatch model of thermal
  boundary conductance using exact phonon dispersion}, {Applied Physics
  Letters} 87~(21) (2005) 211908.
\newblock \href {https://doi.org/10.1063/1.2133890}
  {\path{doi:10.1063/1.2133890}}.

\bibitem{oomm2022}
S.~M. Oommen, L.~Fallarino, J.~Heinze, O.~Hellwig, S.~Pisana, {Role of
  vibrational properties and electron-phonon coupling on thermal transport
  across metal-dielectric interfaces with ultrathin metallic interlayers},
  {Journal of physics. Condensed matter : an Institute of Physics journal}
  (2022).
\newblock \href {https://doi.org/10.1088/1361-648X/ac926a}
  {\path{doi:10.1088/1361-648X/ac926a}}.

\bibitem{herz2022}
M.~Herzog, A.~von Reppert, J.-E. Pudell, C.~Henkel, M.~Kronseder, C.~H. Back,
  A.~A. Maznev, M.~Bargheer, {Phonon--Dominated Energy Transport in Purely
  Metallic Heterostructures}, {Advanced Functional Materials} (2022)
  2206179\href {https://doi.org/10.1002/adfm.202206179}
  {\path{doi:10.1002/adfm.202206179}}.

\bibitem{yeh2005}
P.~Yeh, {Optical waves in layered media}, {Wiley series in pure and applied
  optics}, Wiley-Interscience, Hoboken, NJ, 2005.

\bibitem{leg2013}
L.~{Le Guyader}, A.~Kleibert, F.~Nolting, L.~Joly, P.~M. Derlet, R.~V. Pisarev,
  A.~Kirilyuk, T.~Rasing, A.~V. Kimel, {Dynamics of laser-induced spin
  reorientation in Co/SmFeO 3 heterostructure}, {Physical Review B - Condensed
  Matter and Materials Physics} 87~(5) (2013) 935.
\newblock \href {https://doi.org/10.1103/PhysRevB.87.054437}
  {\path{doi:10.1103/PhysRevB.87.054437}}.

\bibitem{wang2008}
X.~Wang, S.~Nie, J.~Li, R.~Clinite, M.~Wartenbe, M.~Martin, W.~Liang, J.~Cao,
  {Electronic Gr{\"u}neisen parameter and thermal expansion in ferromagnetic
  transition metal}, {Applied Physics Letters} 92~(12) (2008) 121918.
\newblock \href {https://doi.org/10.1063/1.2902170}
  {\path{doi:10.1063/1.2902170}}.

\bibitem{schi2014b}
D.~Schick, M.~Herzog, A.~Bojahr, W.~Leitenberger, A.~Hertwig, R.~Shayduk,
  M.~Bargheer, {Ultrafast lattice response of photoexcited thin films studied
  by X-ray diffraction}, {Structural dynamics (Melville, N.Y.)} 1~(6) (2014)
  064501.
\newblock \href {https://doi.org/10.1063/1.4901228}
  {\path{doi:10.1063/1.4901228}}.

\bibitem{korf2008}
C.~v. {Korff Schmising}, A.~Harpoeth, N.~Zhavoronkov, Z.~Ansari, C.~Aku-Leh,
  M.~Woerner, T.~Elsaesser, M.~Bargheer, M.~Schmidbauer, I.~Vrejoiu, D.~Hesse,
  M.~Alexe, {Ultrafast magnetostriction and phonon-mediated stress in a
  photoexcited ferromagnet}, {Physical Review B} 78~(6) (2008).
\newblock \href {https://doi.org/10.1103/PhysRevB.78.060404}
  {\path{doi:10.1103/PhysRevB.78.060404}}.

\bibitem{bror1987}
Brorson, Fujimoto, Ippen, {Femtosecond electronic heat-transport dynamics in
  thin gold films}, {Physical review letters} 59~(17) (1987) 1962--1965.
\newblock \href {https://doi.org/10.1103/PhysRevLett.59.1962}
  {\path{doi:10.1103/PhysRevLett.59.1962}}.

\bibitem{hohl1997b}
J.~Hohlfeld, J.~G. M{\"u}ller, S.-S. Wellershoff, E.~Matthias, {Time-resolved
  thermoreflectivity of thin gold films and its dependence on film thickness},
  {Applied Physics B} 64~(3) (1997) 387--390.
\newblock \href {https://doi.org/10.1007/s003400050189}
  {\path{doi:10.1007/s003400050189}}.

\bibitem{batt2012}
M.~Battiato, K.~Carva, P.~M. Oppeneer, {Theory of laser-induced ultrafast
  superdiffusive spin transport in layered heterostructures}, {Physical Review
  B} 86~(2) (2012).
\newblock \href {https://doi.org/10.1103/PhysRevB.86.024404}
  {\path{doi:10.1103/PhysRevB.86.024404}}.

\bibitem{mali2018}
G.~Malinowski, N.~Bergeard, M.~Hehn, S.~Mangin, {Hot-electron transport and
  ultrafast magnetization dynamics in magnetic multilayers and nanostructures
  following femtosecond laser pulse excitation}, {Zeitschrift fr Physik B
  Condensed Matter and Quanta} 91~(6) (2018) 3251.
\newblock \href {https://doi.org/10.1140/epjb/e2018-80555-5}
  {\path{doi:10.1140/epjb/e2018-80555-5}}.

\bibitem{hohl2000}
J.~Hohlfeld, S.-S. Wellershoff, J.~G{\"u}dde, U.~Conrad, V.~J{\"a}hnke,
  E.~Matthias, {Electron and lattice dynamics following optical excitation of
  metals}, {Chemical Physics} 251~(1-3) (2000) 237--258.
\newblock \href {https://doi.org/10.1016/S0301-0104(99)00330-4}
  {\path{doi:10.1016/S0301-0104(99)00330-4}}.

\bibitem{nenn2018}
D.~M. Nenno, B.~Rethfeld, H.~C. Schneider, {Particle-in-cell simulation of
  ultrafast hot-carrier transport in Fe/Au heterostructures}, {Physical Review
  B} 98~(22) (2018).
\newblock \href {https://doi.org/10.1103/PhysRevB.98.224416}
  {\path{doi:10.1103/PhysRevB.98.224416}}.

\bibitem{wais2021}
M.~Wais, K.~Held, M.~Battiato, {Numerical solver for the time-dependent
  far-from-equilibrium Boltzmann equation}, {Computer Physics Communications}
  264 (2021) 107877.
\newblock \href {https://doi.org/10.1016/j. cpc.2021.107877}
  {\path{doi:10.1016/j. cpc.2021.107877}}.

\bibitem{chen2001b}
G.~Chen, {Ballistic-diffusive heat-conduction equations}, {Physical Review
  Letters} 86~(11) (2001) 2297--2300.
\newblock \href {https://doi.org/10.1103/PhysRevLett.86.2297}
  {\path{doi:10.1103/PhysRevLett.86.2297}}.

\bibitem{wang2010}
X.~Wang, S.~Nie, J.~Li, R.~Clinite, J.~E. Clark, J.~Cao, {Temperature
  dependence of electron-phonon thermalization and its correlation to ultrafast
  magnetism}, {Physical Review B} 81~(22) (2010).
\newblock \href {https://doi.org/10.1103/PhysRevB.81.220301}
  {\path{doi:10.1103/PhysRevB.81.220301}}.

\bibitem{zahn2021}
D.~Zahn, F.~Jakobs, Y.~W. Windsor, H.~Seiler, T.~Vasileiadis, T.~A. Butcher,
  Y.~Qi, D.~Engel, U.~Atxitia, J.~Vorberger, R.~Ernstorfer, {Lattice dynamics
  and ultrafast energy flow between electrons, spins, and phonons in a 3d
  ferromagnet}, {Physical Review Research} 3~(2) (2021).
\newblock \href {https://doi.org/10.1103/PhysRevResearch.3.023032}
  {\path{doi:10.1103/PhysRevResearch.3.023032}}.

\bibitem{zahn2022}
D.~Zahn, F.~Jakobs, H.~Seiler, T.~A. Butcher, D.~Engel, J.~Vorberger,
  U.~Atxitia, Y.~W. Windsor, R.~Ernstorfer, {Intrinsic energy flow in
  laser-excited 3d ferromagnets}, {Physical Review Research} 4~(1) (2022).
\newblock \href {https://doi.org/10.1103/PhysRevResearch.4.013104}
  {\path{doi:10.1103/PhysRevResearch.4.013104}}.

\bibitem{koop2010}
B.~Koopmans, G.~Malinowski, F.~{Dalla Longa}, D.~Steiauf, M.~F{\"a}hnle,
  T.~Roth, M.~Cinchetti, M.~Aeschlimann, {Explaining the paradoxical diversity
  of ultrafast laser-induced demagnetization}, {Nature materials} 9~(3) (2010)
  259--265.
\newblock \href {https://doi.org/10.1038/nmat2593}
  {\path{doi:10.1038/nmat2593}}.

\bibitem{batt2010}
M.~Battiato, K.~Carva, P.~M. Oppeneer, {Superdiffusive spin transport as a
  mechanism of ultrafast demagnetization}, {Physical review letters} 105~(2)
  (2010) 027203.
\newblock \href {https://doi.org/10.1103/PhysRevLett.105.027203}
  {\path{doi:10.1103/PhysRevLett.105.027203}}.

\bibitem{roth2012}
T.~Roth, A.~J. Schellekens, S.~Alebrand, O.~Schmitt, D.~Steil, B.~Koopmans,
  M.~Cinchetti, M.~Aeschlimann, {Temperature Dependence of Laser-Induced
  Demagnetization in Ni: A Key for Identifying the Underlying Mechanism},
  {Physical Review X} 2~(2) (2012).
\newblock \href {https://doi.org/10.1103/PhysRevX.2.021006}
  {\path{doi:10.1103/PhysRevX.2.021006}}.

\bibitem{frie2015}
B.~Frietsch, J.~Bowlan, R.~Carley, M.~Teichmann, S.~Wienholdt, D.~Hinzke,
  U.~Nowak, K.~Carva, P.~M. Oppeneer, M.~Weinelt, {Disparate ultrafast dynamics
  of itinerant and localized magnetic moments in gadolinium metal}, {Nature
  communications} 6 (2015) 8262.
\newblock \href {https://doi.org/10.1038/ncomms9262}
  {\path{doi:10.1038/ncomms9262}}.

\bibitem{shok2017}
V.~Shokeen, M.~{Sanchez Piaia}, J.-Y. Bigot, T.~M{\"u}ller, P.~Elliott, J.~K.
  Dewhurst, S.~Sharma, E.~K.~U. Gross, {Spin Flips versus Spin Transport in
  Nonthermal Electrons Excited by Ultrashort Optical Pulses in Transition
  Metals}, {Physical review letters} 119~(10) (2017) 107203.
\newblock \href {https://doi.org/10.1103/PhysRevLett.119.107203}
  {\path{doi:10.1103/PhysRevLett.119.107203}}.

\bibitem{sieg2019}
F.~Siegrist, J.~A. Gessner, M.~Ossiander, C.~Denker, Y.-P. Chang, M.~C.
  Schr{\"o}der, A.~Guggenmos, Y.~Cui, J.~Walowski, U.~Martens, J.~K. Dewhurst,
  U.~Kleineberg, M.~M{\"u}nzenberg, S.~Sharma, M.~Schultze, {Light-wave dynamic
  control of magnetism}, {Nature} 571~(7764) (2019) 240--244.
\newblock \href {https://doi.org/10.1038/s41586-019-1333-x}
  {\path{doi:10.1038/s41586-019-1333-x}}.

\bibitem{frie2020}
B.~Frietsch, A.~Donges, R.~Carley, M.~Teichmann, J.~Bowlan, K.~D{\"o}brich,
  K.~Carva, D.~Legut, P.~M. Oppeneer, U.~Nowak, M.~Weinelt, {The role of
  ultrafast magnon generation in the magnetization dynamics of rare-earth
  metals}, {Science Advances} 6~(39) (2020).
\newblock \href {https://doi.org/10.1126/sciadv.abb1601}
  {\path{doi:10.1126/sciadv.abb1601}}.

\bibitem{pude2019}
J.~Pudell, A.~von Reppert, D.~Schick, F.~Zamponi, M.~R{\"o}ssle, M.~Herzog,
  H.~Zabel, M.~Bargheer, {Ultrafast negative thermal expansion driven by spin
  disorder}, {Physical Review B} 99~(9) (2019) 094304.
\newblock \href {https://doi.org/10.1103/PhysRevB.99.094304}
  {\path{doi:10.1103/PhysRevB.99.094304}}.

\bibitem{repp2020}
A.~von Reppert, M.~Mattern, J.-E. Pudell, S.~P. Zeuschner, K.~Dumesnil,
  M.~Bargheer, {Unconventional picosecond strain pulses resulting from the
  saturation of magnetic stress within a photoexcited rare earth layer},
  {Structural dynamics (Melville, N.Y.)} 7~(2) (2020) 024303.
\newblock \href {https://doi.org/10.1063/1.5145315}
  {\path{doi:10.1063/1.5145315}}.

\bibitem{barr1980}
T.~Barron, J.~G. Collins, G.~K. White, {Thermal expansion of solids at low
  temperatures}, {Advances in Physics} 29~(4) (1980) 609--730.
\newblock \href {https://doi.org/10.1080/00018738000101426}
  {\path{doi:10.1080/00018738000101426}}.

\bibitem{whit1993}
G.~K. White, {Solids: Thermal expansion and contraction}, {Contemporary
  Physics} 34~(4) (1993) 193--204.
\newblock \href {https://doi.org/10.1080/00107519308213818}
  {\path{doi:10.1080/00107519308213818}}.

\bibitem{repp2016b}
A.~von Reppert, R.~M. Sarhan, F.~Stete, J.~Pudell, N.~{Del Fatti}, A.~Crut,
  J.~Koetz, F.~Liebig, C.~Prietzel, M.~Bargheer, {Watching the Vibration and
  Cooling of Ultrathin Gold Nanotriangles by Ultrafast X-ray Diffraction}, {The
  Journal of Physical Chemistry C} 120~(50) (2016) 28894--28899.
\newblock \href {https://doi.org/10.1021/acs.jpcc.6b11651}
  {\path{doi:10.1021/acs.jpcc.6b11651}}.

\bibitem{whit1962}
G.~K. White, {Thermal expansion at low temperatures--- V. Dilute alloys of
  manganese in copper}, {Journal of Physics and Chemistry of Solids} 23~(1-2)
  (1962) 169--171.
\newblock \href {https://doi.org/10.1016/0022-3697(62)90077-X}
  {\path{doi:10.1016/0022-3697(62)90077-X}}.

\bibitem{pytt1965}
E.~Pytte, {Spin-phonon interactions in a Heisenberg ferromagnet}, {Annals of
  Physics} 32~(3) (1965) 377--403.
\newblock \href {https://doi.org/10.1016/0003-4916(65)90139-9}
  {\path{doi:10.1016/0003-4916(65)90139-9}}.

\bibitem{argy1967}
B.~E. Argyle, N.~Miyata, T.~D. Schultz, {Magnetoelastic Behavior of
  Single-Crystal Europium Oxide. I. Thermal Expansion Anomaly}, {Physical
  Review} 160~(2) (1967) 413--420.
\newblock \href {https://doi.org/10.1103/PhysRev.160.413}
  {\path{doi:10.1103/PhysRev.160.413}}.

\bibitem{reid2018}
A.~H. Reid, X.~Shen, P.~Maldonado, T.~Chase, E.~Jal, P.~W. Granitzka, K.~Carva,
  R.~K. Li, J.~Li, L.~Wu, T.~Vecchione, T.~Liu, Z.~Chen, D.~J. Higley,
  N.~Hartmann, R.~Coffee, J.~Wu, G.~L. Dakovski, W.~F. Schlotter, H.~Ohldag,
  Y.~K. Takahashi, V.~Mehta, O.~Hellwig, A.~Fry, Y.~Zhu, J.~Cao, E.~E.
  Fullerton, J.~St{\"o}hr, P.~M. Oppeneer, X.~J. Wang, H.~A. D{\"u}rr, {Beyond
  a phenomenological description of magnetostriction}, {Nature Communications}
  9~(1) (2018) 388.
\newblock \href {https://doi.org/10.1038/s41467-017-02730-7}
  {\path{doi:10.1038/s41467-017-02730-7}}.

\bibitem{repp2020b}
A.~von Reppert, L.~Willig, J.-E. Pudell, S.~P. Zeuschner, G.~Sellge, F.~Ganss,
  O.~Hellwig, J.~A. Arregi, V.~{Uhl$\backslash$v r}, A.~Crut, M.~Bargheer,
  {Spin stress contribution to the lattice dynamics of FePt}, {Science
  Advances} 6~(28) (2020).
\newblock \href {https://doi.org/10.1126/sciadv.aba1142}
  {\path{doi:10.1126/sciadv.aba1142}}.

\bibitem{matt2021}
M.~Mattern, J.-E. Pudell, G.~Laskin, A.~von Reppert, M.~Bargheer, {Analysis of
  the temperature- and fluence-dependent magnetic stress in laser-excited SrRuO
  3}, {Structural dynamics (Melville, N.Y.)} 8~(2) (2021) 024302.
\newblock \href {https://doi.org/10.1063/4.0000072}
  {\path{doi:10.1063/4.0000072}}.

\bibitem{ashc2012}
N.~W. Ashcroft, N.~D. Mermin, {Solid state physics}, repr Edition, {Brooks/Cole
  Thomson Learning}, South Melbourne, 2012.

\bibitem{nix1941}
F.~C. Nix, D.~MacNair, {The Thermal Expansion of Pure Metals: Copper, Gold,
  Aluminum, Nickel, and Iron}, {Physical Review} 60~(8) (1941) 597--605.
\newblock \href {https://doi.org/10.1103/PhysRev.60.597}
  {\path{doi:10.1103/PhysRev.60.597}}.

\bibitem{nick1971}
R.~M. Nicklow, N.~Wakabayashi, P.~R. Vijayaraghavan, {Lattice Dynamics of
  Holmium}, {Physical Review B} 3~(4) (1971) 1229--1234.
\newblock \href {https://doi.org/10.1103/PhysRevB.3.1229}
  {\path{doi:10.1103/PhysRevB.3.1229}}.

\bibitem{barr2005}
G.~D. Barrera, J.~A.~O. Bruno, T.~H.~K. Barron, N.~L. Allan, {Negative thermal
  expansion}, {Journal of Physics: Condensed Matter} 17~(4) (2005) R217--R252.
\newblock \href {https://doi.org/10.1088/0953-8984/17/4/R03}
  {\path{doi:10.1088/0953-8984/17/4/R03}}.

\bibitem{repp2016}
A.~von Reppert, J.~Pudell, A.~Koc, M.~Reinhardt, W.~Leitenberger, K.~Dumesnil,
  F.~Zamponi, M.~Bargheer, {Persistent nonequilibrium dynamics of the thermal
  energies in the spin and phonon systems of an antiferromagnet}, {Structural
  dynamics (Melville, N.Y.)} 3~(5) (2016) 054302.
\newblock \href {https://doi.org/10.1063/1.4961253}
  {\path{doi:10.1063/1.4961253}}.

\bibitem{mald2020}
P.~Maldonado, T.~Chase, A.~H. Reid, X.~Shen, R.~K. Li, K.~Carva, T.~Payer,
  M.~{Horn von Hoegen}, K.~Sokolowski-Tinten, X.~J. Wang, P.~M. Oppeneer, H.~A.
  D{\"u}rr, {Tracking the ultrafast nonequilibrium energy flow between
  electronic and lattice degrees of freedom in crystalline nickel}, {Physical
  Review B} 101~(10) (2020).
\newblock \href {https://doi.org/10.1103/PhysRevB.101.100302}
  {\path{doi:10.1103/PhysRevB.101.100302}}.

\bibitem{ritz2020}
U.~Ritzmann, P.~M. Oppeneer, P.~Maldonado, {Theory of out-of-equilibrium
  electron and phonon dynamics in metals after femtosecond laser excitation},
  {Physical Review B} 102~(21) (2020).
\newblock \href {https://doi.org/10.1103/PhysRevB.102.214305}
  {\path{doi:10.1103/PhysRevB.102.214305}}.

\bibitem{boja2015}
A.~Bojahr, M.~Gohlke, W.~Leitenberger, J.~Pudell, M.~Reinhardt, A.~von Reppert,
  M.~Roessle, M.~Sander, P.~Gaal, M.~Bargheer, {Second Harmonic Generation of
  Nanoscale Phonon Wave Packets}, {Physical review letters} 115~(19) (2015)
  195502.
\newblock \href {https://doi.org/10.1103/PhysRevLett.115.195502}
  {\path{doi:10.1103/PhysRevLett.115.195502}}.

\bibitem{schi2014}
D.~Schick, A.~Bojahr, M.~Herzog, R.~Shayduk, C.~von {Korff Schmising},
  M.~Bargheer, {udkm1Dsim---A simulation toolkit for 1D ultrafast dynamics in
  condensed matter}, {Computer Physics Communications} 185~(2) (2014) 651--660.
\newblock \href {https://doi.org/10.1016/j. cpc.2013.10.009}
  {\path{doi:10.1016/j. cpc.2013.10.009}}.

\bibitem{schi2021}
D.~Schick, {udkm1Dsim -- a Python toolbox for simulating 1D ultrafast dynamics
  in condensed matter}, {Computer Physics Communications} 266 (2021) 108031.
\newblock \href {https://doi.org/10.1016/j. cpc.2021.108031}
  {\path{doi:10.1016/j. cpc.2021.108031}}.

\bibitem{herz2012b}
M.~Herzog, D.~Schick, P.~Gaal, R.~Shayduk, C.~{Korff Schmising}, M.~Bargheer,
  {Analysis of ultrafast X-ray diffraction data in a linear-chain model of the
  lattice dynamics}, {Applied Physics A} 106~(3) (2012) 489--499.
\newblock \href {https://doi.org/10.1007/s00339-011-6719-z}
  {\path{doi:10.1007/s00339-011-6719-z}}.

\bibitem{chan1997}
Y.~M. Chang, L.~Xu, H.~W.~K. Tom, {Observation of Coherent Surface Optical
  Phonon Oscillations by Time-Resolved Surface Second-Harmonic Generation},
  {Physical Review Letters} 78~(24) (1997) 4649--4652.
\newblock \href {https://doi.org/10.1103/PhysRevLett.78.4649}
  {\path{doi:10.1103/PhysRevLett.78.4649}}.

\bibitem{meln2003}
A.~Melnikov, I.~Radu, U.~Bovensiepen, O.~Krupin, K.~Starke, E.~Matthias,
  M.~Wolf,
  \href{http://link.aps.org/doi/10.1103/PhysRevLett.91.227403}{{Coherent
  Optical Phonons and Parametrically Coupled Magnons Induced by Femtosecond
  Laser Excitation of the Gd(0001) Surface}}, {Physical Review Letters} 91~(22)
  (2003) 227403.
\newblock \href {https://doi.org/10.1103/physrevlett.91.227403}
  {\path{doi:10.1103/physrevlett.91.227403}}.
\newline\urlprefix\url{http://link.aps.org/doi/10.1103/PhysRevLett.91.227403}

\bibitem{high2007}
M.~Highland, B.~C. Gundrum, Y.~K. Koh, R.~S. Averback, D.~G. Cahill, V.~C.
  Elarde, J.~J. Coleman, D.~A. Walko, E.~C. Landahl,
  \href{https://link.aps.org/doi/10.1103/PhysRevB.76.075337}{{Ballistic-phonon
  heat conduction at the nanoscale as revealed by time-resolved x-ray
  diffraction and time-domain thermoreflectance}}, {Physical Review B} 76~(7)
  (2007) 075337.
\newblock \href {https://doi.org/10.1103/PhysRevB.76.075337}
  {\path{doi:10.1103/PhysRevB.76.075337}}.
\newline\urlprefix\url{https://link.aps.org/doi/10.1103/PhysRevB.76.075337}

\bibitem{mari2012}
S.~O. Mariager, F.~Pressacco, G.~Ingold, A.~Caviezel, E.~M{\"o}hr-Vorobeva,
  P.~Beaud, S.~Johnson, C.~Milne, E.~Mancini, S.~Moyerman, et~al., Structural
  and magnetic dynamics of a laser induced phase transition in ferh, Physical
  Review Letters 108~(8) (2012) 087201.
\newblock \href {https://doi.org/10.1103/PhysRevLett.108.087201}
  {\path{doi:10.1103/PhysRevLett.108.087201}}.

\bibitem{mari2014}
S.~O. Mariager, C.~Dornes, J.~A. Johnson, A.~Ferrer, S.~Gr{\"u}bel, T.~Huber,
  A.~Caviezel, S.~L. Johnson, T.~Eichhorn, G.~Jakob, H.~J. Elmers, P.~Beaud,
  C.~Quitmann, G.~Ingold, {Structural and magnetic dynamics in the magnetic
  shape-memory alloy Ni2MnGa}, {Physical Review B - Condensed Matter and
  Materials Physics} 90~(16) (2014).
\newblock \href {https://doi.org/10.1103/PhysRevB.90.161103}
  {\path{doi:10.1103/PhysRevB.90.161103}}.

\bibitem{matt2022}
M.~Mattern, A.~von Reppert, S.~P. Zeuschner, J.-E. Pudell, F.~K{\"u}hne,
  D.~Diesing, M.~Herzog, M.~Bargheer, {Electronic energy transport in nanoscale
  Au/Fe hetero-structures in the perspective of ultrafast lattice dynamics},
  {Applied Physics Letters} 120~(9) (2022) 092401.
\newblock \href {https://doi.org/10.1063/5.0080378}
  {\path{doi:10.1063/5.0080378}}.

\bibitem{parp2021}
T.~Parpiiev, A.~Hillion, V.~Vlasov, V.~Gusev, K.~Dumesnil, T.~Hauet,
  S.~Andrieu, A.~Anane, T.~Pezeril, {Ultrafast strain excitation in highly
  magnetostrictive terfenol: Experiments and theory}, {Physical Review B}
  104~(22) (2021).
\newblock \href {https://doi.org/10.1103/PhysRevB.104.224426}
  {\path{doi:10.1103/PhysRevB.104.224426}}.

\bibitem{tsun2004}
Y.~Tsunoda, H.~Kobayashi, {Temperature variation of the tetragonality in
  ordered PtFe alloy}, {Journal of Magnetism and Magnetic Materials} 272-276
  (2004) 776--777.
\newblock \href {https://doi.org/10.1016/j. jmmm.2003.11.263}
  {\path{doi:10.1016/j. jmmm.2003.11.263}}.

\bibitem{rasm2005}
P.~Rasmussen, X.~Rui, J.~E. Shield, {Texture formation in FePt thin films via
  thermal stress management}, {Applied Physics Letters} 86~(19) (2005) 191915.
\newblock \href {https://doi.org/10.1063/1.1924889}
  {\path{doi:10.1063/1.1924889}}.

\bibitem{repp2018}
A.~von Reppert, L.~Willig, J.-E. Pudell, M.~R{\"o}ssle, W.~Leitenberger,
  M.~Herzog, F.~Ganss, O.~Hellwig, M.~Bargheer, {Ultrafast laser generated
  strain in granular and continuous FePt thin films}, {Applied Physics Letters}
  113~(12) (2018) 123101.
\newblock \href {https://doi.org/10.1063/1.5050234}
  {\path{doi:10.1063/1.5050234}}.

\bibitem{pude2018}
J.~Pudell, A.~A. Maznev, M.~Herzog, M.~Kronseder, C.~H. Back, G.~Malinowski,
  A.~von Reppert, M.~Bargheer, {Layer specific observation of slow thermal
  equilibration in ultrathin metallic nanostructures by femtosecond X-ray
  diffraction}, {Nature Communications} 9~(1) (2018) 3335.
\newblock \href {https://doi.org/10.1038/s41467-018-05693-5}
  {\path{doi:10.1038/s41467-018-05693-5}}.

\bibitem{lin2008}
Z.~Lin, L.~V. Zhigilei, V.~Celli, {Electron-phonon coupling and electron heat
  capacity of metals under conditions of strong electron-phonon
  nonequilibrium}, {Physical Review B} 77~(7) (2008) 776.
\newblock \href {https://doi.org/10.1103/PhysRevB.77.075133}
  {\path{doi:10.1103/PhysRevB.77.075133}}.

\bibitem{khor2014}
A.~R. Khorsand, M.~Savoini, A.~Kirilyuk, T.~Rasing, {Optical excitation of thin
  magnetic layers in multilayer structures}, {Nature materials} 13~(2) (2014)
  101--102.
\newblock \href {https://doi.org/10.1038/nmat3850}
  {\path{doi:10.1038/nmat3850}}.

\bibitem{esch2014}
A.~Eschenlohr, M.~Battiato, P.~Maldonado, N.~Pontius, T.~Kachel, K.~Holldack,
  R.~Mitzner, A.~F{\"{o}}hlisch, P.~M. Oppeneer, C.~Stamm, {Reply to 'Optical
  excitation of thin magnetic layers in multilayer structures'}, Nature
  Materials 13~(2) (2014) 102--103.
\newblock \href {https://doi.org/10.1038/nmat3851}
  {\path{doi:10.1038/nmat3851}}.

\bibitem{yaku2019}
D.~I. Yakubovsky, Y.~V. Stebunov, R.~V. Kirtaev, G.~A. Ermolaev, M.~S. Mironov,
  S.~M. Novikov, A.~V. Arsenin, V.~S. Volkov, {Ultrathin and Ultrasmooth Gold
  Films on Monolayer MoS 2}, {Advanced Materials Interfaces} 6~(13) (2019)
  1900196.
\newblock \href {https://doi.org/10.1002/admi.201900196}
  {\path{doi:10.1002/admi.201900196}}.

\bibitem{wang2012}
W.~Wang, D.~G. Cahill, {Limits to thermal transport in nanoscale metal bilayers
  due to weak electron-phonon coupling in Au and Cu}, {Physical review letters}
  109~(17) (2012) 175503.
\newblock \href {https://doi.org/10.1103/PhysRevLett.109.175503}
  {\path{doi:10.1103/PhysRevLett.109.175503}}.

\bibitem{choi2014}
G.-M. Choi, B.-C. Min, K.-J. Lee, D.~G. Cahill, {Spin current generated by
  thermally driven ultrafast demagnetization}, {Nature communications} 5 (2014)
  4334.
\newblock \href {https://doi.org/10.1038/ncomms5334}
  {\path{doi:10.1038/ncomms5334}}.

\bibitem{shin2020}
Y.~Shin, Y.~Liu, M.~Vomir, J.-W. Kim, {Higher-order acoustic phonon
  oscillations in Au nanoparticles controlled by a sequence of ultrashort
  strain pulses generated by superdiffusive hot electrons}, {Physical Review B}
  101~(2) (2020).
\newblock \href {https://doi.org/10.1103/PhysRevB.101.020302}
  {\path{doi:10.1103/PhysRevB.101.020302}}.

\bibitem{dume1996}
Dumesnil, Dufour, Mangin, Marchal, Hennion, {Magnetic structure of dysprosium
  in epitaxial Dy films and in Dy/Er superlattices}, {Physical review. B,
  Condensed matter} 54~(9) (1996) 6407--6420.
\newblock \href {https://doi.org/10.1103/PhysRevB.54.6407}
  {\path{doi:10.1103/PhysRevB.54.6407}}.

\bibitem{koc2017}
A.~Koc, M.~Reinhardt, A.~von Reppert, M.~R{\"o}ssle, W.~Leitenberger,
  K.~Dumesnil, P.~Gaal, F.~Zamponi, M.~Bargheer, {Ultrafast x-ray diffraction
  thermometry measures the influence of spin excitations on the heat transport
  through nanolayers}, {Physical Review B} 96~(1) (2017) 429.
\newblock \href {https://doi.org/10.1103/PhysRevB.96.014306}
  {\path{doi:10.1103/PhysRevB.96.014306}}.

\bibitem{pech1996}
V.~K. Pecharsky, K.~A. Gschneidner, D.~Fort, {Superheating and other unusual
  observations regarding the first order phase transition in Dy}, {Scripta
  Materialia} 35~(7) (1996) 843--848.
\newblock \href {https://doi.org/10.1016/1359-6462(96)00225-4}
  {\path{doi:10.1016/1359-6462(96)00225-4}}.

\bibitem{darn1963}
F.~J. Darnell, \href{http://link.aps.org/doi/10.1103/PhysRev.132.1098}{{Lattice
  Parameters of Terbium and Erbium at Low Temperatures}}, {Physical Review}
  132~(3) (1963) 1098--1100.
\newblock \href {https://doi.org/10.1103/physrev.132.1098}
  {\path{doi:10.1103/physrev.132.1098}}.
\newline\urlprefix\url{http://link.aps.org/doi/10.1103/PhysRev.132.1098}

\bibitem{darn1963b}
F.~J. Darnell, {Lattice Parameters of Terbium and Erbium at Low Temperatures},
  {Physical Review} 132~(3) (1963) 1098--1100.
\newblock \href {https://doi.org/10.1103/physrev.132.1098}
  {\path{doi:10.1103/physrev.132.1098}}.

\bibitem{darn1963c}
F.~J. Darnell, {Temperature Dependence of Lattice Parameters for Gd, Dy, and
  Ho}, {Physical Review} 130~(5) (1963) 1825--1828.
\newblock \href {https://doi.org/10.1103/PhysRev.130.1825}
  {\path{doi:10.1103/PhysRev.130.1825}}.

\bibitem{bula1996}
A.~S. Bulatov, V.~F. Dolzenko, A.~V. Kornietz, Temperature dependences of
  thermal expansion and exchange magnetostriction of holmium and dysprosium
  single crystals, Czechoslovak Journal of Physics 46~(4) (1996) 2119--2120.

\bibitem{cher2008}
A.~Chernyshov, Y.~Mudryk, V.~Pecharsky, K.~Gschneidner~Jr, Temperature and
  magnetic field-dependent x-ray powder diffraction study of dysprosium,
  Physical Review B 77~(9) (2008) 094132.

\bibitem{dove2016}
M.~T. Dove, H.~Fang, {Negative thermal expansion and associated anomalous
  physical properties: review of the lattice dynamics theoretical foundation},
  {Reports on Progress in Physics} 79~(6) (2016) 066503.
\newblock \href {https://doi.org/10.1088/0034-4885/79/6/066503}
  {\path{doi:10.1088/0034-4885/79/6/066503}}.

\bibitem{matt2023}
M.~Mattern, J.-E. Pudell, K.~Dumesnil, A.~von Reppert, M.~Bargheer, {Towards
  shaping picosecond strain pulses via magnetostrictive transducers},
  {Photoacoustics} 30 (2023) 100463.
\newblock \href {https://doi.org/10.1016/j. pacs.2023.100463}
  {\path{doi:10.1016/j. pacs.2023.100463}}.

\bibitem{juve2020}
V.~Juv{\'e}, R.~Gu, S.~Gable, T.~Maroutian, G.~Vaudel, S.~Matzen, N.~Chigarev,
  S.~Raetz, V.~E. Gusev, M.~Viret, A.~Jarnac, C.~Laulh{\'e}, A.~A. Maznev,
  B.~Dkhil, P.~Ruello, {Ultrafast light-induced shear strain probed by
  time-resolved x-ray diffraction: Multiferroic BiFeO3 as a case study},
  {Physical Review B} 102~(22) (2020).
\newblock \href {https://doi.org/10.1103/PhysRevB.102.220303}
  {\path{doi:10.1103/PhysRevB.102.220303}}.

\bibitem{soko2003}
K.~Sokolowski-Tinten, C.~Blome, J.~Blums, A.~Cavalleri, C.~Dietrich,
  A.~Tarasevitch, I.~Uschmann, E.~F{\"o}rster, M.~Kammler, M.~Horn-von Hoegen,
  D.~von~der Linde, {Femtosecond X-ray measurement of coherent lattice
  vibrations near the Lindemann stability limit}, {Nature} 422~(6929) (2003)
  287--289.
\newblock \href {https://doi.org/10.1038/nature01490}
  {\path{doi:10.1038/nature01490}}.

\bibitem{john2009b}
S.~L. Johnson, E.~Vorobeva, P.~Beaud, C.~J. Milne, G.~Ingold, {Full
  reconstruction of a crystal unit cell structure during coherent femtosecond
  motion}, {Physical review letters} 103~(20) (2009) 205501.
\newblock \href {https://doi.org/10.1103/PhysRevLett.103.205501}
  {\path{doi:10.1103/PhysRevLett.103.205501}}.

\bibitem{kozi2017}
M.~Kozina, M.~Pancaldi, C.~Bernhard, T.~{van Driel}, J.~M. Glownia, P.~Marsik,
  M.~Radovic, C.~A.~F. Vaz, D.~Zhu, S.~Bonetti, U.~Staub, M.~C. Hoffmann,
  {Local terahertz field enhancement for time-resolved x-ray diffraction},
  {Applied Physics Letters} 110~(8) (2017) 081106.
\newblock \href {https://doi.org/10.1063/1.4977088}
  {\path{doi:10.1063/1.4977088}}.

\bibitem{jong2013}
S.~de~Jong, R.~Kukreja, C.~Trabant, N.~Pontius, C.~F. Chang, T.~Kachel,
  M.~Beye, F.~Sorgenfrei, C.~H. Back, B.~Br{\"a}uer, W.~F. Schlotter, J.~J.
  Turner, O.~Krupin, M.~Doehler, D.~Zhu, M.~A. Hossain, A.~O. Scherz,
  D.~Fausti, F.~Novelli, M.~Esposito, W.~S. Lee, Y.~D. Chuang, D.~H. Lu, R.~G.
  Moore, M.~Yi, M.~Trigo, P.~Kirchmann, L.~Pathey, M.~S. Golden, M.~Buchholz,
  P.~Metcalf, F.~Parmigiani, W.~Wurth, A.~F{\"o}hlisch,
  C.~Sch{\"u}{\ss}ler-Langeheine, H.~A. D{\"u}rr, {Speed limit of the
  insulator-metal transition in magnetite}, {Nature materials} 12~(10) (2013)
  882--886.
\newblock \href {https://doi.org/10.1038/nmat3718}
  {\path{doi:10.1038/nmat3718}}.

\bibitem{mogu2020}
I.~A. Mogunov, S.~Lysenko, F.~Fern{\'a}ndez, A.~R{\'u}a, A.~V. Muratov, A.~J.
  Kent, A.~M. Kalashnikova, A.~V. Akimov, {Photoelasticity of VO2 nanolayers in
  insulating and metallic phases studied by picosecond ultrasonics}, {Physical
  Review Materials} 4~(12) (2020).
\newblock \href {https://doi.org/10.1103/PhysRevMaterials.4.125201}
  {\path{doi:10.1103/PhysRevMaterials.4.125201}}.

\bibitem{schi2013}
D.~Schick, A.~Bojahr, M.~Herzog, P.~Gaal, I.~Vrejoiu, M.~Bargheer, {Following
  strain-induced mosaicity changes of ferroelectric thin films by ultrafast
  reciprocal space mapping}, {Physical review letters} 110~(9) (2013) 095502.
\newblock \href {https://doi.org/10.1103/PhysRevLett.110.095502}
  {\path{doi:10.1103/PhysRevLett.110.095502}}.

\bibitem{chen2016}
F.~Chen, Y.~Zhu, S.~Liu, Y.~Qi, H.~Y. Hwang, N.~C. Brandt, J.~Lu, F.~Quirin,
  H.~Enquist, P.~Zalden, T.~Hu, J.~Goodfellow, M.-J. Sher, M.~C. Hoffmann,
  D.~Zhu, H.~Lemke, J.~Glownia, M.~Chollet, A.~R. Damodaran, J.~Park, Z.~Cai,
  I.~W. Jung, M.~J. Highland, D.~A. Walko, J.~W. Freeland, P.~G. Evans,
  A.~Vailionis, J.~Larsson, K.~A. Nelson, A.~M. Rappe, K.~Sokolowski-Tinten,
  L.~W. Martin, H.~Wen, A.~M. Lindenberg, {Ultrafast terahertz-field-driven
  ionic response in ferroelectric BaTiO3}, {Physical Review B} 94~(18) (2016)
  1--6.
\newblock \href {https://doi.org/10.1103/PhysRevB.94.180104}
  {\path{doi:10.1103/PhysRevB.94.180104}}.

\bibitem{john2010}
S.~L. Johnson, P.~Beaud, E.~Vorobeva, C.~J. Milne, E.~D. Murray, S.~Fahy,
  G.~Ingold, {Non-equilibrium phonon dynamics studied by grazing-incidence
  femtosecond X-ray crystallography}, {Acta crystallographica. Section A,
  Foundations of crystallography} 66~(Pt 2) (2010) 157--167.
\newblock \href {https://doi.org/10.1107/S0108767309053859}
  {\path{doi:10.1107/S0108767309053859}}.

\bibitem{nusk2011}
R.~N{\"u}ske, A.~Jurgilaitis, H.~Enquist, S.~D. Farahani, J.~Gaudin, L.~Guerin,
  M.~Harb, C.~v.~K. Schmising, M.~St{\"o}rmer, M.~Wulff, J.~Larsson,
  {Picosecond time-resolved x-ray refectivity of a laser-heated amorphous
  carbon film}, {Applied Physics Letters} 98~(10) (2011) 101909.
\newblock \href {https://doi.org/10.1063/1.3562967}
  {\path{doi:10.1063/1.3562967}}.

\bibitem{jal2017}
E.~Jal, V.~L{\'o}pez-Flores, N.~Pontius, T.~Fert{\'e}, N.~Bergeard, C.~Boeglin,
  B.~Vodungbo, J.~L{\"u}ning, N.~Jaouen, {Structural dynamics during
  laser-induced ultrafast demagnetization}, {Physical Review B} 95~(18) (2017).
\newblock \href {https://doi.org/10.1103/PhysRevB.95.184422}
  {\path{doi:10.1103/PhysRevB.95.184422}}.

\bibitem{robi2009}
I.~Robinson, R.~Harder, {Coherent X-ray diffraction imaging of strain at the
  nanoscale}, {Nature materials} 8~(4) (2009) 291--298.
\newblock \href {https://doi.org/10.1038/nmat2400}
  {\path{doi:10.1038/nmat2400}}.

\bibitem{newt2014}
M.~C. Newton, M.~Sao, Y.~Fujisawa, R.~Onitsuka, T.~Kawaguchi, K.~Tokuda,
  T.~Sato, T.~Togashi, M.~Yabashi, T.~Ishikawa, T.~Ichitsubo, E.~Matsubara,
  Y.~Tanaka, Y.~Nishino, {Time-resolved coherent diffraction of ultrafast
  structural dynamics in a single nanowire}, {Nano letters} 14~(5) (2014)
  2413--2418.
\newblock \href {https://doi.org/10.1021/nl500072d}
  {\path{doi:10.1021/nl500072d}}.

\bibitem{hols2022}
T.~S. Holstad, L.~E. Dresselhaus-Marais, T.~M. R{\ae}der, B.~Kozioziemski,
  T.~{van Driel}, M.~Seaberg, E.~Folsom, J.~H. Eggert, E.~B. Knudsen, M.~M.
  Nielsen, H.~Simons, K.~Haldrup, H.~F. Poulsen,
  \href{http://arxiv.org/pdf/2211.01042v2}{{Real-time imaging of acoustic waves
  in bulk materials with X-ray microscopy}} (2022).
\newline\urlprefix\url{http://arxiv.org/pdf/2211.01042v2}

\bibitem{als2011}
J.~Als-Nielsen, {Des McMorrow}, {Elements of modern X-ray physics}, second
  edition Edition, Wiley, Chichester, 2011.

\bibitem{warr1990}
B.~E. Warren, {X-ray diffraction}, dover ed. is an unabridged and corr. republ
  Edition, Dover, New York, NY, 1990.

\bibitem{auth2008}
A.~Authier, {Dynamical theory of X-ray diffraction}, rev. ed., reprinted.
  Edition, Vol.~11 of {IUCr monographs on crystallography}, {Oxford Univ.
  Press}, Oxford, 2008.

\bibitem{noya1995}
I.~Noyan, T.~Huang, B.~York, Residual stress/strain analysis in thin films by
  x-ray diffraction, Critical Reviews in Solid State and Material Sciences
  20~(2) (1995) 125--177.

\bibitem{holy1999}
V.~Hol{\'y}, U.~Pietsch, T.~Baumbach, {High-Resolution X-Ray Scattering from
  Thin Films and Multilayers}, Vol. 149 of {Springer Tracts in Modern Physics},
  Springer, Berlin and Heidelberg, 1999.
\newblock \href {https://doi.org/10.1007/BFb0109385}
  {\path{doi:10.1007/BFb0109385}}.

\bibitem{barg2005}
M.~Bargheer, N.~Zhavoronkov, R.~Bruch, H.~Legall, H.~Stiel, M.~Woerner,
  T.~Elsaesser, {Comparison of focusing optics for femtosecond X-ray
  diffraction}, {Applied Physics B} 80~(6) (2005) 715--719.
\newblock \href {https://doi.org/10.1007/s00340-005-1792-7}
  {\path{doi:10.1007/s00340-005-1792-7}}.

\bibitem{hell2007}
J.~R. Helliwell, P.~M. Rentzepis (Eds.), {Time-resolved diffraction}, reprint
  Edition, Vol.~2 of {Oxford series on synchrotron radiation}, {Clarendon
  Press}, Oxford, 2007.

\bibitem{rong2023}
E.~Rongione, O.~Gueckstock, M.~Mattern, O.~Gomonay, H.~Meer, C.~Schmitt,
  R.~Ramos, T.~Kikkawa, M.~Mi{\v{c}}ica, E.~Saitoh, J.~Sinova, H.~Jaffr{\`e}s,
  J.~Mangeney, S.~T.~B. Goennenwein, S.~Gepr{\"a}gs, T.~Kampfrath,
  M.~Kl{\"a}ui, M.~Bargheer, T.~S. Seifert, S.~Dhillon, R.~Lebrun, {Emission of
  coherent THz magnons in an antiferromagnetic insulator triggered by ultrafast
  spin-phonon interactions}, {Nature communications} 14~(1) (2023) 1818.
\newblock \href {https://doi.org/10.1038/s41467-023-37509-6}
  {\path{doi:10.1038/s41467-023-37509-6}}.

\end{thebibliography}
\biboptions{sort&compress}
\end{document}